%% file: main.tex
\documentclass[journal]{IEEEtran}
\IEEEoverridecommandlockouts
\usepackage{cite}
\usepackage{amsmath,amssymb,amsfonts}
\usepackage{algorithmic}
\usepackage{graphicx}
\usepackage{textcomp}
\usepackage{xcolor}
\def\BibTeX{{\rm B\kern-.05em{\sc i\kern-.025em b}\kern-.08em
    T\kern-.1667em\lower.7ex\hbox{E}\kern-.125emX}}

\usepackage[acronyms,nonumberlist,nopostdot,nomain,nogroupskip]{glossaries}
\input{acronyms.tex}
\usepackage{hyperref}

\usepackage[font=small]{caption}
\usepackage[font=footnotesize]{subcaption}
\usepackage{booktabs} 
\usepackage[utf8]{inputenc}
\usepackage{pgfplots}
\DeclareUnicodeCharacter{2212}{−}
\usepgfplotslibrary{groupplots,dateplot}
\usetikzlibrary{patterns,shapes.arrows}
\pgfplotsset{compat=newest}

\usepackage{notation}
\usepackage{multirow}

\usepackage{tabularx}
\usepackage{subcaption}

\usepackage[most]{tcolorbox}
\tcbuselibrary{theorems}

\newtcbtheorem{Definitions}{Summary}%
{colframe=black,colback=white,fonttitle=\bfseries}{}

\usepackage{flushend}
\usetikzlibrary{positioning}

\title{A Joint Reinforcement Learning Scheduling and Compression Framework for Teleoperated Driving}

\author{Giacomo Avanzi,~\IEEEmembership{Student Member,~IEEE}, Marco~Giordani,~\IEEEmembership{Senior Member, IEEE}\\  Michele Zorzi,~\IEEEmembership{Fellow, IEEE}\\
\vspace{0.33cm}
\emph{(Invited Paper)}
\thanks{The authors are with the Department of Information Engineering, University of Padova. Padova, Italy. E-mail: \{giacomo.avanzi, marco.giordani michele.zorzi\}@dei.unipd.it.\\
A preliminary version of this work was presented at the IEEE Vehicular Networking Conference 2025~\cite{avanzi2025multi}.}% <-this % stops a space}
}

\begin{document}
\maketitle

\begin{abstract}
\Gls{td} is envisioned as a key application of future \gls{6g} networks. In this paradigm, connected vehicles transmit sensor-perception data to a remote (software) driver, which returns driving control commands to enhance traffic efficiency and road safety.
This scenario imposes to maintain reliable and low-latency communication between the vehicle and the remote driver.
To this aim, a promising solution is \gls{pqos}, which provides mechanisms to estimate possible \gls{qos} degradation, and trigger timely network corrective actions accordingly. In particular, \gls{rl} agents can be trained to identify the optimal \gls{pqos} configuration. 
In this paper, we develop and implement two integrated \gls{rl} agents that jointly determine (i) the optimal compression configuration for \gls{td} sensor data to balance the trade-off between transmission efficiency and data quality, and (ii) the optimal scheduling configuration to minimize the end-to-end latency by allocating radio resources according to different priority levels. 
We prove via full-stack ns-3 simulations that our integrated agents can deliver superior performance than any standalone model that only optimizes either compression or scheduling, especially in constrained or congested networks.
While these agents can be deployed using either centralized or decentralized learning, we further propose a new meta-learning agent that dynamically selects the most appropriate strategy between the two based on current network conditions and application requirements. 
\end{abstract}

\glsresetall


\glsresetall

\begin{tikzpicture}[remember picture,overlay]
\node[anchor=north,yshift=-10pt] at (current page.north) {\parbox{\dimexpr\textwidth-\fboxsep-\fboxrule\relax}{
\centering\footnotesize This paper is currently under review by IEEE Transactions on Communications.}};
\end{tikzpicture}

\section{Introduction}
\label{sec:intro}

\Gls{6g} networks are being designed to support an increasingly data-centric, data-dependent, and automated society~\cite{giordani6g}.
In particular, the research community identifies \gls{td} as a key \gls{6g} application to enhance traffic efficiency and road safety.
In a \gls{td} scenario, a connected vehicle continuously acquires and transmits multimodal sensor data, including videocamera, \gls{lidar}, and radar measurements, to a remote (typically software) driver via \gls{v2x} communication. The remote driver processes these data to extract critical perception information, such as detecting other vehicles, pedestrians, lane markings, and traffic signals on the road, and generates optimal driving control commands accordingly.
To support this paradigm, the \gls{td} application must satisfy stringent communication requirements.
Notably, connected cars generate large volumes of data~\cite{lidar}, which can lead to network congestion.
At the same time, unanticipated channel degradation, which are likely to occur in dynamic vehicular networks, may pose critical safety risks if control commands from the remote driver are delayed or lost.
According to the \gls{3gpp} specifications, to ensure that both perception data and control commands are transmitted accurately and in a timely manner, the end-to-end latency for TD must be lower than 50 ms in both \gls{ul} and \gls{dl}, while reliability requirements range from 99\% to 99.999\% depending on the automation level~\cite{3GPP_22186}.

%Extreme connectivity represents a key technology in \gls{6g} networks, enabling \gls{urllc} with sub-millisecond latency and extreme reliability up to 99.99999\%~\cite{giordani6g}. 
%\gls{6g} will support machine area networks (e.g., car area networks) with ultra-fast and high reliable communication between hundreds of sensors. 

For this reason, \gls{pqos} was introduced as a mechanism to forecast and communicate potential \gls{qos} degradation in the network, and undertake proper countermeasures to react accordingly~\cite{boban2021predictive}. 
Initially, network dynamics were predicted using filter-based approaches such as Kalman filtering or linear regression. However, these methods rely on prior knowledge of the underlying network and control processes~\cite{kalman}, which is often not available in \gls{td} applications. Moreover, they cannot efficiently handle large-scale input data, and do not capture non-linear dependencies between input and output features.
Recently, \gls{ai} and \gls{ml} techniques have emerged as powerful tools for intelligent network optimization~\cite{AI, AI2} in various domains, including \gls{pqos}.
In particular, \gls{nn} models can be trained to predict the network behavior, and decide appropriate actions to preserve \gls{qos} based on network conditions, available resources, predicted mobility, and other observations \cite{BobanPrediction,QosPrediction,VerdonePrediction, QoSPredictor}.
However, \glspl{nn} require large labeled datasets for training, which are often difficult or costly to obtain in vehicular networks \cite{vehicleDataset}.
Moreover, training is time-consuming, and may not be feasible with the (generally limited) hardware resources available onboard vehicles \cite{vehicleLimitationNN}.
In addition, channel dynamics and mobility may require frequent model updates or retraining, which also introduce non-negligible~delays.

Another valid approach for \gls{pqos} is based on \gls{rl}, where the system learns a predictive and adaptive policy through continuous feedback (reward).
Unlike \glspl{nn}, \gls{rl} does not require a labeled dataset, and can iteratively adapt to time-varying network conditions with minimal retraining.
Specifically, \gls{rl} methods have shown impressive results for \gls{pqos} in \gls{td} scenarios. 
For example, in our previous works~\cite{mason,mason2025prata,ran-ai}, we proposed a simulation framework 
%based on \gls{dql} and \gls{ppo} 
to predict the optimal compression level for \gls{lidar} data to optimize the trade-off between the \gls{e2e} transmission latency and data quality in the event of network resource saturation or channel degradation. 
Specifically, we explored centralized and decentralized \gls{rl} solutions~\cite{bragato}, implemented at the \gls{gnb} or \gls{ue}, respectively,  as well as different \gls{rl} methods~\cite{bragato2024federated}, namely \gls{mab} (stateless), \gls{sarsa} (stateful on-policy), Q-Learning (stateful off-policy), and \gls{dsarsa} and \gls{ddql} (with \gls{nn} approximation). However, compression may inevitably degrade the quality of sensor data, and possibly compromise critical \gls{td} tasks
such as object detection and recognition.

An alternative strategy is to implement \gls{pqos} at the \gls{ran}. Specifically, the RL agent can be designed to
optimize radio parameters, such as the transmission power, the numerology, 
or the scheduler, based on the network conditions,
while preserving data quality.
For example, in ~\cite{avanzi2025multi} we proposed several \gls{marl} scheduling algorithms that allocate radio resources based on latency constraints and the available network capacity. 
However, the main drawback of this method is that it can only be executed at the \gls{gnb} via centralized learning, so it is not directly applicable at the \gls{ue} or edge level.

Based on the above introduction and discussion, in this work we provide the following novel contributions:
\begin{itemize}
    \item Compared to our previous work, where we optimized either data compression or resource allocation/scheduling independently, we now propose an innovative \gls{rl} framework that jointly optimizes both as a \gls{pqos} countermeasure. To do so, we design and implement two \gls{rl} agents. The first, called \acrfull{ca}, selects the optimal \gls{lidar}\footnote{In this paper we focus on LiDAR data, given that LiDAR sensors are widely used in TD applications given their ability to measure distance in different weather and lighting conditions. However, our simulation pipeline can be equivalently applied to different types of sensor-perception~data.} compression level to minimize the \gls{e2e} latency and, at the same time, maximize the quality of the received data, measured in terms of \gls{map}. The second, called \acrfull{sa}, decides how to allocate radio resources for each \gls{ue} based on a dynamic priority metric that accounts for both the \gls{e2e} latency and the available network capacity. The agents are trained either independently or cooperatively.
    \item We evaluate the proposed framework against standalone benchmarks that either select the compression and scheduling decisions statically, or optimize either of the two dimensions separately, as done in the prior literature. We consider key performance metrics such as the \gls{e2e} latency, the \gls{map} of the received data, and the probability of violating the latency requirements of \gls{td} applications. We also compare different \gls{rl} model implementations, i.e., centralized or decentralized (federated) learning.  All algorithms are evaluated via full-stack ns-3 simulations, one of the most advanced software tools for wireless networks, which ensures that our results are both realistic and accurate.  
    We find that 
    %impact of the channel quality and the latency requirements is not negligible. Specifically, 
    federated learning is the most convenient approach in low-quality channels, that is when data collection at the RAN, as required in centralized learning, is impractical. Moreover, we prove that the cooperation between the CA and the SA outperforms any standalone approach that optimizes either compression or scheduling separately, especially in the most constrained or congested environments.
    \item Based on the above results, we further propose a new meta-learning \gls{rl} agent that is able to dynamically select the optimal learning model (centralized or federated) based on real-time network conditions and the application requirements. 
    We compare different meta-learning agent implementations, including stateless, contextual, and stateful models.
    %Therefore, we provide a modular, flexible and responsive system with respect to application-specific requirements and channel conditions.
    We prove via ns-3 simulations that the adaptability features of the meta-learning agent outperforms any static baseline where the learning configuration is selected a priori, and does not depend on the underlying channel~condition. 
\end{itemize}

%Hence, our proposal aims to be compliant with the strict \gls{td} constraints on the latency and, at the same time, to maximize the \gls{qoe} level.
%Independently from the distribution of the learning model (i.e., centralized or decentralized), compression techniques can still be applied to effectively reduce the communication overhead and adapt to time-variant network conditions.
%Conversely, the intelligent resource allocation at the \gls{ran} level can significantly reduce the \gls{e2e} latency, but only in combination with a centralized learning-based compression paradigm at the gNB. 

The rest of the paper is organized as follows.
In Sec.~\ref{sec:soa} we review the state of the art.
In Sec.~\ref{sec:model} we describe our system model.
In Sec.~\ref{sec:agents} we present our RL agents.
In Sec.~\ref{sec:performance} we illustrate our main simulation results.
In Sec.~\ref{sec:conclusions} we conclude the paper with suggestions for future work.

\section{Related Work}
\label{sec:soa}
\gls{ai} is expected to be a fundamental component of \gls{6g} networks, especially to support autonomous network optimization and adaptive decision-making without manual intervention~\cite{11320975}. 
%Notably, \gls{ai} facilitates predictive and adaptive decision-making across a wide range of modern communications applications.
These capabilities are especially promising for \gls{v2x} \gls{td} networks characterized by stringent \gls{urllc} requirements, to dynamically optimize traffic congestion, vehicle navigation, and platooning~\cite{hakeem2025advancing}. 
A common feature of these applications is the need to process and/or transmit huge amounts of sensor-perception data, which may cause delay and packet~loss. 

Along these lines, \gls{pqos} exploits \gls{ai} to predict the network \gls{qos} at the \gls{ran} level, and optimize protocols accordingly. In \cite{BobanPrediction,QosPrediction,VerdonePrediction}, traditional AI/\gls{ml} models (e.g., linear regression, \gls{dnn}, and Random Forest) were used to accurately predict several V2X metrics, such as the uplink throughput and the in-time packet reception probability, under diverse and representative traffic conditions. 
Similarly, Torres-Figueroa \emph{et al.}~\cite{QoSPredictor} proposed to forecast the \gls{e2e} latency by training a complex \gls{rnn} with \gls{lstm}, as well as other basic \gls{ml} algorithms. 
The authors in~\cite{PQoSFramework1} and \cite{PQoSFramework2} also proposed a latency and throughput prediction framework, respectively, for supporting delay-critical \gls{v2x} applications based on \gls{lstm}.

As discussed in Sec.~\ref{sec:intro}, \gls{pqos} can be optimized in the \gls{ran}, specifically at the scheduling level.
In fact, current 5G schedulers, such as \gls{rr}, Proportional Fair, and Earliest Deadline First, were not originally designed to support time-sensitive traffic.
In contrast, (multi-agent) \gls{drl} methods can adapt resource scheduling to the network environment~\cite{drl}, based on, for example, target \gls{qos} requirements or per-UE priorities, as proposed in \cite{drl1}.
A similar strategy, also based on \gls{drl}, was proposed in \cite{drl2} to support \gls{urllc}.

Alternatively, \gls{pqos} can be optimized through efficient data compression at the application layer, immediately after data generation and before transmission. 
%In \gls{td} scenarios, a rich suite of heterogeneous sensors such as radars, cameras and \glspl{lidar}, continuously captures the surrounding environment to build a reliable and accurate digital representation of the real world. Among these, \glspl{lidar} generate large volumes of data points (i.e., point clouds), which require high transmission data rates and pose significant challenges to meet stringent \gls{qos} constraints in \gls{v2x} networks. 
For example, \cite{mason} proposed an \gls{rl} agent based on \gls{dql} and \gls{ppo} 
that dynamically selects the optimal compression level for \gls{lidar} perceptions (combining both semantic segmentation using RangeNet++~\cite{rangenet} and point-cloud compression using Draco~\cite{draco}), to guarantee timely data transmission with minor degradation of the quality of the received compressed data. 
%Along these lines, in~\cite{bragato2025teleoperateddrivingnewchallenge} we showed that Draco is the best compression algorithm for delay-constrained networks, such as for \gls{td} applications, given its low compression (encoding) time.

As far as the learning model is concerned, \gls{pqos} is generally implemented at the \gls{gnb} via centralized learning, e.g., as in~\cite{mason}. 
This approach permits to operate directly at the RAN, including at the scheduler level.
However, centralized learning also introduces non-negligible delays to transmit and aggregate (receive) data (control signals and commands) to (from) the RAN, which may not be fully compatible with \gls{td} applications~\cite{patriciello2018ran, popowski2018ran, pase2022mab, pase2023disnets}.
 This observation motivates the adoption of decentralized learning (either distributed or federated), where decision-making is executed locally at the \glspl{ue} or at the network edge, thereby reducing the latency and signaling overhead compared to centralized learning in the \gls{ran}~\cite{bragato}.
 For example, the authors in~\cite{decentralizationRaV2X} developed a decentralized \gls{drl} algorithm for optimizing resource allocation in a rapidly changing network environment.
 Decentralized learning can promote privacy, as user-sensitive data, e.g., sensor measurements, are processed locally and are not directly shared over the network. 
Moreover, decentralized models can operate even with no or limited \gls{gnb} connectivity or feedback, and tolerate network delays or partial~disconnections. 

%However, centralization implies a larger data availability at the agent, aggregated from the global status of the network. 
%This characteristic is a key design requirement in safety-critical \gls{v2x} systems, where data exposure has to be minimized. 

Even though the research on \gls{pqos} is quite mature, several research questions are still unanswered. First, the state of the art has primarily focused on optimizing data compression or resource scheduling for PQoS independently, rather than integrating them as correlated components. % within a modular framework. 
Moreover, the literature generally implements centralized or decentralized learning, while the design of an adaptive learning model able to dynamically switch between the two, e.g., based on the temporal evolution of the network conditions, has not yet been explored for TD~applications.

\section{System Model}
\label{sec:model}
In this section we describe our simulation scenario (Sec.~\ref{sub:scenario}) and optimization framework (Sec.~\ref{sub:optframework}).

\subsection{Simulation Scenario}
\label{sub:scenario}

%\begin{figure}
 %   \centering
  %  \includegraphics[width=0.5\textwidth]{simulationscenario.PNG}
   % \caption{Simulation Scenario}
   % \label{fig:simulationscenario}
%\end{figure}

Our simulation scenario, based on~\cite{mason}, 
%and illustrated in Fig.~\ref{fig:simulationscenario}, 
 is mainly implemented in ns-3, one of the most popular and accurate discrete-event simulators for wireless networks.
 It 
consists of a \gls{gnb}, a remote host (i.e., the teleoperator or driving software), $N$ vehicles (i.e., the \glspl{ue}), and the following modules. 

\paragraph{Network}
The remote host is connected to the \gls{gnb} through a wired backhaul link. The \gls{gnb} and the UEs exchange data via the 5G \gls{nr} protocol stack~\cite{38300}, which is simulated in ns-3 via the open-source \texttt{mmwave} module~\cite{mmwave}.

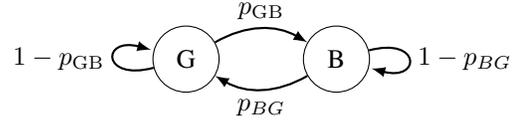
\begin{figure}
    \centering
    \input{mc}
    \caption{Two-state Markov Chain representing the wireless channel. The transition probabilities $p_{GB}$ and $p_{\rm BG}$ are simulation parameters.}
    \label{fig:channel_model}
\end{figure}

\paragraph{Channel and mobility} 
The mobility of the vehicles is simulated in Simulation of Urban MObility
(SUMO)~\cite{sumo}, in an urban area of the city of Bologna. The path loss state (\gls{los} or \gls{nlos}) and the large- and small-scale  fading components of the channel are evaluated via the GEMV$^2$ simulator~\cite{gemv}, and then used in ns-3 to derive the received power. 
Notably, the wireless channel is modeled as a discrete-time two-state Markov Chain, as illustrated in Fig.~\ref{fig:channel_model}, where the Good (G) and Bad (B) states represent high-quality and low-quality channel conditions, respectively. 
The transition probabilities $p_{GB}$ and $p_{BG}$ between these states capture the channel temporal correlation, and can be tuned in ns-3 to simulate different levels of link degradation. 

%We assumed the channel to be in state G when almost ideal conditions hold, i.e., a mean \gls{sinr} of approximately 30 dB and an outage probability basically equal to zero (setting the threshold to 0 dB). Conversely, the channel is considered to be in state B when the conditions are degraded, characterized by a mean \gls{sinr} around 5 dB and an outage probability of nearly 20\%.

\paragraph{Application}
Each vehicle runs a \gls{udp} application that transmits, without loss of generality, \gls{lidar} point cloud data at a constant frame rate~$f$.
Before transmission, the data is compressed using Google Draco~\cite{draco}.\footnote{In~\cite{bragato2025teleoperateddrivingnewchallenge}, we showed that Draco is the best compression algorithm for delay-constrained networks, such as for \gls{td} applications, given its low compression (encoding) time.} The corresponding frame size is determined by the resulting compression configuration $(q, c)$, defined by the number of quantization bits $q \in \{1,\dots,31\}$ and the compression level $c \in \{0,\dots,10\}$.

\paragraph{RAN-AI}
\label{par:ran-ai}
The RAN-AI~\cite{ran-ai} acts as an intelligent centralized network controller for PQoS, deployed at the \gls{gnb}. It collects measurements and performance metrics from the \gls{ran}, 
%such as the \gls{e2e} latency and \gls{sinr}, 
relative to all UEs.
Specifically, the RAN-AI is trained to optimize both data compression and resource allocation based on a centralized single-agent or multi-agent \gls{rl} framework, as described in Sec. \ref{sec:agents}.

%and dynamically optimizes network operations to satisfy \gls{pqos} requirements. 
%Specifically, the RAN-AI is trained based on a centralized single-agent or multi-agent \gls{rl} approach to optimize both data compression and resource allocation, respectively, as described in Sec. \ref{sec:agents}.

\paragraph{UE-AI}
The UE-AI~\cite{bragato} acts an intelligent distributed network controller for \gls{pqos}, deployed at each \gls{ue}.
It collects local measurements and performance metrics from each UE. 
%such as the \gls{rlc} buffer size, number of transmitted bytes, and the \gls{sinr}. 
%and dynamically optimizes UE operations to satisfy \gls{pqos} requirements. 
Specifically, the UE-AI is trained to only optimize data compression based on a decentralized single-agent \gls{rl} framework, as described in Sec. \ref{sec:agents}.

\subsection{Optimization Framework}
\label{sub:optframework}

%\gls{pqos} aims at forecasting QoS degradation and reacting accordingly.
%In our previous works~\cite{mason,bragato}, we proposed an intelligent PQoS agent to optimize the compression configuration for point clouds.
%However, compression may be inevitably compromise data quality, and degrade the quality of the \gls{td} application. 
%Lately, in~\cite{avanzi2025multi} we focused on the RAN, and proposed another intelligent PQoS agent to optimize resource scheduling for the transmission of point clouds.
%However, this approach was demonstrated less effective than compression, that directly determines the size of the point clouds, and 

%Therefore, we designed an intellegent agent to optimize (limited) radio resource allocation at the \gls{ran}, enabling a dynamic scheduling of the users based on current network conditions, while preseving the maximum level of \gls{qoe}  (i.e., data quality).
%Our main idea is to degrade the data quality by enabling a compression mode at the application layer only when it is necessary. 
%For instance, when a decentralized learning framework is enabled due to unstable channel conditions. Otherwise, we operate only at the scheduler level for optimizing radio resource allocation

In our framework, we propose and implement two interacting \gls{rl} agents for \gls{td} applications, namely the \gls{ca} to manage point cloud compression, and the \gls{sa} to manage radio resource scheduling.

The \gls{ca} (Sec.~\ref{sub:ca}) is designed for selecting the optimal compression configuration, i.e., compression level and quantization bits, of the LiDAR data.
While this approach can reduce the size of the data to send, it may inevitably compromise data quality, and degrade the performance of the \gls{td} application. 

The \gls{sa} (Sec.~\ref{sub:sa}) only operates at the RAN, and governs how the gNB assigns radio transmission resources to the UEs in order to satisfy latency constraints. Notably, resource allocation is driven by a priority parameter $k \in \{1, \dots, K \}$, which is related to the network conditions of a given UE. 
Hence, lower-priority UEs ($k\rightarrow 1$) receive fewer resources, as latency requirements are easier to satisfy, whereas higher-priority UEs ($k \rightarrow K$) require more resources.
The number of priority levels $K$ determines the granularity of the agent~decision.

We explore four possible operating modes, depending on how the two agents are integrated:
    \begin{itemize}
        \item \textit{Only Compression (C):} In this mode, only the \gls{ca} is active, while the \gls{sa} is disabled. 
        %The system focuses exclusively on optimizing point-cloud compression without dynamically reallocating channel resouces to improve user scheduling. 
        %This approach is suitable for scenarios where latency requirements are less critical, thus, modulating the volume of data injected into the network is sufficient to meet the target \gls{qos}, though sacrifycing the data quality. 
        This approach is feasible in both centralized and decentralized learning, depending on whether the decision is taken at the \gls{ran} or at the UE, respectively, even though data quality may be sacrificed after compression.
        This mode is based on~\cite{mason2025prata,bragato2024federated}.
        \item \textit{Only Scheduling (S):} In this mode, only the \gls{sa} is active, while the \gls{ca} is disabled. %The system focuses exclusively on optimizing radio resource allocation, tuning a priority level which controls the amount of (limited) resources to assign to a user. 
        This approach is only feasible in centralized learning, since radio resources are assigned at the RAN level, i.e., at the gNB. This approach preserves
    the integrity and quality of the transmitted data.
    This mode is based on~\cite{avanzi2025multi}.
        \item \textit{\gls{ics}:} In this mode, both agents are active but operate independently. Clearly, this is the simplest integration approach to combine the \gls{ca} and the \gls{sa}, as no coordination is required. 
        On the other hand, the learning phase may be unstable because the two agents are not aware of each other's decisions. For example, when a QoS constraint is violated, the agents cannot determine whether the issue derives from inappropriate compression (CA) or scheduling (SA).
        \item \textit{\gls{ccs}:} In this mode, both agents are active and cooperate to find the optimal equilibrium between the compression and scheduling configurations. Hence, a collaborative model must be designed to ensure stability and convergence, which may increase the training complexity.
    \end{itemize}

\section{Proposed RL Agents}
\label{sec:agents}

    An \gls{rl} framework is mathematically modeled as a \gls{mdp} \cite{sutton2018reinforcement}, defined by the tuple ${<\mathcal{S}, \mathcal{A}, \mathcal{P}, \mathcal{R}, \mathcal{\gamma}>,}$ where $\mathcal{S}$ represents the set of states, $\mathcal{A}$ represents the set of actions, $\mathcal{P}$ represents the state transition probability matrix with elements ${\mathcal{P}}^a_{ss'} = P[S_{t+1}=s' | S_t=s, A_t=a]$, $\mathcal{R}$ represents the reward function with $\mathcal{R}^a_s = E[R_{t+1} | S_t=s, A_t=a]$, and $\gamma \in (0, 1)$ represents the discount factor.
    %The agent follows a policy can be deterministic, i.e., $A_t = \pi(S_t)$, or stochastic, i.e., $\pi(a | s) = P[A_t=a , S_t=s]$. 
    At each time step $t$, the agent observes state $S_t$, chooses action $A_t$, receives a reward $R_{t+1}$, and moves to state $S_{t+1}$ according to ${\mathcal{P}}$.

    The agent's goal is to find the optimal policy $\pi^*$ that maximizes the expected return $G_t$, defined as the sum of the discounted rewards from time $t$. Specifically, $G_t$ is defined as
    \begin{equation}
    	G_t=\sum_{\tau=0}^{+ \infty} \gamma^\tau R_{t+\tau}.
    \end{equation}
    In certain cases, the agent is only able to perceive an observation $O_t$ of $S_t$, which does not provide complete information about the environment and the underlying state $S_t$. This problem is formalized as a \gls{pomdp} \cite{astrom1965pomdp}.

    Various algorithms have been developed to determine $\pi^*$, and can be classified as value-based or policy-based, and as single-agent or multi-agent. 
    Value-based methods try to estimate the optimal action-value function and learn a policy based on that. Policy-based methods, instead, directly learn the optimal policy to maximize the expected return. This latter approach is particularly effective in high-dimensional and continuous action spaces, where the value function is difficult to estimate and generalize.
    In a single-agent setup, the size of the state-action space grows exponentially with the number of \glspl{ue}. To address this challenge, the problem is decomposed into a smaller and more tractable multi-agent decision problem.
    In this case, we consider a multi-agent extension of a \gls{pomdp}, i.e., a \gls{dec-pomdp}~\cite{dec-pomdp}, defined by the tuple $<\mathcal{N}, \mathcal{S}, \{\mathcal{A}_i \}_{i \in N}, \{\mathcal{O}_i \}_{i \in N}, \mathcal{P}, \mathcal{R}, \mathcal{\gamma}>$, where $\mathcal{N}$ represents the set of agents, $\mathcal{A}_i$ represents the set of actions for agent $i \in \mathcal{N}$, $\mathcal{O}_i$ represents the the set of observations for agent $i \in \mathcal{N}$, ${\mathcal{P}}^a_{s,s'}: \mathcal{S} \times \mathcal{A}_\mathcal{N} \times \mathcal{S} \rightarrow \left[0, 1\right]$ represents the state transition probability function, and $\mathcal{R}: \mathcal{S} \times \mathcal{A}_\mathcal{N} \rightarrow \mathbb{R}$ represents the reward function. In this setting, each agent interacts with the shared environment and learns a decentralized policy based on its own local observations, without any prior knowledge of the global state or the actions of the other agents.
    
  \emph{Notation:} In the considered multi-agent frameworks, all agents are implemented using the same model, architecture, parameters, and training configuration. Accordingly, for simplicity, the agent index $i$ is omitted from the model notation throughout the remainder of this paper.

    \subsection{\acrfull{ca}}
    \label{sub:ca}
    
        The \gls{ca} selects the optimal compression configuration for \gls{lidar} data to minimize the \gls{e2e} latency, subject to a constraint on the resulting data quality for reliable \gls{td} operations (e.g., object detection).

        \subsubsection{State}
        The state of the environment depends on the \gls{ca} implementation, i.e., 
        %is represented by a multidimensional vector that consists of the following measurements: the \gls{sinr},  the \gls{mcs} index, the number of \gls{ofdm} symbols required to transmit the data in the \gls{ul} buffer, and the mean, standard deviation, minimum and maximum latency and \gls{prr} at the \gls{phy}, \gls{rlc}, \gls{pdcp}, and application layers.
        centralized or decentralized.
        \begin{itemize}
            \item \emph{Centralized state}: The RAN-AI in the gNB collects the following measurements from all \glspl{ue} in the RAN: the \gls{sinr},  the \gls{mcs} index, the number of \gls{ofdm} symbols required to transmit the data in the \gls{ul} buffer, and the mean, standard deviation, minimum and maximum latency and \gls{prr} at the \gls{phy}, \gls{rlc}, \gls{pdcp}, and application layers.
            Data collection at the gNB may lead to overhead and channel congestion, and expose the network to security threats, since sensitive data must be shared with the RAN. 
            \item \emph{Decentralized state}: Each UE-AI implements an independent and autonomous agent, which learns solely from local data and measurements. Compared to the centralized state, the \gls{rlc} and \gls{pdcp} measurements are not available at the UEs, and are replaced by the mean, standard deviation, minimum and maximum occupancy of the UL buffers at the \gls{mac} layer.
             While this approach eliminates the need to send raw data and measurements to the gNB and promotes data privacy, convergence may be slower, since each agent has access to limited (and only local) training data.
        To accelerate convergence, we train the agent in a \gls{fl} setup~\cite{qi2021federated}. Accordingly, UEs periodically share learning updates with the gNB, which aggregates them into a global model. The resulting model is then redistributed to the federated agents, which further adapt it using their local data.
        \end{itemize}
        
        %    At the vehicle, the state derives from the centralized state at the \gls{ran}, where RLC and PDCP layer statistics are replaced by mean, standard deviation, minimum and maximum of volumes of data inside the Transmission Buffer and the Transmitted \glspl{pdu} Buffer.

        \subsubsection{Action}
            The discrete action space $\mathcal{A}_{\rm CA}$ is defined as the set of possible compression configurations for LiDAR data.
        
        \begin{comment}
        \begin{table}
            \centering
            \caption{Compression quality and time of DRACO configurations}
            \label{tab:mAP}
            \begin{tabular}{c|c|c|c|c}
                \hline
                Compression mode & $q$ & $c$ & mAP & Compression time (ms)\\
                \hline
                $C_1$ & 8 & 10 & 0.257 & 8.21 \\
                $C_2$ & 9 & 5 & 0.572 & 6.97 \\
                $C_3$ & 10 & 0 & 0.686 & 5.50 \\
                \hline
            \end{tabular}
        \end{table}
        \end{comment}
        
        \subsubsection{Reward}
        \label{reward-ca}
            The reward function $\mathcal{R}$ is designed to optimize both the \gls{e2e} latency $\ell$ associated with the transmission process, and the \gls{map} $m_a$ of the point cloud after compression (using compression configuration $a\in \mathcal{A}_{CA}$, i.e., the action). Therefore, the latter represents the quality of the received~data.
            
           % Indeed, \gls{qos} ensure the satisfaction of \gls{td} requirements in terms of \gls{kpi}; in this work, only the \gls{e2e} delay $\ell$ is considered. $\tau$ is the maximum tolerated \gls{e2e} delay, up to 50 ms from the \gls{5gaa}. 
            % The \gls{map} $m_a$ achieved by the object detector PointPillars \cite{pointPillars} on the SELMA \cite{selma} dataset, with compression mode $a$, is considered as metric to evaluate the accuracy of the object detection. The average compression quality (\gls{map}) and time are reported in Tab. \ref{tab:mAP}.
            	
           The reward is a piecewise function, whose definition depends on the operating mode of the framework (Sec.~\ref{sub:optframework}).
           \begin{itemize}
               \item In the C and ICS modes, the CA operates independently of the \gls{sa}. So, the reward is defined as:

            \begin{equation}
            	R_{\rm CA}^{\{\rm C,ICS\}} = \begin{cases}
            		m_a&\text{if } \ell \leq \tau,\\
            		-\ell/{100}&\text{otherwise}. \\
            	\end{cases}
                \label{R_CA_C,ICS}
            \end{equation}
            Specifically, a positive reward equal to $m_a$ is returned if the \gls{e2e} latency $\ell$ is lower than or equal to a predefined threshold $\tau$, which depends on the requirements of the TD application.
Otherwise, the reward is a penalization proportional to $\ell$.

            \item  In the CCS mode, the \gls{ca} cooperates with the \gls{sa}. Now, the reward function should depend on both the compression configuration (for the CA) and the  priority level $k$ for resource allocation (for the SA). 
            Specifically, when latency constraints are not satisfied, the reward is a penalization inversely proportional to $k$, so $k$ acts as a discount factor on the latency. In this case, the agent should apply more aggressive compression (i.e., the action) when a UE is allocated fewer resources. 
            %On the contrary, high latencies originated from poorly aggressive compression modes should be better tolerated in case of users with large priority. 
            So, the reward is defined~as:
            \begin{equation}
            	R_{\rm CA}^{\rm CCS} = \begin{cases}
            		m_a&\text{if } \ell \leq \tau,\\
            		-\ell/({100\cdot k)}&\text{otherwise}. \\
            	\end{cases}
                \label{R_CA_CCS}
            \end{equation}

            \end{itemize}

        \subsubsection{Algorithm} 
        \label{sub:ddlq}
        The CA is implemented as a \gls{ddql} model. It is a value-based strategy that extends the classical \gls{dql} model by approximating the Q-function, i.e., the expected cumulative reward when taking an action in a given state, using two \glspl{dnn}: 
        the Q-Network with parameters ${\theta_Q}$, that selects the best action for the next state based on the Q-values $Q_{\theta_Q}$,
        and the Target-Network with parameters ${\theta_T}$, that provides stable target Q-values $Q_{\theta_T}$ for learning.
    The Q-Network is trained using Adam by minimizing the Mean Squared Error (MSE) $L (\theta_Q)$ between $Q_{\theta_Q}$ and a target Q-value $y_t$ computed using the Target-Network at time $t$, i.e., 
    \begin{align}
            L (\theta_Q) &= \mathbb{E}_\mathcal{B} \left[ \left(y_t - Q_{\theta_Q}(S_t, A_t)\right)^2 \right],\label{eq:L}\\
            y_t &= R_t + \gamma \max_{A'} Q_{\theta_T}(S_{t+1}, A').
        \end{align}
        The MSE in Eq.~\eqref{eq:L} is computed over a mini-batch $\mathcal{B}$. Moreover, while the Q-Network is updated at every step, the parameters of the Target-Network used to compute $y_t$ are kept fixed for a predefined number of training iterations $T_{\rm DDQL}$, to improve the stability of the training process. 
    Finally,  to balance exploration vs. exploitation, we adopt an $\epsilon$-greedy policy, where the exploration rate $\epsilon$ decreases linearly over~time.

        %The two \glspl{nn} have an identical structure, but the Target network is not updated at each step as the Q Network, but only after some steps they are synchronized. Therefore, the algorithm ensures that the Target Q-values, which are used in the computation of the loss function, are kept stable for a certain period, so the target does not change continuously decreasing the stability of the learning. After a number of steps, the Target network is updated for improving the output estimation.
        
        %An experience replay buffer is exploited to collect experience tuples gathered by the agent interaction with the environment at each step. Hence, the stability of the learning is improved (uncorrelated samples) and the algorithm implementatation is more data efficient. Furthermore, an $\epsilon$-greedy strategy is considered to deal with the usual exploration vs. exploitation dilemma, using a linearly decreasing rate of exploration $\epsilon$. 
             
    \subsection{\acrfull{sa}}
    \label{sub:sa}

        The \gls{sa} selects the optimal priority level for UEs, which determines the corresponding number of \gls{ofdm} symbols per slot to use for transmission to minimize the \gls{e2e} latency. %subject to a constraint on the available network capacity.
    
        \subsubsection{State} The RAN-AI in the gNB collects the following measurements from all UEs in the RAN: the mean \gls{sinr}, the mean \gls{mcs} index, the mean \gls{ul} buffer size, the number of \gls{ofdm} symbols required to transmit the data in the UL buffer given the \gls{mcs}, the mean \gls{e2e} latency, and the number of transmitted bytes at the application~layer.
        
        %state (observation) is representative of the network status by the point of view of all UEs (from a single UE). To build the state, the RAN-AI entity in the \gls{gnb} gathers several measurements by dedicated control signals during data transmissions. In this work, the following metrics are selected for the state definition: the mean \gls{sinr}, the mean \gls{mcs} index, the mean \gls{ul} buffer size, the mean number of \gls{ofdm} symbols required for tranmsmitting the entire \gls{ul} buffer (given the \gls{mcs}), and the mean \gls{e2e} latency and number of bytes transmitted at the application~layer.

        \subsubsection{Action} The discrete action space $\mathcal{A}_{\rm SA}$ is defined as the
    set of possible priority levels $K$ for resource allocation.
    %    A suitable value $k$ representing the priority level parameter has to be assigned to a given user (i.e., \gls{ue}, at each resource allocation opportunity. 
    Specifically, a UE is assigned a higher priority $k\in\{1,\dots,K\}$ if it necessitates more radio resources to satisfy its \gls{qos} requirements under the current network conditions.
    
        \subsubsection{Reward} 
        \label{reward-sa}
        The reward is a piecewise function that depends on the operating mode of the framework.
        \begin{itemize}
            \item In the S and ICS modes, the SA operates independently of the CA. In this case, the reward is a penalization proportional to the violation of $\ell$ with respect to $\tau$, i.e., 
            %is only expressed as a function of $\ell$ and $\tau$, since resource allocation has no impact on the quality of the transmitted data, i.e.,
            \begin{equation}
            R_{\rm SA}^{\{\rm S, ICS\}} = \begin{cases}
                1&\text{if } \ell \leq \tau,\\
                -{(\ell - \tau)}/{100}&\text{otherwise}. \\
            \end{cases}
            \label{eq:S}
        \end{equation}
        \item In the CCS mode, the reward also depends on the quality of the compressed data, measured in terms of the \gls{map} $m_a$. Specifically, when latency constraints are violated (i.e., $\ell>\tau$), the penalty is directly proportional to $m_a$. In this case, the network is encouraged to apply more aggressive compression, even at the cost of some data quality degradation, to reduce the data size and transmit faster. In parallel, the penalty is inversely proportional to the priority level $k$. In this case, the network is encouraged to schedule higher-priority UEs, which are allocated more radio resources and are less likely to violate latency constraints. Therefore, the reward is defined as 
        \begin{equation}
            R_{\rm SA}^{\rm CCS} = \begin{cases}
                1&\text{if } \ell \leq \tau,\\
                -{((\ell - \tau) \cdot m_a)}/{(100 \cdot k)}{}&\text{otherwise}.
            \end{cases}
        \end{equation}

        \end{itemize}

        \subsubsection{Algorithm} 
        The SA is implemented as a \gls{ppo} model, which is more effective than traditional methods such as Q-Learning to handle complex, non-stationary, large-scale, multi-agent networks \cite{ppo, mappo}.
        \gls{ppo} is a model-free, policy-based approach derived from the \gls{trpo} algorithm~\cite{trpo}, where a policy $\pi_\theta$ is iteratively improved using \gls{sgd} by interacting with the environment and optimizing (in multiple epochs) a clipped surrogate objective function $L(\theta,\phi)$, given~by
        \begin{equation}
        \label{eq:L_theta-phi}
            L(\theta, \phi) = L^{\text{CL}}(\theta) - c_1 L^{\text{VF}}(\phi) + c_2 S(\pi_\theta).
        \end{equation}
        In particular, $L^{\text{VF}}(\phi)$ is the MSE between a state-value function approximator $V_\phi$ and the target return $G_t$, $S(\pi_\theta)$ is an entropy bonus to encourage the exploration of $\pi_\theta$, and $c_1$ and $c_2$ are hyperparameters. Then, $L^{\text{CL}}(\theta)$ can be written~as 
\begin{equation}
L^{\text{CL}}(\theta)=\mathbb{E}_t\!\big[\min(r_t(\theta)\hat{A}_t,\text{clip}(r_t(\theta),1-\varepsilon,1+\varepsilon)\hat{A}_t)\big].
\label{eq:LCL}
\end{equation}
        where $\varepsilon$ is a hyperparameter, $r_t(\theta)$ measures the divergence between the updated policy $\pi_\theta$ and the original policy  $\pi_{\theta_{{o}}}$ before the most recent parameters update based on observation $o_t$ at time $t$, that is
        \begin{equation}
            r_t(\theta) = \frac{\pi_\theta (a_t | o_t)}{\pi_{\theta_{o}} (a_t | o_t)},
        \end{equation}
        and $\hat{A}_t$ is an  estimator of the advantage function at time $t$, defined as the difference between the Q-function and the state-value function. Notably, $\hat{A}_t$ is approximated using the \gls{gae}~\cite{gae} technique as 
        \begin{equation}
            \hat{A}_t = \sum_{l=0}^{T_{\rm PPO}-t} (\gamma \lambda)^l \delta_{t+l},
        \end{equation}
        where $\delta_{t}$ is the temporal difference error at time $t$, and $\lambda \in (0, 1]$ is a control parameter. %which controls the trade-off between bias in advantange function estimation and variance in long trajectories. 

        %The implementation of this model involves two \glspl{nn} with two hidden layers of $n_N$ neurons each. The former NN (actor) consists of $|\mathcal{O}_{i}|$ input neurons and $|\mathcal{A}_{\rm SA}|=K$ output neurons, and is used to represent policy $\pi_\theta$.  It receives as input the state of the SA, and returns as output a probability distribution over the action space.
        %The second NN (critic) consists of $|\mathcal{O}_{\rm in}|$ input neurons and a single output neuron, and is used to represent the state-value function approximator $V_\phi$ and reduce the variance of the advantage function.
        The implementation of this model involves two \glspl{nn}. The former NN (actor) is used to represent policy $\pi_\theta$.  It receives as input the state of the SA, and returns as output a probability distribution over the action space.
        The second NN (critic) is used to represent the state-value function approximator $V_\phi$ and reduce the variance of the advantage function.
        Both NNs are trained using Adam with a learning rate $\alpha$, from trajectories of $T_{\rm PPO}$ steps, and over mini-batches of size $M$ to improve~stability.
        
        %A policy network $\pi_\theta$ (actor) and a value network $V_\phi$ (critic) are implemented in the model. The state of \gls{ca} is received in input by $\pi_\theta$ and a probability distribution over the action space is returned. $V_\phi$ approximates the state value function and contributes to decrease the variance of the advantage function. Two fully connected \gls{nn} are exploited with two hidden layers of $n_N$ neurons each and the hyperbolic tangent as activation function. The architecture of $\pi_\theta$ involves $|\mathcal{O}_i|$ input neurons and $|\mathcal{A}_i|$ output neurons preceded by the softmax function; $V_\phi$ has $|\mathcal{O}_i|$ input neurons and a single output neuron. 

        %In this paper, the SA is implemented via two \gls{ppo} algorithms.
        For the determination of the priority level $k$, we use \gls{mappo} \cite{mappo}. Notably, \gls{mappo} is a \gls{ctde} framework \cite{ctde}, where a single actor and a single critic are updated centrally using local data gathered from $N$ decentralized agents.
        For the determination of the number of \gls{ofdm} symbols per slot to use for transmission, we use a \gls{ga} algorithm, which is greedy with respect to the priority level $k$. 
        Specifically, the available \gls{ofdm} symbols per slot $U$ are assigned to the UEs in decreasing order of~priority.
        
    \subsection{Meta-Learning Agent}
    \label{sub:meta}

        %The CA and SA can be implemented according to a either centralized or decentralized model.
        In the centralized approach, a controller collects global data from the network at the gNB, which 
        enables more accurate and representative learning, at the cost of significant signaling overhead for data collection %as well as limited scalability and generalization 
        as the network size increases.
        In a federated approach, the learning model is distributed across multiple agents, which improves scalability and convergence speed, and can operate even with no or limited \gls{gnb} connectivity.      
        However, the agents may take suboptimal actions since only local data is available.
        % Moreover, as we discuss in Sec. XYZ, the simulation results highlights how the decentralized (federated) \gls{ca} agent outperforms the centralized counterpart, with \gls{ca} enabled, when the channel is more degraded and latency requirements are not critical.

        In this paper, we propose and implement a new meta-learning agent that dynamically selects the optimal learning model (centralized or federated) based on the network conditions. 
        Specifically, centralized learning (based on the ICS mode) is selected if \gls{sinr} $\geq \Gamma$, where $\Gamma$ is a predefined channel quality threshold, as sufficient channel capacity is available to support global data collection at the RAN. 
        Otherwise, the meta-learning agent converges to federated learning (based on the C mode) to preserve radio resources and mitigate network congestion in case of channel degradation.
        The meta-learning agent is implemented as a single \gls{rl} agent, so a single global threshold $\Gamma$ is used.

        \subsubsection{State}  The state of the environment is represented by a multidimensional vector that consists of the following measurements from all \glspl{ue} in the RAN: the mean \gls{sinr}, CA reward, \gls{e2e} latency $\ell$, and \gls{map} $m_a$. These measurements are collected every time a new point cloud is received, and averaged over a window of $W$ receptions.

        \subsubsection{Action} The discrete action space $\mathcal{A}_{\rm MLA}$ is defined as the set of possible \gls{sinr} thresholds $\Gamma$, which determines the corresponding learning model (centralized or federated).
    %Specifically, the action is defined as a scalar value $\Gamma \in \{10, 20, 30\}$ dB. These values are derived experimentally from our simulated channel. Surely, the values of $\Gamma$ need to be designed and tuned with respect to the specific simulated scenario and the desired level of granularity for the agent's decisions. The larger the range of values $\Gamma$ spans, the more accurate the agent's ability in distinguishing between different channel regimes, but this also increases the complexity of the learning.

        \subsubsection{Reward} The reward of the meta-learning agent $R_{\rm MLA}$ depends on the reward of the \gls{ca} $R^{w}_{\rm CA}$, with $w\in\{\rm C,ICS\}$, defined in Eq.~\eqref{R_CA_C,ICS} for the C and ICS modes. 
        This choice depends on the fact that the CA, unlike the SA, can be implemented both as a centralized and as a federated agent.
        %optimizes both $\ell$ and $m_a$. 
        Specifically, $R_{\rm MLA}$ is defined as the average CA reward of all $N$ UEs in the network, over a window of $W$ receptions.
            %\begin{equation}
           % 	R_{\rm MLA}^w = \left({\sum_{k=1}^{N} R_{\rm CA}^{w}}\right)/{N}, \: w\in\{\rm C,ICS\}.
           % \end{equation}
            
        \subsubsection{Algorithms} 
        We propose and compare three different meta-learning agent implementations, i.e., stateless \acrfull{mab}, contextual \gls{mab}, and stateful DDQL.

        \paragraph{Stateless}
        In this class of algorithms, the concept of state is not defined. Specifically, we formalize a \gls{mab} problem~\cite{intro_mab} where actions are selected based only on immediate rewards from previous learning steps, and with no explicit knowledge of the underlying state.
        We evaluate two models.
        
        In Random (R) \gls{mab}, the meta-learning agent uniformly~selects a random action, i.e., the threshold $\Gamma$, so $\pi = \mathcal{U}(\mathcal{A_{\rm MLA}})$.
        
        In \gls{eg} \gls{mab}, the meta-learning agent approximates the action-value function $Q(a)$, $\forall a \in \mathcal{A_{\rm MLA}}$. At each learning step, the agent usually selects the action that maximizes $Q(a)$ (exploitation) but, in some occasions, it explores by selecting (uniformly) a random action among all the available actions. Specifically, the policy is defined as
            \begin{equation}
                \pi = \begin{cases}
                    \mathcal{U}(\mathcal{A_{\rm MLA}})&\text{w.p. } \epsilon_{\rm EG},\\
                    \arg\max_{a\in \mathcal{A_{\rm MLA}}} Q(a)&\text{w.p. }1-\epsilon_{\rm EG}.\\
                \end{cases}
            \end{equation}
            
            The hyperparameter $\epsilon_{\rm EG}$ is selected to control the trade-off between exploration and exploitation.
            Since the environment is non-stationary, $Q(a)$ is updated as
            \begin{equation}
                Q_{t+1} (a)= Q_t (a) + \alpha ( R_t - Q_t (a)),
            \end{equation}
            where $\alpha$ is the learning rate.

            \paragraph{Contextual}
            We formalize a Contextual Combinatorial Multi-Armed Bandit (CC-MAB) problem, where now the reward is modeled as a function of both the action and the context~\cite{pase2023disnets}. 
            We evaluate two models.

            \Gls{lts}~\cite{tutorial_thompson_sampling} assumes that the average reward behind each action $a \in \mathcal{A_{\rm MLA}}$ is a linear function of the context $x$ and an unknown action parameter vector $\theta_a$, i.e., $x^T \theta_a + \epsilon_t$, where $\epsilon_t \sim \mathcal{N}(0, v_t^2)$ represents the noise at time $t$. The standard deviation of the noise is defined as $v_t = R_M \sqrt{9\ |x| \ln \left({t}/{\rho}\right)}$, where $R_M$ is an upper bound of the noise of the reward, and $\rho$ is a confidence parameter~\cite{noise_LTS}.

	        In this case, estimating the most accurate vector $\theta_a$ turns out to be an {online} Bayesian linear regression problem. 
            To solve this problem, \gls{lts} assumes that the rewards, given the action $a$ and the context $x$, are modeled as a Gaussian random variable, i.e., $\mathcal{N}(x^T\theta_a,v_t^2)$. Notably, the distribution of $\theta_a$ at time~$t$ can be modeled as $\mathcal{N}\left(\hat{\theta}_a(t),v_t^2 [\Phi_a(t)]^{-1} \right)$, where
            \begin{align}
            \Phi_a(t) = I_d + \sum_{\tau=1}^{t-1} x_\tau x_\tau^T \cdot \mathbf{1}_{a_t=a},
            \label{LTS_sigma}\\
            \hat{\theta}_a(t) = [\Phi_a(t)]^{-1} \sum_{\tau=1}^{t-1} x_\tau {R}_\tau^T \cdot \mathbf{1}_{a_t=a},
            \label{LTS_theta}
            \end{align}
            where $R_\tau$ is the realization of the reward at time $\tau$, $I_d$ is the identity matrix of size $d$, $x_t$ is the context realization at time $t$, and $\mathbf{1}_{a_t = a}$ is the indicator function and is equal to 1 if $a_t = a$ or 0 otherwise.
            After the observation of context $x_t$ and the selection of action $a_t$,  the agent receives a reward and updates the distribution of $\theta_a$ at time $t+1$~as
            \begin{equation}
                \mathcal{N}\left(\hat{\theta}_a(t+1),v_{t+1}^2 [\Phi_a(t+1)]^{-1} \right).
                \label{LTS}
            \end{equation}
            In the algorithm, the agent samples a vector $\theta_a$ for each $a \in \mathcal{A_{\rm MLA}}$, selects $a_t = \arg\max_{a\in\mathcal{A_{\rm MLA}}}(x_t^T\theta_a)$, and updates the posterior distribution according to Eq. \eqref{LTS}.
            \Gls{lts} achieves a good trade-off between exploration and exploitation.

            LTS assumes a linear relation between the context and the reward, which does not describe real-world scenarios.
            To overcome this limitation, we also consider a \gls{nlts} algorithm~\cite{Riquelme2018}  where the expected reward is modeled as a linear function of the action parameter vector $\theta_a$ and a non-linear representation of the context $\phi(x)$, i.e., $\phi(x)^T\theta_a +\epsilon_t$. 
            Specifically, \gls{nlts} involves an initial step where the context $x$ is mapped into a non-linear representation $\phi(x)$ using a \gls{dnn}, parameterized by weights $\omega\in\Omega$. Then, NLTS uses \gls{lts} to select the optimal action $a$ by applying Bayesian linear regression as per Eqs. \eqref{LTS_sigma}-\eqref{LTS_theta}, where the original context $x$ is now replaced by $\phi(x)$.
            Notice that the LTS posteriors are updated based on Eq.~\eqref{LTS} at each learning step, while the weights $\omega$ of the DNN are updated at fixed intervals of $T_{\rm NLTS}$ steps.

            \paragraph{Stateful} 
            In this class of algorithms, the meta-learning agent exploits an explicit representation of the environment (i.e., the state) to estimate the action-value function, which captures the expected cumulative return resulting from both immediate rewards and the future state evolution of the environment.
            Specifically, we employ the \gls{ddql} algorithm, where a \gls{dnn} is trained to approximate the action-value function $Q(s,a)$ for each state $s$ and each action $a$, and a policy is derived by selecting the action that maximizes $Q(s,a)$. %For a more formal description of \gls{ddql}, we remind the interested readers to Sec.~\ref{sub:ddlq}.

\section{Performance Evaluation}
\label{sec:performance}

    \begin{table}
        \centering
        \caption{Simulation parameters.}
        \label{tab:simulation_parameters}
        \begin{tabular}{l|l}
            \hline
            Parameter & Value\\
            \hline
            Number of UEs/agents/vehicles ($N$) & $5$\\
            Carrier frequency ($f_c$) & 28 GHz \\
            Bandwidth ($B$) & 50 MHz \\
            Available OFDM symbols/slot ($U$) & 12 \\
            %\hline
            LiDAR frame rate ($f$) & 30 fps\\
            %Compression $(q,c)$ & $\{$(8, 0), (9, 5), (10, 10)\} \\
            %\hline
            Latency threshold ($\tau$) & $\{30, 50\}$ ms\\
            Number of priority levels ($K$) & 3\\
            %\hline
            Discount factor ($\gamma$) & 0.95 \\
            GAE control parameter ($\lambda$) & 0.95 \\
            %Number of neurons in hidden layers ($n_N$) & 64 \\
            Learning rate ($\alpha$) & $10^{-4}$\\
            PPO hyperparameters ($\{\varepsilon, c_1,  c_2\}$) & \{0.2, 0.5, 0.01\}\\
            %Length of a trajectory ($T$) & 512 steps\\
            Mini-batch size ($M$) & 64 steps \\
            Meta-learning agent window size ($W$) & 1 \\
            Update interval ($\{T_{\rm DDQL},T_{\rm PPO},T_{\rm NLTS}\}$) & \{8\,000, 512, 128\} steps\\
            LTS noise parameters ($\{R_M,\rho\}$) & \{1, 0.5\}\\
            %Duty cycle of state B ($T^B_{DC}$) & $\{\frac{1}{6}, \frac{1}{2}\}$ \\
            \hline
        \end{tabular}
    \end{table}
    
    In this section we describe our simulation parameters (Sec.~\ref{sub:params}) and numerical results relative to both the CA and the SA (Sec.~\ref{sub:results1}) and the meta-learning agent (Sec.~\ref{sub:results2}).

    \subsection{Simulation Parameters}
    \label{sub:params}

        Numerical results are obtained in ns-3. Specifically, we use the ns-3 PRATA framework, described in~\cite{mason2025prata}, which integrates the GEMV$^2$ and  SUMO simulators for the simulation of the channel and mobility models, respectively, and the \texttt{ns3-mmwave} module for the simulation of the 5G \gls{nr} protocol stack for data transmission, the automotive application, and the RAN-AI and UE-AI entities for training and testing the RL agents. Notably, we extended the baseline ns-3 PRATA framework to implement the novel RL algorithms proposed in this paper, namely the \gls{ca}, \gls{sa} and meta-learning agent. Simulation and learning parameters are reported in Tab. \ref{tab:simulation_parameters}, and described below.
    
        \subsubsection{Network} We consider a \gls{gnb} and $N=5$ \glspl{ue}/vehicles that communicate via a 5G NR network at a carrier frequency $f_c =$ 28 GHz and with a bandwidth $B =$ 50 MHz, using numerology 3.
        Specifically, the symbol duration is $8.92$ $\mu$s, and each slot consists of $U=12$ useful OFDM symbols, given that the first 2 symbols are
        reserved for control in UL and DL.
        The \gls{gnb} (\gls{ue}) transmit power is set to 30 (23) dBm.
        The propagation delay and the (ideal) bit rate of the wired channel between the remote host and the gNB are 10 ms and 100 Gbps, respectively.

        As far as the two-state Markov channel described in Sec.~\ref{sub:scenario} is concerned, the mean \gls{sinr} in the Good (Bad) state is approximately 30 (5) dB, while the outage probability is equal to 0 (0.2).

        \subsubsection{Application} The application layer generates and transmits \gls{lidar} point clouds at a frame rate $f = 30$ fps. 
        Data is compressed using  Google Draco~\cite{draco}, which introduces a non-negligible encoding delay. For simplicity, we consider only three representative compression configurations, i.e.,  $(q,c) \in\{C_1,C_2,C_3\}=\{(8, 10), (9, 5), (10, 0)\}$.
        The resulting data rates are $\{21.25,31.35,41.35\}$ Mbps, while the encoding delays are $\{8.21,6.97,5.50\}$ ms, respectively. 
        Object detection on the received (compressed) data is based on PointPillars \cite{pointPillars}, and the resulting \gls{map}, measured on the SELMA dataset~\cite{selma}, is $\{0.257,0.572,0.686\}$, respectively.
        Finally, the \gls{e2e} latency threshold $\tau$ of the TD application is set to $30$ or $50$ ms, depending on the TD application~\cite{3GPP_22186}.
    
        \subsubsection{Learning algorithms} For the \gls{ca}, the Q/Target-Network involves 16 input neurons equal to the size of the centralized and decentralized/federated state, and 3 output neurons equal to the size of the action space, i.e., the number of (available) Draco compression configurations. The Q/Target-Network has two hidden layers of 64 and 32 neurons, respectively.
        For the \gls{sa}, the actor/critic NN has 6 input neurons equal to the size of the observation space, and $\mathcal{A}_{\rm SA}=K$ output neurons equal to the size of the SA action space, i.e., the number of priority levels. We fixed $K=3$ based on offline simulations, representing a good trade-off between learning complexity and stability. 
        Both NNs have two hidden layers of 64 neurons each. %sufficiently differentiating \glspl{ue} and for reducing complexity and unstability of the learning environment.
        Regarding the meta-learning agent, for contextual learning the context/state is a vector in 
        $\mathbb{R}^{4|\mathcal{N}|}$, meaning that the agent collects and uses 4 measurements for each UE, as described in Sec.~\ref{sub:meta}. For the \gls{nlts} and \gls{ddql} models, the \gls{dnn} involves $4|\mathcal{N}|$ input neurons equal to the size of the context/state, and 3 output neurons equal to the size of $\mathcal{A_{\rm MLA}}$, i.e., the number of SINR thresholds $\Gamma$. The DNN has two hidden layers of 64 and 32 neurons, respectively.
        Specifically, we set $\Gamma\in\{10,20,30\}$ dB, which have been derived experimentally on simulated channel traces to consider different and representative  channel quality~regimes.
        %Finally,  we set $\{T_{\rm DDQL},T_{\rm PPO},T_{\rm NLTS}\}= \{8\,000, 512, 128\}$ training iterations.        
        The rest of the learning hyperparameters are reported in Tab.~\ref{tab:simulation_parameters}.

        Our \gls{rl} algorithms have been trained on 250 episodes, where each episode is an independent ns-3 simulation. Each agent performs 400 learning steps per episode, where a learning step is performed every time a new point cloud is~received. 
        %For the meta-learning agent, the agent performs a learning step every time a new point cloud is generated.
        %For the meta-learning agent, the agent performs a learning step at every point cloud generation, deciding for a single action to be used from al the \glspl{ue}.
    
        \subsubsection{Benchmarks} We compare the four operating modes described in Sec.~\ref{sub:optframework}: ICS and CCS, where both the CA and the SA are active and operate independently or cooperatively, respectively; C and S, where only the CA or the SA is active, respectively. 
        Notice that the S, ICS and CCS modes are only centralized models. In fact, they involve the SA, which is only feasible in centralized learning for collecting data from the RAN.
        Instead, the C mode is only based on the CA, so we compare centralized vs. federated~learning. 

  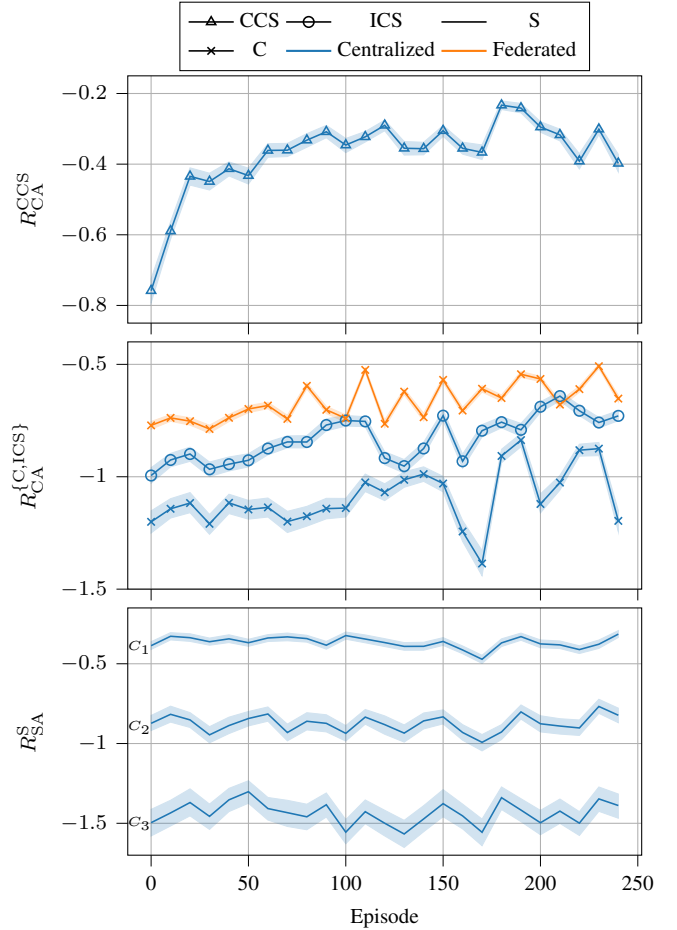
\begin{figure}
            \centering
            \input{r_lineplot.tex}
            \vspace{-0.5cm}
            \caption{Average rewards during training in the Bad (B) channel state, vs. the optimization framework modes.}
            \label{r_lineplot}
        \end{figure}
        
        \begin{figure*}[t!]
            \centering
                \begin{subfigure}[t]{\linewidth}
                    \centering
                    \input{legend_G}
                \end{subfigure}
                \centering
                \vskip 0.25 cm
                \subfloat[][Average reward.]
                 {
                    \label{G_r}
                    \input{G_r}        
                }\hspace{-0.8cm}
                \subfloat[][Average \gls{map}.]
                 {
                    \label{G_mAP}
                    \input{G_mAP}        
                }\\[2ex]
            \subfloat[][\gls{e2e} latency statistics, with $\tau = 50$ ms.]
             {
                \label{G_50_d}
                \input{G_50_d}        
            }\hspace{-0.8cm}
            \subfloat[][\gls{e2e} latency statistics, with $\tau = 30$ ms.]
             {
                \label{G_30_d}
                \input{G_30_d}        
            }
            \caption{Average reward, average mAP, and \gls{e2e} latency statistics in Good (G) channel conditions, vs. the optimization framework modes.}
            \label{fig:good}
        \end{figure*}
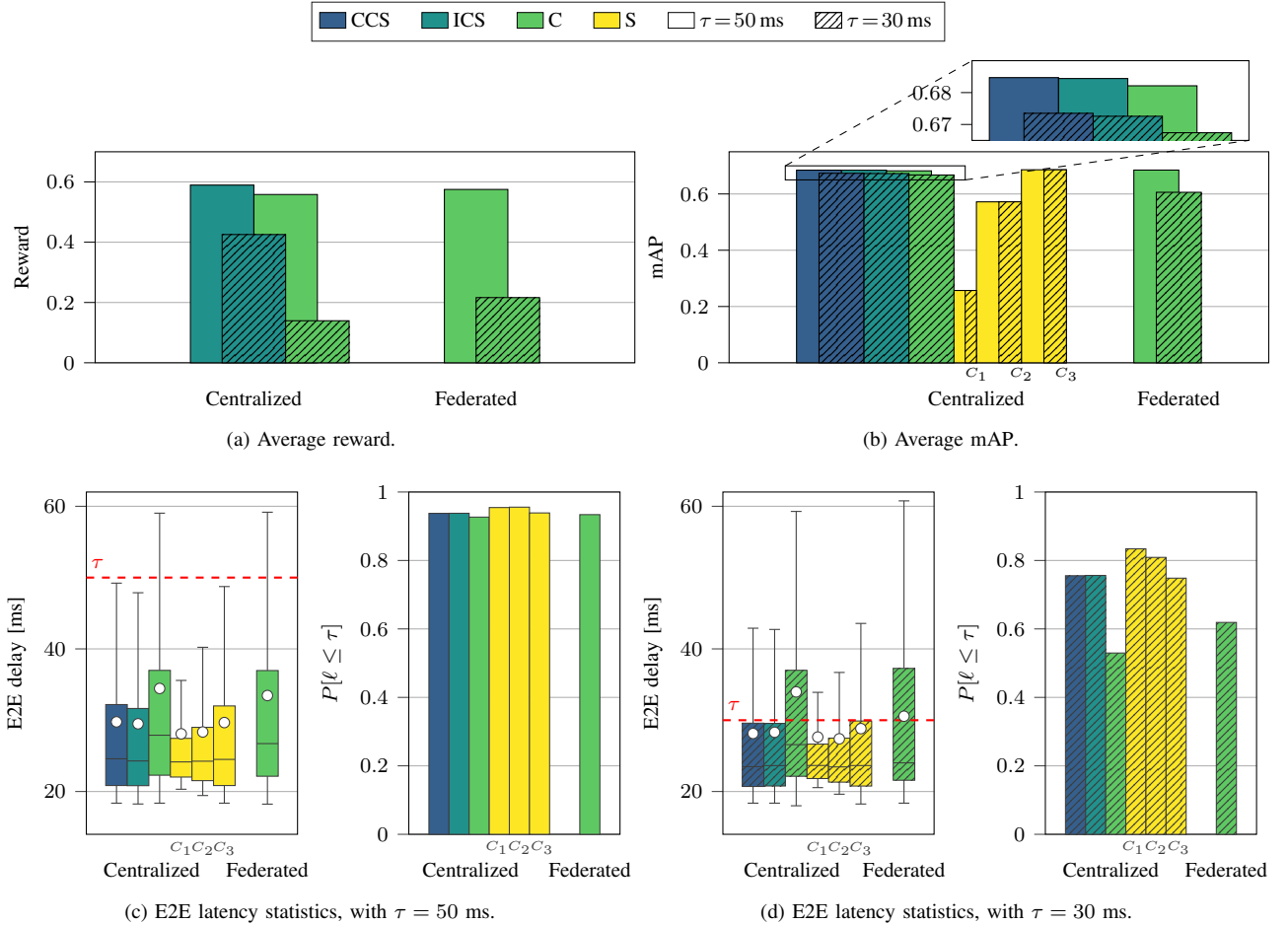

        \begin{figure*}[t!]
        \centering
            \begin{subfigure}[t]{\linewidth}
                \centering
                \input{legend_B}
            \end{subfigure}
            \centering
            \vskip 0.25 cm
            \subfloat[][Average reward.]
             {
                \label{B_r}
                \input{B_r}        
            }\hspace{-0.8cm}
            \subfloat[][Average \gls{map}.]
             {
                \label{B_mAP}
                \input{B_mAP}        
            }\\[2ex]
            \subfloat[][Average \gls{e2e} latency, with $\tau$ = 50 ms.]
             {
                \label{B_50_d}
                \input{B_50_d}        
            }\hspace{-0.8cm}
            \subfloat[][Average \gls{e2e} latency, with $\tau$ = 30 ms.]
             {
                \label{B_30_d}
                \input{B_30_d}        
            }
            \caption{Average reward, mAP, and \gls{e2e} latency in Bad (B) channel conditions, vs. the optimization framework modes.}
            \label{fig:bad}
        \end{figure*}
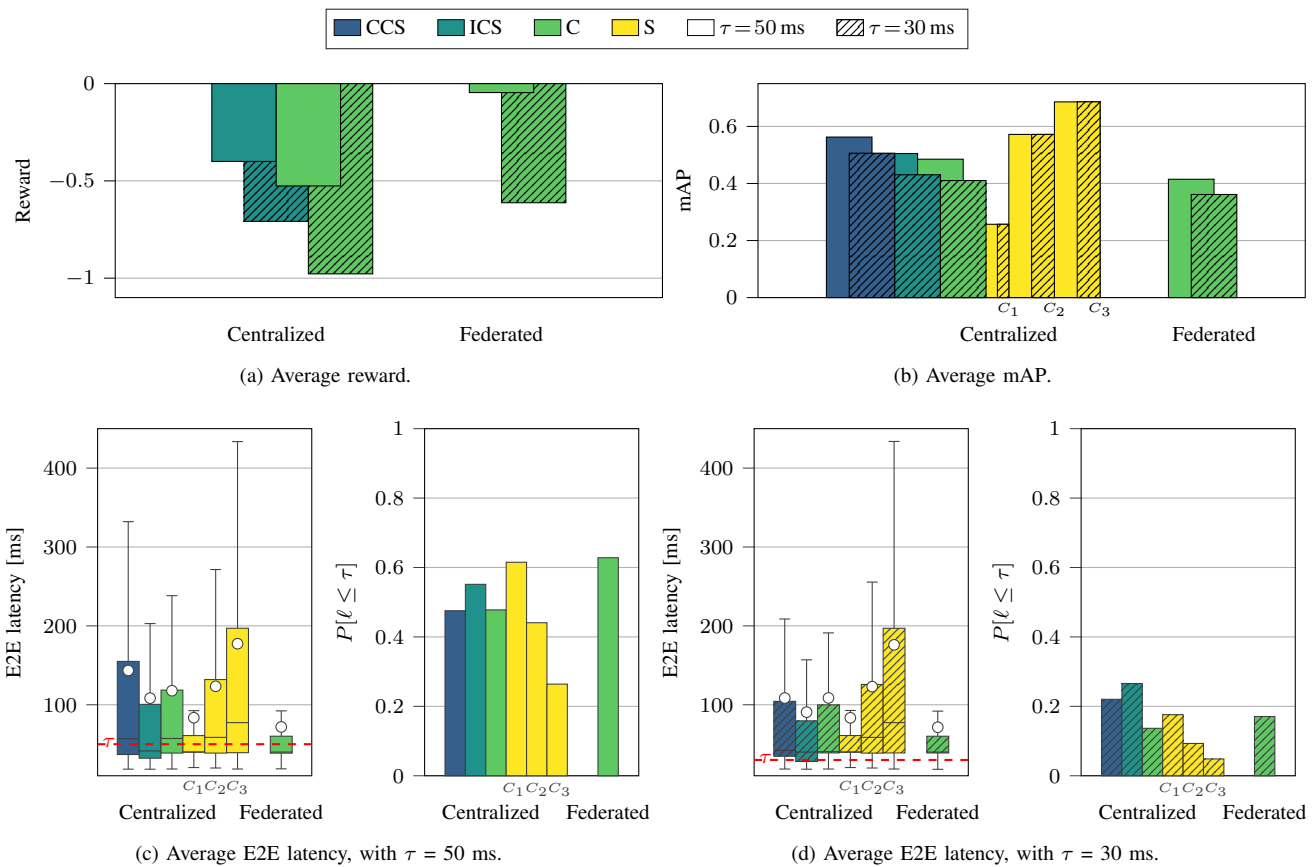
        
        \subsubsection{Metrics} We consider the following evaluation metrics: (i) the average reward per episode; (ii) the average \gls{e2e} latency $\ell$ at the application layer per episode,  measured from the time at which a LiDAR point cloud is generated at the transmitter to the time it is received, and including the  encoding delay; and (iii) the average \gls{map} $m_a$ measured on the received point cloud per episode. 
        Simulation results are given as a function of the \gls{e2e} latency threshold $\tau$, the channel quality state (Good or Bad), and the compression configuration ($\{C_1,C_2,C_3\}$).
        %As far as the meta-learning agents is concerned, the impact of the average quality of the communication channel is explored.
    
    \subsection{\gls{ca} and \gls{sa} Numerical Results}
    \label{sub:results1}

        %  \begin{figure}
        %     \centering
        %     \input{P_state_collection}
        %     \caption{Probability of state collection}
        %     \label{r}
        % \end{figure}

        %\subsubsection{Benchmarks} We compare the performance of \gls{ics} vs. \gls{ccs} operating modes, where both compression and scheduling optimization are activated for maximizing jointly \gls{qos} and \gls{qoe}. As benchmark, we consider operating modes where \gls{ca} and \gls{sa} operates in a mutual exclusive fashion, i.e., only compression or scheduling optimization is active. We denote configuration modes $\{(8, 10), (9, 5), (10, 0)\}$ as $\{C_1, C_2, C_3 \}$. For completeness, the reward of \gls{ccs} is not reported because a different reward function is defined, so \gls{e2e} delay and \gls{map} metrics are exploited in our analysis of \gls{ccs}.

        \subsubsection{Convergence and complexity} In Fig. \ref{r_lineplot} we compare the learning performance of the operating modes in terms of reward in the Bad (B) channel state. This condition represents the most challenging environment, where adverse channel quality makes the learning task less obvious and more complex, accentuating the differences among the considered strategies. 
        By design, we defined different reward functions based on the operating mode, as described in Sec.~\ref{sec:agents}. Specifically, in Fig. \ref{r_lineplot} we plot the reward of CCS based on Eq.~\eqref{R_CA_CCS} (top), the rewards of C and ICS for both centralized and federated learning based on Eq.~\eqref{R_CA_C,ICS} (center), and the reward of S based on Eq.~\eqref{eq:S} (bottom), for different compression configurations.
        We observe that all agents successfully learn an optimal policy after approximately 180 episodes, while for the S agent the learning is much faster, meaning that scheduling is easier to optimize than~compression.
        Notice that the rewards may oscillate due to poor channel conditions in the B state, causing agents to frequently select sub-optimal actions, and to take longer to learn a stable policy.

        From a computational complexity perspective, the C and S modes are the most lightweight, as only a single agent is active. Conversely, in the ICS and CCS modes both the CA and the SA operate simultaneously, increasing the overall computational cost. 
        Specifically, the CA (SA) is implemented as a DDQL (PPO) model, and requires $ P_{\rm{DDQL}} = 32|\mathcal{S^{\rm CA}_i}| + 611 $ ($ P_{\rm{PPO}} = 9476 $) parameters, where $|\mathcal{S^{\rm CA}_i}|$ is the size of the state space of the CA.
        At each learning step, DDQL updates based on a single mini-batch of size  $|\mathcal{B}|$, with complexity $ O(|\mathcal{B}|P_{\rm DDQL})$. In contrast, PPO updates based on trajectories of length $T_{\rm PPO}$ and over mini-batches of size $M$, so the complexity is $O(ETP_{\rm PPO}/M)$, where $E$ is the number of training epochs, set to 10 in our simulations.
        
        \subsubsection{Impact of the learning paradigm} %Compression is the only feasible approach in the decentralized federated framework for \gls{ca}, since unavailability of a direct access to the \gls{ran} parameters. In contrast, the centralized case offers more degrees of freedom, especially integrating an optimized scheduling of channel resources. 

        In Fig.~\ref{fig:good} we focus on the Good channel scenario (G), and plot
        %\footnote{In Figs.~\ref{G_r} and \ref{B_r} we omit the curve of the average reward relative to the CCS mode, because the CCS reward function is significantly different from those of the other modes, which would make this comparison~unfair.} 
        the average mAP, the E2E latency statistics (in the form of boxplots), and the probability that the E2E latency requirement is satisfied (i.e.,  $P[{\ell} \leq \tau]$), considering different optimization framework modes and latency thresholds~$\tau$.
        We also plot the average reward of the ICS and C modes according to Eq.~\eqref{R_CA_C,ICS}.\footnote{In Figs.~\ref{G_r} and \ref{B_r}, our intention is to compare centralized vs. federated learning, so we plot the reward of C since it is the only mode that can be implemented via both learning models. For comparison, we also plot the reward of ICS, since ICS and C have the same reward function as per Eq.~\eqref{R_CA_C,ICS}. Conversely, the other modes are defined according to different reward functions, as described in Sec.~\ref{sec:agents}, which would make the comparison unfair.}

        When $\tau = 50$ ms, all operating modes have similar performance, and both the median and the average \gls{e2e} latency generally satisfy the latency constraints, i.e., $P[{\ell} \leq \tau]\simeq 1$. This is because, 
        under good channel conditions, the network has sufficient capacity to handle data transmissions efficiently, regardless of the underlying optimization policy.
        
        Focusing on the C mode, which is the only feasible approach in federated learning, we observe that federated learning slightly outperforms centralized learning (as it appears from the average reward in Fig.~\ref{G_r}), given that the former does not require resource-consuming metric collection at the RAN, thereby reducing the control~overhead. 
       
       In general, S achieves superior performance compared to C in terms of \gls{e2e} latency (Fig.~\ref{G_50_d}), indicating that, in the G state, latency can be optimized at the scheduling level more effectively than via compression.
       In terms of \gls{map}, the performance of S strongly depends on the underlying compression configuration ($C_1$, $C_2$ or $C_3$), which is fixed~a~priori:
         %the average reward is inversely proportional to the level of data compression, dropping below 0.2 for $C_1$, compared to around 0.6 for $C_3$. In fact, 
        $C_1$ applies very aggressive compression, and the resulting \gls{map} is below 0.3, compared to around 0.7 for $C_3$ (Fig. \ref{G_mAP}).

        %In fact, the SA in S can dynamically allocate additional network resources to the most constrained UEs, thereby reducing the \gls{e2e} latency and mitigating the outliers, as depicted in Fig.~\ref{G_50_d}.
        
        Finally, ICS and CCS optimize both data compression and scheduling simultaneously, and achieve a similar performance as S using $C_3$. Actually, CCS slightly outperforms ICS in terms of mAP, given that both the CA and the SA are active and cooperate together (rather than working independently). %to find the optimal compression and scheduling configurations. 
        In particular, the compression configuration is selected based on the underlying scheduling priority, and the agent is incentivized to reduce the level of compression when the UE is assigned more network resources, despite some violations of~$\tau$.

        \subsubsection{Impact of the latency threshold} 
        From Fig.~\ref{G_r}, we observe that the reward decreases as $\tau$ decreases since the network is more constrained, and compression alone (C) is insufficient.
        For example, when $\tau=30$ ms, $P[{\ell} \leq \tau]=0.78$ for ICS/CCS vs. $0.53$ for C, and the average E2E latency decreases by more than 6 ms in ICS/CCS (Fig.~\ref{G_30_d}), with no degradation in terms of mAP. 
        %The resulting reward improves by around 3 times compared to C, and by around 7\% compared to the best S configuration.      
        These results demonstrate that combining compression and scheduling, either independently (ICS) or cooperatively (CCS), is more effective for network optimization than any standalone approach. 
        
        In this scenario, centralized learning enables better network coordination by collecting global data at the RAN, and is generally more convenient than federated learning. Specifically, for $\tau=30$ ms, the average reward improves by more than 40\% using C.

        \subsubsection{Impact of the channel state}
        In Fig.~\ref{fig:bad} we focus on the Bad channel scenario (B), which is characterized by poor network conditions and lower available capacity compared to the G scenario (Fig.~\ref{fig:good}).         
        Now, the agents are forced to adopt more aggressive compression to reduce the data rate and consume fewer radio resources, which has negative effects in terms of mAP. For this reason, the average reward drops from up to 0.6 in the G state to approximately $-1$ in the B~state.
        
        %all network metrics significantly deteriorates given the lower available channel capacity. In this case, 
        
        In this scenario, federated learning becomes a more attractive option than  centralized learning, increasing the average reward from $-0.52$ to $-0.04$ in the C mode, for $\tau=50$ ms (Fig.~\ref{B_r}). This is due to the fact that centralized learning requires an active and reliable connection with the RAN: in the B state, data collection and control signals at and from the RAN can be delayed or lost, respectively, which may increase the \gls{e2e} latency. On the contrary, federated learning relies primarily on local data, and only requires model parameter updates to be shared with the network, making the system more robust to possible channel disconnections.
        Considering $\tau=50$ ms, federated learning in the C mode outperforms even the more advanced centralized ICS mode: the reward improves from $-0.4$ to $-0.04$ (Fig.~\ref{B_r}), and the 75th percentile (upper quartile) of the E2E latency decreases from around 100 and 150 ms in ICS and CCS, respectively, to 65 ms in federated C (Fig.~\ref{B_50_d}), with very few outliers. 
        This improvement comes at the cost of a tolerable degradation of the mAP (Fig.~\ref{B_mAP}).

        Compared to the G state, compression (C) is generally more effective than resource allocation (S), since the latter is mostly implemented at the RAN, which may be unavailable in case of network disconnections in the B state. 
        Interestingly, $C_3$ (low compression) is the best S configuration in the G state, while it is the worst in the B state. This is because $C_3$ produces high data rates, that may exceed the available (limited) channel capacity under poor network conditions. Now, $C_1$ (severe compression) becomes the only S configuration capable of reducing the median latency below $\tau=50$ ms (Fig.~\ref{B_50_d}).
        %On the other hand, ICS and CCS learn to adapt the compression and scheduling configurations based on the channel conditions, and stand out as the best optimization frameworks even in the most challenging environments.

        Finally, when $\tau=30$ ms in the B state, the network is always unable to satisfy the E2E latency requirement (Fig.~\ref{B_30_d}), indicating that additional countermeasures are required to improve network performance. Nevertheless, ICS and CCS stand out as the best learning modes, given that compression and scheduling are optimized simultaneously, achieving $P[{\ell} \leq \tau]\simeq 0.22$ and $0.26$, respectively, whereas all the competitors remain below 0.2.

    \subsection{Meta-Learning Agent Numerical Results}
    \label{sub:results2}

        \begin{figure}
            \centering
            \input{G_MA_r}
            \vspace{0.3cm}
            \caption{Average reward of the meta-learning agents vs. static baselines (ICS and F), in a channel with $T^{\rm B}_{\rm DC}=1/6$ (top) and $1/2$ (bottom).}
            \label{MA_r}
        \end{figure}
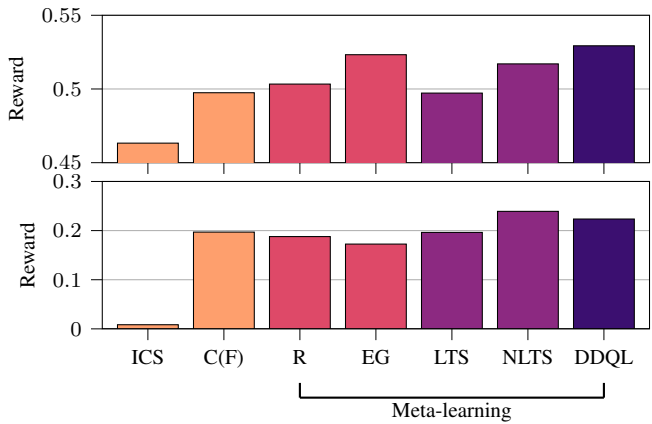

        \begin{figure}
            \centering
            \input{G_MA_mAP}
            \caption{Average mAP of the meta-learning agents vs. static baselines (ICS and F), in a channel with $T^{\rm B}_{\rm DC}=1/6$ (top) and 1/2 (bottom).}
            \label{MA_map}
        \end{figure}
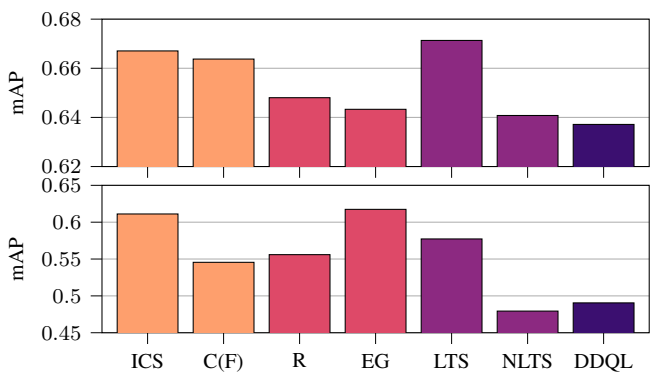

        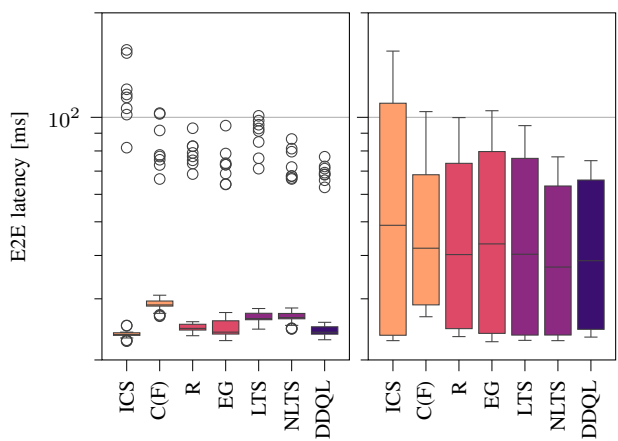
\begin{figure}
            \centering
            \input{G_MA_d}
            \caption{E2E latency of the meta-learning agents vs. static baselines (ICS and F), in a channel with $T^{\rm B}_{\rm DC}=1/6$ (left) and $1/2$ (right).}
            \label{MA_d}
        \end{figure}

        \begin{figure}
            \centering
            \input{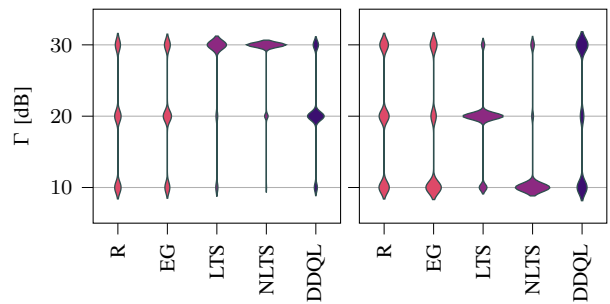}
            \caption{SINR threshold distribution of the meta-learning agents, in a channel with $T^{\rm B}_{\rm DC}=1/6$ (left) and $1/2$ (right).}
            \label{MA_tau}
        \end{figure}
        
        In this section we analyze the performance of the meta-learning agent, capable of dynamically selecting the optimal learning model (centralized or federated) based on the network conditions. We compare all the implementations described in Sec.~\ref{sub:meta}, namely R and EG (stateless), LTS and NLTS (contextual), and DDQL (stateful). We also consider two baselines, where the learning configuration is fixed: ICS (always centralized) and C(F) (always federated). 
        In our analysis, we consider two channel regimes. 
        In the first regime, we set a duty cycle $T^{\rm B}_{\rm DC} = 1/6$ (i.e., the channel transition parameters are set to $p_{GB} = 0.2$ and $p_{BG} = 1$), meaning that the channel is in the G state for most of the time, with short and sporadic transitions to the B state.
        In the second regime, the duty cycle is $T^{\rm B}_{\rm DC} = 1/2$ (i.e., the channel transition parameters are set to $p_{GB} =p_{BG} = 1$), representing a more dynamic and unstable channel (e.g., due to mobility, interference, or blockage) that frequently alternates between the G and B~states.

        We observe that all meta-learning agents outperform the baselines 
        since, unlike ICS and C(F),
        %baseline is penalized by the several low mAP (Fig.~\ref{MA_map}) and the F baseline is penalized by a high E2E latency (Fig.~\ref{MA_d}), 
        the meta-learning agents can dynamically adapt the learning model based on the network conditions. %thereby achieving the highest overall rewards (Fig.~\ref{MA_r}).
        Notice that C(F) outperforms ICS, especially when $T^{\rm B}_{\rm DC} = 1/2$, which is consistent with our previous results in Sec.~\ref{sub:results2}, i.e., federated learning is more convenient than centralized learning under poor channel conditions.
       Focusing on the case $T^{\rm B}_{\rm DC} = 1/6$, the average reward of the meta-learning agents increases with respect to the C(F) baseline (Fig.~\ref{MA_r}), while the median E2E latency decreases by up to $17\%$ (Fig.~\ref{MA_d}).
       In this context, meta-learning \gls{ddql} achieves the highest reward (0.53) while minimizing the number of E2E latency outliers, at the expense of a small degradation in terms of \gls{map} (Fig.~\ref{MA_map}).
       In fact, as illustrated in Fig.~\ref{MA_tau}, \gls{ddql} produces a more concentrated distribution of the SINR thresholds $\Gamma$ (i.e., the meta-learning agent actions) within the action space, suggesting that it learns a more stable and adaptive policy rather than oscillating between extreme values.    
       %Moreover, \gls{ddql} is able to recognize recurring patterns, and refine $\Gamma$ so as to maximize the long-term reward.
       In contrast, contextual models such as \gls{lts} and \gls{nlts} have a slightly lower reward than \gls{ddql}, since the agent makes decisions based solely on the current context, and ignores the impact of those decisions on future states. Still, \gls{nlts} outperforms \gls{lts} since the former leverages a non-linear representation of the context, which is able to capture the (inherently non-linear) evolution of the channel more accurately.
       Notice that, when $T^{\rm B}_{\rm DC} = 1/6$, the channel is predominantly stable in the G state; consequently, even a simple stateless approach such as \gls{eg} achieves similar performance as DDQL, because the reward distribution is nearly stationary during exploration.

       Next, we focus on the case $T^{\rm B}_{\rm DC} = 1/2$. In this more unstable channel regime, the average reward decreases by more than 60\% compared to $T^{\rm B}_{\rm DC} = 1/6$ (Fig.~\ref{MA_r}), while the variability of the \gls{e2e} latency and the number of extreme outliers significantly increase (Fig.~\ref{MA_d}).
       Although most meta-learning agents continue to outperform the static baselines, the optimization becomes more challenging. This is because the channel changes frequently between the G and B states, giving the agents limited time to learn a stable policy before the environment changes again.
        In this regime, stateless algorithms (R and EG) are unable to learn effectively, since random exploration fails to capture the rapid and complex  dynamics of the channel, and cannot anticipate state transitions.
        Interestingly, NLTS achieves the highest average reward (0.24) and the lowest median E2E latency (around 15 ms).
        Conversely, the DNN in DDQL implicitly relies on some degree of temporal regularity in the environment, which is not verified when the channel state changes rapidly. In this case, the learned action–value function may overfit to non-representative channel patterns observed during training, thereby degrading the resulting reward.

% \begin{figure*}[h!]
% \centering
%     \begin{subfigure}[t]{\linewidth}
%         \centering
%         \input{legend}
%     \end{subfigure}
%     \centering
%     \vskip 0.5cm
%     \subfloat[][Average Reward]
%      {
%         \label{G_50_r}
%         \input{G_50_r}        
%     }
%     \subfloat[][Average mAP]
%      {
%         \label{G_50_mAP}
%         \input{G_50_mAP}        
%     }
%     \caption{cap}
%     \label{fig}
% \end{figure*}

% \begin{figure*}[h!]
% \centering
%     \subfloat[][$\tau$ = 50 ms]
%      {
%         \label{G_50_d}
%         \input{G_50_d}        
%     }
%     \subfloat[][$\tau$ = 25 ms]
%      {
%         \label{G_25_d}
%         \input{G_25_d}        
%     }
%     \caption{cap}
%     \label{fig}
% \end{figure*}

\section{Conclusions and Future Work}
\label{sec:conclusions}
In this paper we investigated possible \gls{pqos} strategies for \gls{td} applications, and propose novel RL-based frameworks that jointly optimize data compression and radio resource scheduling. Specifically, we designed two interacting and cooperative agents capable of balancing the stringent \gls{e2e} latency requirements of \gls{td} applications with the quality of the transmitted sensory data. In addition, we introduced a meta-learning agent that dynamically selects between centralized and decentralized (federated) learning paradigms based on real-time network conditions.
Extensive full-stack ns-3 simulations demonstrated that the proposed cooperative compression-and-scheduling strategy (ICS and CCS) consistently outperforms any other standalone solution, particularly in constrained scenarios characterized by strict latency requirements or poor channel capacity.
Under favorable channel conditions, centralized learning enables effective coordination at the RAN, and improves the reward by roughly $3\times$ compared to compression-only solutions, and by about 7\% compared to the best scheduling-only configuration.
In more unstable channels, federated learning proves more robust as it primarily relies on local data and requires minimum interactions with the RAN, and the average rewards improves by up to 75\% compared to centralized ICS.
These complementary behaviors motivate the adoption of the proposed meta-learning agent, which successfully adapts the learning strategy to the underlying channel conditions.
Specifically, DDQL is the best meta-learning implementation in good and stable channels, while NLTS is more effective in highly dynamic channels.

As part of our future work, we plan to evaluate the proposed meta-learning agent in different (non-TD) scenarios, such as in robotic or industrial control systems, and extend the framework to optimize conventional network metrics based on the performance of the underlying control application.

\bibliography{ref}{}
\bibliographystyle{IEEEtran}

\end{document}

%% file: acronyms.tex
\newacronym{quic}{QUIC}{Quick UDP Internet Connections}
\newacronym{3gpp}{3GPP}{3rd Generation Partnership Project}
\newacronym{adc}{ADC}{Analog to Digital Converter}
\newacronym{5g}{5G}{5th Generation}
\newacronym{5gaa}{5GAA}{5G Automotive Association}
\newacronym{aimd}{AIMD}{Additive Increase Multiplicative Decrease}
\newacronym{am}{AM}{Acknowledged Mode}
\newacronym{amc}{AMC}{Adaptive Modulation and Coding}
\newacronym{aqm}{AQM}{Active Queue Management}
\newacronym{awgn}{AGWN}{Additive White Gaussian Noise}
\newacronym{balia}{BALIA}{Balanced Link Adaptation}
\newacronym{bdp}{BDP}{Bandwidth-Delay Product}
\newacronym{bf}{BF}{Beamforming}
\newacronym{cc}{CC}{Congestion Control}
\newacronym{pdf}{PDF}{Probability Density Function}
\newacronym{cdf}{CDF}{Cumulative Distribution Function}
\newacronym{icdf}{Inverse CDF}{Inverse Cumulative Distribution Function}
\newacronym{c-v2x}{C-V2X}{Cellular Vehicle-To-Everything}
\newacronym{ci}{CI}{Close-in free space reference}
\newacronym{cn}{CN}{Core Network}
\newacronym{cqi}{CQI}{Channel Quality Information}
\newacronym{cp}{CP}{Control Plane}
\newacronym{csirs}{CSI-RS}{Channel State Information - Reference Signal}
\newacronym{d2d}{D2D}{Device-to-Device}
\newacronym{dc}{DC}{Dual Connectivity}
\newacronym{dce}{DCE}{Direct Code Execution}
\newacronym{dci}{DCI}{Downlink Control Information}
\newacronym{dl}{DL}{downlink}
\newacronym{dmr}{DMR}{Deadline Miss Ratio}
\newacronym{dmrs}{DMRS}{DeModulation Reference Signal}
\newacronym{dsrc}{DSRC}{Dedicated Short-Range Communication}
\newacronym{e2e}{E2E}{end-to-end}
\newacronym{ecn}{ECN}{Explicit Congestion Notification}
\newacronym{edf}{EDF}{Earliest Deadline First}
\newacronym{enb}{eNB}{evolved Node Base}
\newacronym{epc}{EPC}{Evolved Packet Core}
\newacronym{es}{ES}{Edge Server}
\newacronym{fdma}{FDMA}{Frequency Division Multiple Access}
\newacronym{fdd}{FDD}{Frequency Division Duplexing}
\newacronym[firstplural=Radio Access Technologies (RATs)]{rat}{RAT}{Radio Access Technology}
\newacronym{fs}{FS}{Fast Switching}
\newacronym{ftp}{FTP}{File Transfer Protocol}
\newacronym{gnb}{gNB}{Next Generation Node Base}
\newacronym{harq}{HARQ}{Hybrid Automatic Repeat reQuest}
\newacronym{hetnet}{HetNet}{Heterogeneous Network}
\newacronym{hh}{HH}{Hard Handover}
\newacronym{hol}{HOL}{Head-of-Line}
\newacronym{ia}{IA}{Initial Access}
\newacronym{imt}{IMT}{International Mobile Telecommunication}
\newacronym{iot}{IoT}{Internet of Things}
\newacronym{lidar}{LiDAR}{Light Detection and Ranging}
\newacronym{los}{LOS}{Line of Sight}
\newacronym{lte}{LTE}{Long Term Evolution}
\newacronym{m2m}{M2M}{Machine to Machine}
\newacronym{mac}{MAC}{Medium Access Control}
\newacronym{mc}{MC}{Multi-Connectivity}
\newacronym{mcs}{MCS}{Modulation and Coding Scheme}
\newacronym{mec}{MEC}{Mobile Edge Cloud}
\newacronym{mi}{MI}{Mutual Information}
\newacronym{mimo}{MIMO}{Multiple Input, Multiple Output}
\newacronym{mmwave}{mmWave}{millimeter wave}
\newacronym{ml}{ML}{machine learning}
\newacronym{mr}{MR}{Maximum Rate}
\newacronym{mss}{MSS}{Maximum Segment Size}
\newacronym{mtd}{MTD}{Machine-Type Device}
\newacronym{mtu}{MTU}{Maximum Transmission Unit}
\newacronym{nn}{NN}{Neural Network}
\newacronym{nsf}{NSF}{National Science Foundation}
\newacronym{nfv}{NFV}{Network Function Virtualization}
\newacronym{nlos}{NLOS}{Non Line of Sight}
\newacronym{nr}{NR}{New Radio}
\newacronym{ofdm}{OFDM}{Orthogonal Frequency Division Multiplexing}
\newacronym{pc}{PC}{Point Cloud}
\newacronym{pdcch}{PDCCH}{Physical Downlonk Control Channel}
\newacronym{pdcp}{PDCP}{Packet Data Convergence Protocol}
\newacronym{pdsch}{PDSCH}{Physical Downlink Shared Channel}
\newacronym{pdu}{PDU}{Packet Data Unit}
\newacronym{pf}{PF}{Proportional Fair}
\newacronym{pgw}{PGW}{Packet Gateway}
\newacronym{phy}{PHY}{Physical}
\newacronym{pbch}{PBCH}{Physical Broadcast Channel}
\newacronym[plural=\gls{mme}s,firstplural=Mobility Management Entities (MMEs)]{mme}{MME}{Mobility Management Entity}
\newacronym{prb}{PRB}{Physical Resource Block}
\newacronym{pss}{PSS}{Primary Synchronization Signal}
\newacronym{pucch}{PUCCH}{Physical Uplink Control Channel}
\newacronym{pusch}{PUSCH}{Physical Uplink Shared Channel}
\newacronym{qos}{QoS}{Quality of Service}
\newacronym{rach}{RACH}{Random Access Channel}
\newacronym{ran}{RAN}{Radio Access Network}
\newacronym{red}{RED}{Random Early Detection}
\newacronym{rf}{RF}{Radio Frequency}
\newacronym{rlc}{RLC}{Radio Link Control}
\newacronym{rlf}{RLF}{Radio Link Failure}
\newacronym{rrc}{RRC}{Radio Resource Control}
\newacronym{rrm}{RRM}{Radio Resource Management}
\newacronym{rr}{RR}{Round Robin}
\newacronym{rs}{RS}{Remote Server}
\newacronym{rsrp}{RSRP}{Reference Signal Received Power}
\newacronym{rss}{RSS}{Received Signal Strength}
\newacronym{rtt}{RTT}{Round Trip Time}
\newacronym{rw}{RW}{Receive Window}
\newacronym{rx}{RX}{Receiver}
\newacronym{sack}{SACK}{Selective Acknowledgment}
\newacronym{sap}{SAP}{Service Access Point}
\newacronym{sch}{SCH}{Secondary Cell Handover}
\newacronym{scoot}{SCOOT}{Split Cycle Offset Optimization Technique}
\newacronym{sdma}{SDMA}{Spatial Division Multiple Access}
\newacronym{sinr}{SINR}{Signal to Interference plus Noise Ratio}
\newacronym{sm}{SM}{Saturation Mode}
\newacronym{snr}{SNR}{Signal to Noise Ratio}
\newacronym{son}{SON}{Self-Organizing Network}
\newacronym{ss}{SS}{Synchronization Signal}
\newacronym{srs}{SRS}{Sounding Reference Signal}
\newacronym{sss}{SSS}{Secondary Synchronization Signal}
\newacronym{tb}{TB}{Transport Block}
\newacronym{tcp}{TCP}{Transmission Control Protocol}
\newacronym{udp}{UDP}{User Datagram Protocol}
\newacronym{fr1}{FR1}{Frequency Range 1}
\newacronym{fr2}{FR2}{Frequency Range 2}
\newacronym{tdd}{TDD}{Time Division Duplexing}
\newacronym{tdma}{TDMA}{Time Division Multiple Access}
\newacronym{tfl}{TfL}{Transport for London}
\newacronym{thz}{THz}{Terahertz}
\newacronym{tm}{TM}{Transparent Mode}
\newacronym{trp}{TRP}{Transmitter Receiver Pair}
\newacronym{tti}{TTI}{Transmission Time Interval}
\newacronym{ttt}{TTT}{Time-to-Trigger}
\newacronym{tx}{TX}{Transmitter}
\newacronym{ue}{UE}{User Equipment}
\newacronym{ul}{UL}{uplink}
\newacronym{uml}{UML}{Unified Modeling Language}
\newacronym{um}{UM}{Unacknowledged Mode}
\newacronym{utc}{UTC}{Urban Traffic Control}
\newacronym{v2v}{V2V}{Vehicle-to-Vehicle}
\newacronym{vm}{VM}{Virtual Machine}
\newacronym{rsrq}{RSRQ}{Reference Signal Received Quality}
\newacronym{rssi}{RSSI}{Received Signal Strength Indicator}
\newacronym{crs}{CRS}{Cell Reference Signal}
\newacronym{comp}{CoMP}{Coordinated Multi-Point}
\newacronym{cran}{C-RAN}{Cloud \acrlong{ran}}
\newacronym{cco}{CC}{Carrier Component}
\newacronym{nsa}{NSA}{Non Stand Alone}
\newacronym{embb}{eMBB}{Enhanced Mobility Broadband}
\newacronym{bsr}{BSR}{Buffer Status Report}
\newacronym{srb}{SRB}{Service Radio Bearer}
\newacronym{scm}{SCM}{Spatial Channel Model}
\newacronym{sctp}{SCTP}{Stream Control Transmission Protocol}
\newacronym{mptcp}{MPTCP}{Multi-path TCP}
\newacronym{ietf}{IETF}{Internet Engineering Task Force}
\newacronym{os}{OS}{Operating System}
\newacronym{tls}{TLS}{Transport Layer Security}
\newacronym{rfc}{RFC}{Request for Comments}
\newacronym{http}{HTTP}{HyperText Transfer Protocol}
\newacronym{nat}{NAT}{Network Address Translation}
\newacronym{api}{API}{Application Programming Interface}
\newacronym{rto}{RTO}{Retransmission Timeout}
\newacronym{psc}{PSC}{Public Safety Communication}
\newacronym{rpgm}{RPGM}{Reference Point Group Mobility}
\newacronym{ic}{IC}{Incident Command}
\newacronym{rsu}{RSU}{Road Side Unit}
\newacronym{uav}{UAV}{Unmanned Aerial Vehicle}
\newacronym{usa}{U.S.}{United States}
\newacronym{vr}{VR}{Virtual Reality}
\newacronym{iab}{IAB}{Integrated Access and Backhaul}
\newacronym{wlan}{WLAN}{Wireless Local Area Network}
\newacronym{cots}{COTS}{Commercial Off-the-Shelf}
\newacronym{fpga}{FPGA}{Field Programmable Gate Array}
\newacronym{rcn}{RCN}{Research Coordination Network}
\newacronym{abg}{ABG}{Alpha-Beta-Gamma}
\newacronym{fi}{FI}{Floating Intercept}
\newacronym{uas}{UAS}{Unmanned Aerial System}
\newacronym{gps}{GPS}{Global Positioning System}
\newacronym{a2g}{A2G}{air-to-ground}
\newacronym{a2a}{A2A}{air-to-air}
\newacronym{uma}{UMa}{Urban Macro}
\newacronym{umi}{UMi}{Urban Micro}
\newacronym{upa}{UPA}{Uniform Planar Array}
\newacronym{rma}{RMa}{Rural Macro}
\newacronym{inoo}{InOo}{Indoor Open Office}
\newacronym{ple}{PLE}{path loss exponent}
\newacronym{aoa}{AoA}{Angle of Arrival}
\newacronym{aod}{AoD}{Angle of Departure}
\newacronym{toa}{ToA}{Time of Arrival}
\newacronym{mpc}{MPC}{Multi-path Component}
\newacronym{cir}{CIR}{Channel Impulse Response}
\newacronym{rt}{RT}{Ray-tracing}
\newacronym{tc}{TC}{Time Cluster}
\newacronym{sl}{SL}{Spatial Lobe}
\newacronym{svd}{SVD}{Singular Value Decomposition}
\newacronym{6g}{6G}{sixth generation}
\newacronym{ns3}{ns-3}{Network Simulator 3}
\newacronym{fsc}{FS}{Fully Stochastic}
\newacronym{hbc}{HB}{Hybrid}
\newacronym{hpbw}{HPBW}{Half Power Beamwidth}
\newacronym{hsc}{HSC}{Hybrid Semantic Compression}
\newacronym{prr}{PRR}{Packet Receipt Rate}
\newacronym{v2x}{V2X}{Vehicle-To-Everything}

\newacronym{ai}{AI}{artificial intelligence}

\newacronym{pqos}{PQoS}{Predictive Quality of Service}
\newacronym{rl}{RL}{Reinforcement Learning}
\newacronym{marl}{MARL}{Multi-Agent Reinforcement Learning}
\newacronym{mdp}{MDP}{Markov Decision Process}
\newacronym{pomdp}{POMDP}{Partially Observable MDP}
\newacronym{dec-pomdp}{Dec-POMDP}{Decentralized POMDP}
\newacronym{fl}{FL}{federated learning}
\newacronym{sgd}{SGD}{Stochastic Gradient Descent}
\newacronym{ddqn}{DDQN}{Double Deep Q-Network}
\newacronym{dql}{DQL}{Deep Q-Learning}
\newacronym{td}{TD}{teleoperated driving}
\newacronym{map}{mAP}{mean Average Precision}
\newacronym{kpi}{KPI}{Key Performance Indicator}
\newacronym{mab}{MAB}{Multi-Armed Bandit}
\newacronym{dsarsa}{DSARSA}{Deep SARSA}
\newacronym{sarsa}{SARSA}{State–Action–Reward–State–Action}
\newacronym{qoe}{QoE}{Quality of Experience}
\newacronym{ppo}{PPO}{Proximal Policy Optimization}
\newacronym{trpo}{TRPO}{Trust Region Policy Optimization}
\newacronym{gae}{GAE}{Generalized Advantage Estimation}
\newacronym{ctde}{CTDE}{centralized training with decentralized execution}
\newacronym{ippo}{IPPO}{Independent PPO}
\newacronym{mappo}{MAPPO}{Multi-Agent PPO}
\newacronym{pa}{PA}{Proportional Allocation}
\newacronym{ga}{GA}{Greedy Allocation}
\newacronym{ca}{CA}{Compression Agent}
\newacronym{sa}{SA}{Scheduling Agent }
\newacronym{ddql}{DDQL}{Double Deep Q-Learning}
\newacronym{ics}{ICS}{Independent Compression and Scheduling}
\newacronym{ccs}{CCS}{Cooperative Compression and Scheduling}
\newacronym{ts}{TS}{Thompson Sampling}
\newacronym{lts}{LTS}{Linear Thompson Sampling}
\newacronym{nlts}{NLTS}{Neural Linear Thompson Sampling}
\newacronym{dnn}{DNN}{Deep Neural Network}
\newacronym{lstm}{LSTM}{Long Short-Term Memory}
\newacronym{rnn}{RNN}{Recurrent Neural Network}
\newacronym{urllc}{URLLC}{Ultra-Reliable Low-Latency Communication}
\newacronym{drl}{DRL}{Deep Reinforcement Learning}
\newacronym{eg}{EG}{Epsilon Greedy}

%% file: mc.tex
\usetikzlibrary{arrows,automata}

\begin{tikzpicture}[node distance=2cm,->,>=latex,auto,
  every edge/.append style={thick}]
  \node[state] (1) {G};
  \node[state] (2) [right of=1] {B};  
  \path (1) edge[loop left]  node{$1-p_{\rm GB}$} (1)
            edge[bend left]  node{$p_{\rm GB}$}   (2)
        (2) edge[loop right] node{$1-p_{BG}$}  (2)
            edge[bend left] node{$p_{BG}$}     (1);
\end{tikzpicture}

%% file: r_lineplot.tex
% This file was created with tikzplotlib v0.10.1.
\pgfplotsset{
tick label style={font=\footnotesize},
label style={font=\footnotesize},
legend  style={font=\footnotesize}
}

\begin{tikzpicture}

\definecolor{darkgray176}{RGB}{176,176,176}
\definecolor{darkorange25512714}{RGB}{255,127,14}
\definecolor{lightgray204}{RGB}{204,204,204}
\definecolor{steelblue31119180}{RGB}{31,119,180}

\begin{groupplot}[
    group style={
        group size=1 by 3, % One column, two rows
        vertical sep=0.25cm, % Vertical spacing between the plots
    },
    width=0.95\columnwidth, % Adjust the width of the plots
    height=0.55\columnwidth % Adjust the height of the plots
]

\nextgroupplot[
legend style={at={(0.5,1.3)},
legend entries={CCS,
                ICS,
                S,
                C,
                Centralized,
                Federated},
anchor=north,legend columns=3},
tick align=outside,
tick pos=left,
x grid style={darkgray176},
xlabel={},
xmajorgrids,
xmin=-1.2, xmax=25.2,
xtick style={color=black},
xtick={0,5,10,15,20,25},
xticklabels={},
y grid style={darkgray176},
ylabel={$R_{\rm CA}^{\rm CCS}$},
ymajorgrids,
ymin=-0.85, ymax=-0.15,
ytick style={color=black}
]

\addlegendimage{black, semithick, mark = triangle}
\addlegendimage{black, semithick, mark = o}
\addlegendimage{black, semithick}
\addlegendimage{black, semithick, mark = x}
\addlegendimage{steelblue31119180, thick}
\addlegendimage{darkorange25512714, thick}

\path [draw=steelblue31119180, fill=steelblue31119180, opacity=0.2]
(axis cs:0,-0.724478679646783)
--(axis cs:0,-0.792509554929403)
--(axis cs:1,-0.612564135834755)
--(axis cs:2,-0.459282487586355)
--(axis cs:3,-0.473260934538814)
--(axis cs:4,-0.433368983765236)
--(axis cs:5,-0.456286396736512)
--(axis cs:6,-0.3804791748062)
--(axis cs:7,-0.378543629838446)
--(axis cs:8,-0.352202138856224)
--(axis cs:9,-0.325190079406165)
--(axis cs:10,-0.367330481882057)
--(axis cs:11,-0.337366970075306)
--(axis cs:12,-0.305391676799392)
--(axis cs:13,-0.374293097490966)
--(axis cs:14,-0.373581149540248)
--(axis cs:15,-0.321160991306431)
--(axis cs:16,-0.371675130158169)
--(axis cs:17,-0.386906329174876)
--(axis cs:18,-0.24542609138393)
--(axis cs:19,-0.254754222630885)
--(axis cs:20,-0.309571285862707)
--(axis cs:21,-0.334134408227541)
--(axis cs:22,-0.413699253167413)
--(axis cs:23,-0.316837090870296)
--(axis cs:24,-0.421669943281519)
--(axis cs:24,-0.372144608732069)
--(axis cs:24,-0.372144608732069)
--(axis cs:23,-0.287196835599662)
--(axis cs:22,-0.370469073574575)
--(axis cs:21,-0.300946508100853)
--(axis cs:20,-0.279826677669057)
--(axis cs:19,-0.228020909885108)
--(axis cs:18,-0.220990242749311)
--(axis cs:17,-0.345551243392127)
--(axis cs:16,-0.337806386516831)
--(axis cs:15,-0.290549881405135)
--(axis cs:14,-0.338070371422157)
--(axis cs:13,-0.33595949943209)
--(axis cs:12,-0.275439777453878)
--(axis cs:11,-0.307542759979349)
--(axis cs:10,-0.327266863968167)
--(axis cs:9,-0.290333384115394)
--(axis cs:8,-0.313784333995675)
--(axis cs:7,-0.341152812021012)
--(axis cs:6,-0.342128807938678)
--(axis cs:5,-0.411294472118432)
--(axis cs:4,-0.393734172032126)
--(axis cs:3,-0.425647334306242)
--(axis cs:2,-0.410432701547125)
--(axis cs:1,-0.563191314879027)
--(axis cs:0,-0.724478679646783)
--cycle;

\addplot [semithick, steelblue31119180, mark=triangle, forget plot]
table {%
0 -0.758257552381271
1 -0.589236631778676
2 -0.434606697316089
3 -0.449148859485308
4 -0.412944732256151
5 -0.432439286918769
6 -0.361116328773431
7 -0.360473392537362
8 -0.332132027946155
9 -0.308294211638932
10 -0.345944411284906
11 -0.322652693674137
12 -0.290008636451122
13 -0.355049737679891
14 -0.356263177400873
15 -0.305244973057301
16 -0.355247171099938
17 -0.366629740892228
18 -0.23303717294684
19 -0.241232399023823
20 -0.294939148452136
21 -0.317063402983152
22 -0.391507175476309
23 -0.301039070171708
24 -0.397785941256479
};

\nextgroupplot[
tick align=outside,
tick pos=left,
x grid style={darkgray176},
xlabel={},
xmajorgrids,
xmin=-1.2, xmax=25.2,
xtick style={color=black},
xtick={0,5,10,15,20,25},
xticklabels={},
y grid style={darkgray176},
ylabel={$R_{\rm CA}^{\rm\{ C, ICS\}}$},
ymajorgrids,
ymin=-1.5, ymax=-0.4,
ytick style={color=black}
]

\path [draw=steelblue31119180, fill=steelblue31119180, opacity=0.2]
(axis cs:0,-0.959897653002298)
--(axis cs:0,-1.02847021823369)
--(axis cs:1,-0.955679437543415)
--(axis cs:2,-0.93150161726392)
--(axis cs:3,-0.999081610024734)
--(axis cs:4,-0.974299151049038)
--(axis cs:5,-0.956107858068773)
--(axis cs:6,-0.902302507417242)
--(axis cs:7,-0.871417972029676)
--(axis cs:8,-0.871598847332878)
--(axis cs:9,-0.795737841995531)
--(axis cs:10,-0.774770732373743)
--(axis cs:11,-0.777070226838958)
--(axis cs:12,-0.94622237880897)
--(axis cs:13,-0.983608064643311)
--(axis cs:14,-0.906503818973599)
--(axis cs:15,-0.750697587464854)
--(axis cs:16,-0.962183822643317)
--(axis cs:17,-0.822807103998261)
--(axis cs:18,-0.782427346214157)
--(axis cs:19,-0.819932972809435)
--(axis cs:20,-0.713045735869727)
--(axis cs:21,-0.661704142731348)
--(axis cs:22,-0.727935017170398)
--(axis cs:23,-0.782830699910545)
--(axis cs:24,-0.750733794411971)
--(axis cs:24,-0.706102494541603)
--(axis cs:24,-0.706102494541603)
--(axis cs:23,-0.736168025105583)
--(axis cs:22,-0.682052117314548)
--(axis cs:21,-0.621144152994348)
--(axis cs:20,-0.666310216687378)
--(axis cs:19,-0.762852712799831)
--(axis cs:18,-0.732625662557494)
--(axis cs:17,-0.768308617678299)
--(axis cs:16,-0.899233066245818)
--(axis cs:15,-0.703981758656389)
--(axis cs:14,-0.846860505161883)
--(axis cs:13,-0.923338408502798)
--(axis cs:12,-0.888777685742481)
--(axis cs:11,-0.732899319670262)
--(axis cs:10,-0.727501760509945)
--(axis cs:9,-0.742102388579239)
--(axis cs:8,-0.817088485068686)
--(axis cs:7,-0.818268753336888)
--(axis cs:6,-0.845473929635789)
--(axis cs:5,-0.897621470717659)
--(axis cs:4,-0.912114721955504)
--(axis cs:3,-0.934434964481288)
--(axis cs:2,-0.868506221540319)
--(axis cs:1,-0.895931972691866)
--(axis cs:0,-0.959897653002298)
--cycle;

\path [draw=steelblue31119180, fill=steelblue31119180, opacity=0.2]
(axis cs:0,-1.15184155476012)
--(axis cs:0,-1.2503030982983)
--(axis cs:1,-1.18365268059699)
--(axis cs:2,-1.15990674836733)
--(axis cs:3,-1.2560457520106)
--(axis cs:4,-1.16390595300598)
--(axis cs:5,-1.18864549129232)
--(axis cs:6,-1.17922818468217)
--(axis cs:7,-1.24818528714718)
--(axis cs:8,-1.21977988307785)
--(axis cs:9,-1.18728957509376)
--(axis cs:10,-1.17899022795076)
--(axis cs:11,-1.06581769578373)
--(axis cs:12,-1.10603460884703)
--(axis cs:13,-1.0468355574469)
--(axis cs:14,-1.02063407900761)
--(axis cs:15,-1.06705386924061)
--(axis cs:16,-1.28748627546654)
--(axis cs:17,-1.44048048920031)
--(axis cs:18,-0.937795619479427)
--(axis cs:19,-0.866734984726196)
--(axis cs:20,-1.16098158446689)
--(axis cs:21,-1.05952795752546)
--(axis cs:22,-0.908417215826902)
--(axis cs:23,-0.906002411612633)
--(axis cs:24,-1.24341900107957)
--(axis cs:24,-1.15074138947484)
--(axis cs:24,-1.15074138947484)
--(axis cs:23,-0.847451223315006)
--(axis cs:22,-0.855102719664072)
--(axis cs:21,-0.993856217946735)
--(axis cs:20,-1.07664629698148)
--(axis cs:19,-0.80767489993451)
--(axis cs:18,-0.87778436369733)
--(axis cs:17,-1.32829767739218)
--(axis cs:16,-1.19461856954945)
--(axis cs:15,-0.998301414537078)
--(axis cs:14,-0.955944402338504)
--(axis cs:13,-0.978089590258249)
--(axis cs:12,-1.03125261003066)
--(axis cs:11,-0.988173018884634)
--(axis cs:10,-1.09643756895525)
--(axis cs:9,-1.09600406166282)
--(axis cs:8,-1.13182716588295)
--(axis cs:7,-1.15382052357657)
--(axis cs:6,-1.09451351881727)
--(axis cs:5,-1.10512120817437)
--(axis cs:4,-1.07770491547105)
--(axis cs:3,-1.167713855627)
--(axis cs:2,-1.07035958302515)
--(axis cs:1,-1.09754237575011)
--(axis cs:0,-1.15184155476012)
--cycle;

\path [draw=darkorange25512714, fill=darkorange25512714, opacity=0.2]
(axis cs:0,-0.7541903493079)
--(axis cs:0,-0.7888381447185)
--(axis cs:1,-0.753348495819225)
--(axis cs:2,-0.770200182282637)
--(axis cs:3,-0.806105946482247)
--(axis cs:4,-0.75450225329365)
--(axis cs:5,-0.714569369920612)
--(axis cs:6,-0.697245839599338)
--(axis cs:7,-0.75946347591445)
--(axis cs:8,-0.608604476031938)
--(axis cs:9,-0.718769408051839)
--(axis cs:10,-0.757170351384775)
--(axis cs:11,-0.535488165671438)
--(axis cs:12,-0.781511505698546)
--(axis cs:13,-0.634335922800787)
--(axis cs:14,-0.75237884993962)
--(axis cs:15,-0.581223164107713)
--(axis cs:16,-0.720525565435033)
--(axis cs:17,-0.623342589954075)
--(axis cs:18,-0.663591551525562)
--(axis cs:19,-0.557574034311825)
--(axis cs:20,-0.576451338669238)
--(axis cs:21,-0.695261940278262)
--(axis cs:22,-0.623661718348713)
--(axis cs:23,-0.518415208647237)
--(axis cs:24,-0.667152113381487)
--(axis cs:24,-0.63723547098655)
--(axis cs:24,-0.63723547098655)
--(axis cs:23,-0.49876377569425)
--(axis cs:22,-0.5971723123219)
--(axis cs:21,-0.663526033488175)
--(axis cs:20,-0.552754656757425)
--(axis cs:19,-0.532175080811025)
--(axis cs:18,-0.63591103485755)
--(axis cs:17,-0.595136038856825)
--(axis cs:16,-0.690592944272477)
--(axis cs:15,-0.556714741415687)
--(axis cs:14,-0.717551013969732)
--(axis cs:13,-0.608603693913913)
--(axis cs:12,-0.747494042276148)
--(axis cs:11,-0.514675575545188)
--(axis cs:10,-0.72452269615855)
--(axis cs:9,-0.68672601052884)
--(axis cs:8,-0.582040028653737)
--(axis cs:7,-0.726486871045438)
--(axis cs:6,-0.668449240255637)
--(axis cs:5,-0.683419886745013)
--(axis cs:4,-0.722842696440538)
--(axis cs:3,-0.770670152258608)
--(axis cs:2,-0.7373711410523)
--(axis cs:1,-0.723361630247425)
--(axis cs:0,-0.7541903493079)
--cycle;

\addplot [semithick, steelblue31119180, mark = o, forget plot]
table {%
0 -0.994485964465657
1 -0.925298617788604
2 -0.89888790513986
3 -0.966791433068087
4 -0.944160203240185
5 -0.92673058599028
6 -0.874382631926937
7 -0.845132265408633
8 -0.845393144087336
9 -0.769870045960721
10 -0.749824885291966
11 -0.753774739639048
12 -0.916950014777862
13 -0.95278514762053
14 -0.874269741833333
15 -0.728275360480206
16 -0.9311534790004250
17 -0.795059144571645
18 -0.757059142711665
19 -0.791278296482109
20 -0.688676408255822
21 -0.641194949245455
22 -0.706380431384992
23 -0.758767501167483
24 -0.729198853334348
};
\addplot [semithick, steelblue31119180, mark = x, forget plot]
table {%
0 -1.20056836181256
1 -1.14248188339503
2 -1.11695311238063
3 -1.20936740810051
4 -1.11612406655814
5 -1.14561293651397
6 -1.13632418449438
7 -1.19960447303972
8 -1.17558920603376
9 -1.14181550990277
10 -1.13918781239055
11 -1.02514426757436
12 -1.06919966525854
13 -1.01361230588381
14 -0.988205858619139
15 -1.03015056897729
16 -1.24354244742088
17 -1.38602744109877
18 -0.907486619515458
19 -0.836946064336124
20 -1.12106945101133
21 -1.02584954519092
22 -0.881781821864097
23 -0.874973694941874
24 -1.19632094588328
};

\addplot [semithick, darkorange25512714, mark = x, forget plot]
table {%
0 -0.771508945606
1 -0.738086055504
2 -0.752721053928
3 -0.787510556680139
4 -0.737769606871
5 -0.6985054642425
6 -0.683024900716
7 -0.742680472662
8 -0.5956324957695
9 -0.702238480949914
10 -0.740184660288
11 -0.5246954491125
12 -0.764372420903054
13 -0.6213039950205
14 -0.735207044361042
15 -0.5690903524905
16 -0.705582942178916
17 -0.6084972506425
18 -0.6497775299955
19 -0.5439500505115
20 -0.565024620704
21 -0.6790360473635
22 -0.610142442201
23 -0.508275805531
24 -0.6523109005145
};

\nextgroupplot[
tick align=outside,
tick pos=left,
x grid style={darkgray176},
xlabel={Episode},
xmajorgrids,
xmin=-1.2, xmax=25.2,
xtick style={color=black},
xtick={0,5,10,15,20,25},
xticklabels={0,50,100,150,200,250},
y grid style={darkgray176},
ylabel={$R_{\rm SA}^{\rm S}$},
ymajorgrids,
ymin=-1.7, ymax=-0.15,
ytick style={color=black}
]
\path [draw=steelblue31119180, fill=steelblue31119180, opacity=0.2]
(axis cs:0,-0.361530145470204)
--(axis cs:0,-0.410191374571201)
--(axis cs:1,-0.350553483242364)
--(axis cs:2,-0.359718455144294)
--(axis cs:3,-0.38381568636896)
--(axis cs:4,-0.367168280846108)
--(axis cs:5,-0.390444093197951)
--(axis cs:6,-0.360462252229478)
--(axis cs:7,-0.355938910635293)
--(axis cs:8,-0.364387200988749)
--(axis cs:9,-0.408536065301863)
--(axis cs:10,-0.345936325617686)
--(axis cs:11,-0.36853470587044)
--(axis cs:12,-0.391981972096605)
--(axis cs:13,-0.415245008547722)
--(axis cs:14,-0.415033260865482)
--(axis cs:15,-0.38481922495281)
--(axis cs:16,-0.436984723714476)
--(axis cs:17,-0.497685468460554)
--(axis cs:18,-0.392028706150721)
--(axis cs:19,-0.352609487280419)
--(axis cs:20,-0.399574870840083)
--(axis cs:21,-0.404633101890618)
--(axis cs:22,-0.435838268333918)
--(axis cs:23,-0.403686138006126)
--(axis cs:24,-0.335362714027769)
--(axis cs:24,-0.292282464858699)
--(axis cs:24,-0.292282464858699)
--(axis cs:23,-0.351251107732552)
--(axis cs:22,-0.387155721200399)
--(axis cs:21,-0.357188861478051)
--(axis cs:20,-0.35065630366405)
--(axis cs:19,-0.307602352541326)
--(axis cs:18,-0.343717207014743)
--(axis cs:17,-0.446848911931992)
--(axis cs:16,-0.3883107496308)
--(axis cs:15,-0.336215939742418)
--(axis cs:14,-0.365212687735703)
--(axis cs:13,-0.367329305550878)
--(axis cs:12,-0.342593546901603)
--(axis cs:11,-0.319267940482647)
--(axis cs:10,-0.301000595246807)
--(axis cs:9,-0.359611443485076)
--(axis cs:8,-0.321558555685871)
--(axis cs:7,-0.307407696553195)
--(axis cs:6,-0.317210302561148)
--(axis cs:5,-0.346600085657545)
--(axis cs:4,-0.318398458603547)
--(axis cs:3,-0.338990455281675)
--(axis cs:2,-0.311890667252794)
--(axis cs:1,-0.304954590222831)
--(axis cs:0,-0.361530145470204)
--cycle;

\path [draw=steelblue31119180, fill=steelblue31119180, opacity=0.2]
(axis cs:0,-0.824138018832487)
--(axis cs:0,-0.918504190531497)
--(axis cs:1,-0.865907282272893)
--(axis cs:2,-0.899339468464895)
--(axis cs:3,-0.995706245598831)
--(axis cs:4,-0.937727404524409)
--(axis cs:5,-0.889597473636364)
--(axis cs:6,-0.86010769662774)
--(axis cs:7,-0.983160967855869)
--(axis cs:8,-0.914682069233102)
--(axis cs:9,-0.927067989204637)
--(axis cs:10,-0.984574151396317)
--(axis cs:11,-0.878725950009825)
--(axis cs:12,-0.941867355627664)
--(axis cs:13,-0.990946490084454)
--(axis cs:14,-0.908431374229366)
--(axis cs:15,-0.876877496812171)
--(axis cs:16,-0.979098527067124)
--(axis cs:17,-1.04612512819707)
--(axis cs:18,-0.975087115023285)
--(axis cs:19,-0.846803148101178)
--(axis cs:20,-0.92719658622834)
--(axis cs:21,-0.940830510291015)
--(axis cs:22,-0.94934833055524)
--(axis cs:23,-0.810826896924033)
--(axis cs:24,-0.869518760155341)
--(axis cs:24,-0.778292011049564)
--(axis cs:24,-0.778292011049564)
--(axis cs:23,-0.72265175137445)
--(axis cs:22,-0.85240339867016)
--(axis cs:21,-0.844406239576762)
--(axis cs:20,-0.825859705032751)
--(axis cs:19,-0.758196580251856)
--(axis cs:18,-0.881482334592911)
--(axis cs:17,-0.94370733654168)
--(axis cs:16,-0.881432229872571)
--(axis cs:15,-0.78912270534726)
--(axis cs:14,-0.808490129955993)
--(axis cs:13,-0.883358422496004)
--(axis cs:12,-0.826709614255888)
--(axis cs:11,-0.786642900424491)
--(axis cs:10,-0.887744259261533)
--(axis cs:9,-0.819652701212555)
--(axis cs:8,-0.806775215564916)
--(axis cs:7,-0.87837415563485)
--(axis cs:6,-0.770702912752928)
--(axis cs:5,-0.799188008848909)
--(axis cs:4,-0.834665972802031)
--(axis cs:3,-0.894358830324435)
--(axis cs:2,-0.806300569922478)
--(axis cs:1,-0.765729024979293)
--(axis cs:0,-0.824138018832487)
--cycle;

\path [draw=steelblue31119180, fill=steelblue31119180, opacity=0.2]
(axis cs:0,-1.41292833690633)
--(axis cs:0,-1.57941616998917)
--(axis cs:1,-1.516977867173)
--(axis cs:2,-1.4535340317369)
--(axis cs:3,-1.5384098028434)
--(axis cs:4,-1.42662269180749)
--(axis cs:5,-1.37441840391012)
--(axis cs:6,-1.47964066149299)
--(axis cs:7,-1.51987322349019)
--(axis cs:8,-1.54053376578948)
--(axis cs:9,-1.46070796601154)
--(axis cs:10,-1.62797513154755)
--(axis cs:11,-1.50267713572697)
--(axis cs:12,-1.57928472845292)
--(axis cs:13,-1.65158432370492)
--(axis cs:14,-1.55298394715134)
--(axis cs:15,-1.45886825032632)
--(axis cs:16,-1.53489827600795)
--(axis cs:17,-1.64083121596678)
--(axis cs:18,-1.41361187224806)
--(axis cs:19,-1.48963186876439)
--(axis cs:20,-1.57177410673394)
--(axis cs:21,-1.50068278191534)
--(axis cs:22,-1.57653110357801)
--(axis cs:23,-1.42504219450778)
--(axis cs:24,-1.46821423076308)
--(axis cs:24,-1.31650357889762)
--(axis cs:24,-1.31650357889762)
--(axis cs:23,-1.27245093572217)
--(axis cs:22,-1.4268506070892)
--(axis cs:21,-1.34791918039452)
--(axis cs:20,-1.41911240240697)
--(axis cs:19,-1.34595138358888)
--(axis cs:18,-1.27221831580838)
--(axis cs:17,-1.47366199799549)
--(axis cs:16,-1.37957768587764)
--(axis cs:15,-1.28903602692951)
--(axis cs:14,-1.38971040338392)
--(axis cs:13,-1.48142234354918)
--(axis cs:12,-1.41844175348742)
--(axis cs:11,-1.35370626841106)
--(axis cs:10,-1.47688084184616)
--(axis cs:9,-1.31001842832432)
--(axis cs:8,-1.38041150381387)
--(axis cs:7,-1.35601948159096)
--(axis cs:6,-1.33828337836799)
--(axis cs:5,-1.23352128488675)
--(axis cs:4,-1.28107982944486)
--(axis cs:3,-1.37959312143816)
--(axis cs:2,-1.28421286337089)
--(axis cs:1,-1.35569685860244)
--(axis cs:0,-1.41292833690633)
--cycle;

\addplot [semithick, steelblue31119180, forget plot]
table {%
0 -0.386634053418803
1 -0.327666536129564
2 -0.336537648779356
3 -0.361847480489763
4 -0.343703666305982
5 -0.368145119890462
6 -0.338303424313987
7 -0.331452089811707
8 -0.342161369763089
9 -0.383733720427885
10 -0.323248998305494
11 -0.345013527169492
12 -0.3671777245752
13 -0.391055439911371
14 -0.390677669291566
15 -0.360065953271973
16 -0.413473775695238
17 -0.471946474083778
18 -0.369236644830733
19 -0.329745694673995
20 -0.375423739206677
21 -0.381565757537222
22 -0.411154001360701
23 -0.377577562326729
24 -0.314099693004461
};

\addplot [semithick, steelblue31119180, forget plot]
table {%
0 -0.873400651674699
1 -0.816924384841017
2 -0.851846123788761
3 -0.945618759722525
4 -0.887270920717873
5 -0.843173663112727
6 -0.814936974460961
7 -0.931129872454959
8 -0.860389956370664
9 -0.873413817046369
10 -0.9363362408125
11 -0.834254376742358
12 -0.881599424241121
13 -0.934711977352826
14 -0.858143051188356
15 -0.832632622166548
16 -0.930657623802508
17 -0.992554085672026
18 -0.927619162646341
19 -0.801795189321913
20 -0.876243838838428
21 -0.890431282692486
22 -0.90234836073357
23 -0.767607539006823
24 -0.822739811255814
};

\addplot [semithick, steelblue31119180, forget plot]
table {%
0 -1.49613587996458
1 -1.434964631645
2 -1.36984410854937
3 -1.45656016028269
4 -1.35292245007426
5 -1.3018190773075
6 -1.40772155129454
7 -1.43340194785018
8 -1.45940069553351
9 -1.38380841828876
10 -1.55577851642857
11 -1.42720683703589
12 -1.49965859670648
13 -1.56706122387962
14 -1.47266192099767
15 -1.37671158951944
16 -1.4534925334236
17 -1.55645495977984
18 -1.3394381808615
19 -1.41719242428773
20 -1.49671494731287
21 -1.42372303723221
22 -1.49893854577053
23 -1.34720765216222
24 -1.38863887699756
};

\node [black] at (-0.6, -0.4) {\tiny$C_1$};
\node [black] at (-0.6, -0.9) {\tiny$C_2$};
\node [black] at (-0.6, -1.5) {\tiny$C_3$};

\end{groupplot}

\end{tikzpicture}

%% file: legend_G.tex
\definecolor{brown1926061}{RGB}{192,60,61}
\definecolor{darkgray176}{RGB}{176,176,176}
\definecolor{darkslategray66}{RGB}{66,66,66}
\definecolor{lightgray204}{RGB}{204,204,204}
\definecolor{mediumpurple147113178}{RGB}{147,113,178}
\definecolor{peru22412844}{RGB}{224,128,44}
\definecolor{seagreen5814558}{RGB}{58,145,58}
\definecolor{steelblue49115161}{RGB}{49,115,161}

\definecolor{viridis1}{RGB}{53, 95, 141}
\definecolor{viridis2}{RGB}{33, 145, 140}
\definecolor{viridis3}{RGB}{94, 201, 98}
\definecolor{viridis4}{RGB}{253, 231, 37}

\pgfplotsset{
tick label style={font=\footnotesize},
label style={font=\footnotesize},
legend  style={font=\footnotesize}
}
\begin{tikzpicture}
\pgfplotsset{every tick label/.append style={font=\scriptsize}}

\pgfplotsset{compat=1.11,
  /pgfplots/ybar legend/.style={
    /pgfplots/legend image code/.code={%
      \draw[##1,/tikz/.cd,yshift=-0.25em]
      (0cm,0cm) rectangle (10pt,0.6em);},
  },
}

\begin{axis}[%
width=0,
height=0,
at={(0,0)},
scale only axis,
xmin=0,
xmax=0,
xtick={},
ymin=0,
ymax=0,
ytick={},
axis background/.style={fill=white},
legend style={legend cell align=left,
              align=center,
              draw=white!15!black,
              at={(0.5, 1.3)},
              anchor=center,
              /tikz/every even column/.append style={column sep=1em}},
legend columns=7,
]
\addplot[ybar,ybar legend,draw=black,fill=viridis1,line width=0.08pt]
table[row sep=crcr]{%
  0  0\\
};
\addlegendentry{CCS}

\addplot[ybar legend,ybar,draw=black,fill=viridis2,fill opacity=1.0,line width=0.08pt]
  table[row sep=crcr]{%
  0  0\\
};
\addlegendentry{ICS}

\addplot[ybar legend,ybar,draw=black,fill=viridis3,fill opacity=1.0,line width=0.08pt]
  table[row sep=crcr]{%
  0  0\\
};
\addlegendentry{C}

\addplot[ybar legend,ybar,draw=black,fill=viridis4,fill opacity=1.0,line width=0.08pt]
  table[row sep=crcr]{%
  0  0\\
};
\addlegendentry{S}

\addplot[ybar legend,ybar,draw=black,fill=white,fill opacity=1.0,line width=0.08pt]
  table[row sep=crcr]{%
  0  0\\
};
\addlegendentry{$\tau\!=\!50\,$ms}

\addplot[postaction={pattern=north east lines}, ybar legend,ybar,draw=black,fill=white,fill opacity=1.0,line width=0.08pt]
  table[row sep=crcr]{%
  0  0\\
};
\addlegendentry{$\tau\!=\!30\,$ms}

\end{axis}
\end{tikzpicture}%

%% file: G_r.tex
% This file was created with tikzplotlib v0.10.1.
\pgfplotsset{
tick label style={font=\footnotesize},
label style={font=\footnotesize},
legend  style={font=\footnotesize}
}
\begin{tikzpicture}

\definecolor{viridis1}{RGB}{53, 95, 141}
\definecolor{viridis2}{RGB}{33, 145, 140}
\definecolor{viridis3}{RGB}{94, 201, 98}
\definecolor{viridis4}{RGB}{253, 231, 37}

\definecolor{brown1926061}{RGB}{192,60,61}
\definecolor{darkgray176}{RGB}{176,176,176}
\definecolor{darkslategray66}{RGB}{66,66,66}
\definecolor{lightgray204}{RGB}{204,204,204}
\definecolor{mediumpurple147113178}{RGB}{147,113,178}
\definecolor{peru22412844}{RGB}{224,128,44}
\definecolor{seagreen5814558}{RGB}{58,145,58}
\definecolor{steelblue49115161}{RGB}{49,115,161}
\definecolor{antiquewhite239222208}{RGB}{239,222,208}
\definecolor{antiquewhite240226214}{RGB}{240,226,214}
\definecolor{antiquewhite240229220}{RGB}{240,229,220}
\definecolor{antiquewhite241231223}{RGB}{241,231,223}
\definecolor{bisque239218200}{RGB}{239,218,200}
\definecolor{brown1926061}{RGB}{192,60,61}
\definecolor{burlywood233184141}{RGB}{233,184,141}
\definecolor{burlywood235196161}{RGB}{235,196,161}
\definecolor{cadetblue119161191}{RGB}{119,161,191}
\definecolor{crimson2143940}{RGB}{214,39,40}
\definecolor{darkgray144177201}{RGB}{144,177,201}
\definecolor{darkgray166139191}{RGB}{166,139,191}
\definecolor{darkgray176}{RGB}{176,176,176}
\definecolor{darkorange25512714}{RGB}{255,127,14}
\definecolor{darksalmon231169116}{RGB}{231,169,116}
\definecolor{darkseagreen123181123}{RGB}{123,181,123}
\definecolor{darkseagreen146193146}{RGB}{146,193,146}
\definecolor{darkseagreen165204165}{RGB}{165,204,165}
\definecolor{darkslategray61}{RGB}{61,61,61}
\definecolor{dimgray114}{RGB}{114,114,114}
\definecolor{forestgreen4416044}{RGB}{44,160,44}
\definecolor{gainsboro209221229}{RGB}{209,221,229}
\definecolor{gainsboro213230213}{RGB}{213,230,213}
\definecolor{gainsboro215225232}{RGB}{215,225,232}
\definecolor{gainsboro217208225}{RGB}{217,208,225}
\definecolor{gainsboro218232218}{RGB}{218,232,218}
\definecolor{gainsboro220228234}{RGB}{220,228,234}
\definecolor{gainsboro222215229}{RGB}{222,215,229}
\definecolor{gainsboro222234221}{RGB}{222,234,221}
\definecolor{gainsboro224236224}{RGB}{224,236,224}
\definecolor{gainsboro226220231}{RGB}{226,220,231}
\definecolor{gainsboro227237227}{RGB}{227,237,227}
\definecolor{gainsboro229224233}{RGB}{229,224,233}
\definecolor{gainsboro234212212}{RGB}{234,212,212}
\definecolor{gainsboro235217217}{RGB}{235,217,217}
\definecolor{gainsboro237222222}{RGB}{237,222,222}
\definecolor{indianred2029798}{RGB}{202,97,98}
\definecolor{lavender224230236}{RGB}{224,230,236}
\definecolor{lavender227232237}{RGB}{227,232,237}
\definecolor{lavender230234238}{RGB}{230,234,238}
\definecolor{lavender231227235}{RGB}{231,227,235}
\definecolor{lavender233230236}{RGB}{233,230,236}
\definecolor{lavender235232237}{RGB}{235,232,237}
\definecolor{lavender236234238}{RGB}{236,234,238}
\definecolor{lightgray201223201}{RGB}{201,223,201}
\definecolor{lightgray202216226}{RGB}{202,216,226}
\definecolor{lightgray204}{RGB}{204,204,204}
\definecolor{lightgray208227208}{RGB}{208,227,208}
\definecolor{lightgray232205205}{RGB}{232,205,205}
\definecolor{lightpink226183184}{RGB}{226,183,184}
\definecolor{lightsteelblue164191210}{RGB}{164,191,210}
\definecolor{lightsteelblue179201216}{RGB}{179,201,216}
\definecolor{lightsteelblue191209222}{RGB}{191,209,222}
\definecolor{linen238228228}{RGB}{238,228,228}
\definecolor{linen239231231}{RGB}{239,231,231}
\definecolor{linen241233226}{RGB}{241,233,226}
\definecolor{linen241234229}{RGB}{241,234,229}
\definecolor{mediumpurple147113178}{RGB}{147,113,178}
\definecolor{mediumpurple148103189}{RGB}{148,103,189}
\definecolor{mediumseagreen9416594}{RGB}{94,165,94}
\definecolor{mistyrose238226226}{RGB}{238,226,226}
\definecolor{palevioletred210126127}{RGB}{210,126,127}
\definecolor{peru22412844}{RGB}{224,128,44}
\definecolor{rosybrown217150150}{RGB}{217,150,150}
\definecolor{sandybrown22815183}{RGB}{228,151,83}
\definecolor{seagreen5814558}{RGB}{58,145,58}
\definecolor{silver180212180}{RGB}{180,212,180}
\definecolor{silver181160201}{RGB}{181,160,201}
\definecolor{silver191218191}{RGB}{191,218,191}
\definecolor{silver194176209}{RGB}{194,176,209}
\definecolor{steelblue31119180}{RGB}{31,119,180}
\definecolor{steelblue49115161}{RGB}{49,115,161}
\definecolor{steelblue88141177}{RGB}{88,141,177}
\definecolor{tan222169169}{RGB}{222,169,169}
\definecolor{thistle203190216}{RGB}{203,190,216}
\definecolor{thistle211200221}{RGB}{211,200,221}
\definecolor{thistle229195195}{RGB}{229,195,195}
\definecolor{wheat237205177}{RGB}{237,205,177}
\definecolor{wheat238212190}{RGB}{238,212,190}

\begin{axis}[
height=0.50\columnwidth,
width=\columnwidth,
legend style={legend cell align=left,
              align=center,
              draw=white!15!black,
              at={(0.5,1.45)},
              anchor=center,
              /tikz/every even column/.append style={column sep=1em},
legend columns=1,
anchor=north,legend rows=2},
tick align=outside,
tick pos=left,
x grid style={darkgray176},
xmin=-0.52, xmax=0.16,
xtick style={color=black},
xtick={-0.32,-0.04},
xticklabels = {Centralized, Federated},
clip=false,
y grid style={darkgray176},
xticklabel shift={3pt},
ylabel={Reward},
ymajorgrids,
ymin=0, ymax=0.7,
xtick style={draw=none},
ytick style={color=black},
]
% \draw[draw=black,fill=viridis1] (axis cs:-0.48,0) rectangle (axis cs:-0.40,0.588576500020125);
\draw[draw=black,fill=viridis2] (axis cs:-0.4,0) rectangle (axis cs:-0.32,0.589505963846);
\draw[draw=black,fill=viridis3] (axis cs:-0.32,0) rectangle (axis cs:-0.24,0.558254485750523);
\draw[draw=black,fill=viridis3] (axis cs:-0.08,0) rectangle (axis cs:0,0.575044944309);

% \draw[postaction={pattern=north east lines},draw=black,fill=viridis1] (axis cs:-0.44,0) rectangle (axis cs:-0.36,0.424639863169146);
\draw[postaction={pattern=north east lines},draw=black,fill=viridis2] (axis cs:-0.36,0) rectangle (axis cs:-0.28,0.425417210598287);
\draw[postaction={pattern=north east lines},draw=black,fill=viridis3] (axis cs:-0.28,0) rectangle (axis cs:-0.2,0.139436742286779);
\draw[postaction={pattern=north east lines},draw=black,fill=viridis3] (axis cs:-0.04,0) rectangle (axis cs:0.04,0.216232938008833);

% \addplot [line width=0.9pt, darkslategray66, forget plot]
% table {%
% -0.36 0.564530176814455
% -0.36 0.56975396959421
% };
% \addplot [line width=0.9pt, darkslategray66, forget plot]
% table {%
% 0.64 0.59153763790257
% 0.64 0.595511039887238
% };
% \addplot [line width=0.9pt, darkslategray66, forget plot]
% table {%
% -0.20 0.603215667488685
% -0.20 0.606723860231615
% };
% \addplot [line width=0.9pt, darkslategray66, forget plot]
% table {%
% -0.04 0.228725206063478
% -0.04 0.230499560890395
% };
% \addplot [line width=0.9pt, darkslategray66, forget plot]
% table {%
% 0.12 0.530760004829933
% 0.12 0.533011537248994
% };
% \addplot [line width=0.9pt, darkslategray66, forget plot]
% table {%
% 0.28 0.601263766599824
% 0.28 0.605041629595162
% };

\end{axis}

\end{tikzpicture}

%% file: G_mAP.tex
% This file was created with tikzplotlib v0.10.1.
\pgfplotsset{
tick label style={font=\footnotesize},
label style={font=\footnotesize},
legend  style={font=\footnotesize}
}
\usetikzlibrary{calc}
\begin{tikzpicture}

\definecolor{viridis1}{RGB}{53, 95, 141}
\definecolor{viridis2}{RGB}{33, 145, 140}
\definecolor{viridis3}{RGB}{94, 201, 98}
\definecolor{viridis4}{RGB}{253, 231, 37}

\definecolor{brown1926061}{RGB}{192,60,61}
\definecolor{darkgray176}{RGB}{176,176,176}
\definecolor{darkslategray66}{RGB}{66,66,66}
\definecolor{lightgray204}{RGB}{204,204,204}
\definecolor{mediumpurple147113178}{RGB}{147,113,178}
\definecolor{peru22412844}{RGB}{224,128,44}
\definecolor{seagreen5814558}{RGB}{58,145,58}
\definecolor{steelblue49115161}{RGB}{49,115,161}
\definecolor{antiquewhite239222208}{RGB}{239,222,208}
\definecolor{antiquewhite240226214}{RGB}{240,226,214}
\definecolor{antiquewhite240229220}{RGB}{240,229,220}
\definecolor{antiquewhite241231223}{RGB}{241,231,223}
\definecolor{bisque239218200}{RGB}{239,218,200}
\definecolor{brown1926061}{RGB}{192,60,61}
\definecolor{burlywood233184141}{RGB}{233,184,141}
\definecolor{burlywood235196161}{RGB}{235,196,161}
\definecolor{cadetblue119161191}{RGB}{119,161,191}
\definecolor{crimson2143940}{RGB}{214,39,40}
\definecolor{darkgray144177201}{RGB}{144,177,201}
\definecolor{darkgray166139191}{RGB}{166,139,191}
\definecolor{darkgray176}{RGB}{176,176,176}
\definecolor{darkorange25512714}{RGB}{255,127,14}
\definecolor{darksalmon231169116}{RGB}{231,169,116}
\definecolor{darkseagreen123181123}{RGB}{123,181,123}
\definecolor{darkseagreen146193146}{RGB}{146,193,146}
\definecolor{darkseagreen165204165}{RGB}{165,204,165}
\definecolor{darkslategray61}{RGB}{61,61,61}
\definecolor{dimgray114}{RGB}{114,114,114}
\definecolor{forestgreen4416044}{RGB}{44,160,44}
\definecolor{gainsboro209221229}{RGB}{209,221,229}
\definecolor{gainsboro213230213}{RGB}{213,230,213}
\definecolor{gainsboro215225232}{RGB}{215,225,232}
\definecolor{gainsboro217208225}{RGB}{217,208,225}
\definecolor{gainsboro218232218}{RGB}{218,232,218}
\definecolor{gainsboro220228234}{RGB}{220,228,234}
\definecolor{gainsboro222215229}{RGB}{222,215,229}
\definecolor{gainsboro222234221}{RGB}{222,234,221}
\definecolor{gainsboro224236224}{RGB}{224,236,224}
\definecolor{gainsboro226220231}{RGB}{226,220,231}
\definecolor{gainsboro227237227}{RGB}{227,237,227}
\definecolor{gainsboro229224233}{RGB}{229,224,233}
\definecolor{gainsboro234212212}{RGB}{234,212,212}
\definecolor{gainsboro235217217}{RGB}{235,217,217}
\definecolor{gainsboro237222222}{RGB}{237,222,222}
\definecolor{indianred2029798}{RGB}{202,97,98}
\definecolor{lavender224230236}{RGB}{224,230,236}
\definecolor{lavender227232237}{RGB}{227,232,237}
\definecolor{lavender230234238}{RGB}{230,234,238}
\definecolor{lavender231227235}{RGB}{231,227,235}
\definecolor{lavender233230236}{RGB}{233,230,236}
\definecolor{lavender235232237}{RGB}{235,232,237}
\definecolor{lavender236234238}{RGB}{236,234,238}
\definecolor{lightgray201223201}{RGB}{201,223,201}
\definecolor{lightgray202216226}{RGB}{202,216,226}
\definecolor{lightgray204}{RGB}{204,204,204}
\definecolor{lightgray208227208}{RGB}{208,227,208}
\definecolor{lightgray232205205}{RGB}{232,205,205}
\definecolor{lightpink226183184}{RGB}{226,183,184}
\definecolor{lightsteelblue164191210}{RGB}{164,191,210}
\definecolor{lightsteelblue179201216}{RGB}{179,201,216}
\definecolor{lightsteelblue191209222}{RGB}{191,209,222}
\definecolor{linen238228228}{RGB}{238,228,228}
\definecolor{linen239231231}{RGB}{239,231,231}
\definecolor{linen241233226}{RGB}{241,233,226}
\definecolor{linen241234229}{RGB}{241,234,229}
\definecolor{mediumpurple147113178}{RGB}{147,113,178}
\definecolor{mediumpurple148103189}{RGB}{148,103,189}
\definecolor{mediumseagreen9416594}{RGB}{94,165,94}
\definecolor{mistyrose238226226}{RGB}{238,226,226}
\definecolor{palevioletred210126127}{RGB}{210,126,127}
\definecolor{peru22412844}{RGB}{224,128,44}
\definecolor{rosybrown217150150}{RGB}{217,150,150}
\definecolor{sandybrown22815183}{RGB}{228,151,83}
\definecolor{seagreen5814558}{RGB}{58,145,58}
\definecolor{silver180212180}{RGB}{180,212,180}
\definecolor{silver181160201}{RGB}{181,160,201}
\definecolor{silver191218191}{RGB}{191,218,191}
\definecolor{silver194176209}{RGB}{194,176,209}
\definecolor{steelblue31119180}{RGB}{31,119,180}
\definecolor{steelblue49115161}{RGB}{49,115,161}
\definecolor{steelblue88141177}{RGB}{88,141,177}
\definecolor{tan222169169}{RGB}{222,169,169}
\definecolor{thistle203190216}{RGB}{203,190,216}
\definecolor{thistle211200221}{RGB}{211,200,221}
\definecolor{thistle229195195}{RGB}{229,195,195}
\definecolor{wheat237205177}{RGB}{237,205,177}
\definecolor{wheat238212190}{RGB}{238,212,190}

\begin{axis}[
height=0.50\columnwidth,
width=\columnwidth,
legend style={legend cell align=left,
              align=center,
              draw=white!15!black,
              at={(0.5,1.45)},
              anchor=center,
              /tikz/every even column/.append style={column sep=1em},
legend columns=1,
anchor=north,legend rows=2},
tick align=outside,
tick pos=left,
x grid style={darkgray176},
xmin=-0.6, xmax=0.36,
xtick style={color=black},
xtick={-0.16,0.20},
xticklabels = {Centralized, Federated},
y grid style={darkgray176},
xticklabel shift={3pt},
ylabel={mAP},
ymajorgrids,
ymin=0, ymax=0.75,
xtick style={draw=none},
clip=false,
name=ax1,
ytick style={color=black},
]
\draw[draw=black,fill=viridis1] (axis cs:-0.48,0) rectangle (axis cs:-0.40,0.684695225);
\draw[draw=black,fill=viridis2] (axis cs:-0.4,0) rectangle (axis cs:-0.32,0.684410225);
\draw[draw=black,fill=viridis3] (axis cs:-0.32,0) rectangle (axis cs:-0.24,0.682130031364349);
\draw[draw=black,fill=viridis4, fill opacity=1.0] (axis cs:-0.24,0) rectangle (axis cs:-0.16,0.257);
\draw[draw=black,fill=viridis4, fill opacity=1.0] (axis cs:-0.16,0) rectangle (axis cs:-0.08,0.572);
\draw[draw=black,fill=viridis4, fill opacity=1.0] (axis cs:-0.08,0) rectangle (axis cs:0,0.686);
\draw[draw=black,fill=viridis3] (axis cs:0.12,0) rectangle (axis cs:0.20,0.6849275);

\draw[postaction={pattern=north east lines},draw=black,fill=viridis1] (axis cs:-0.44,0) rectangle (axis cs:-0.36,0.673500916666667);
\draw[postaction={pattern=north east lines},draw=black,fill=viridis2] (axis cs:-0.36,0) rectangle (axis cs:-0.28,0.672573489820892);
\draw[postaction={pattern=north east lines},draw=black,fill=viridis3] (axis cs:-0.28,0) rectangle (axis cs:-0.2,0.667409068053749);
\draw[postaction={pattern=north east lines},draw=black,fill=viridis4, fill opacity=1.0] (axis cs:-0.18,0) rectangle (axis cs:-0.16,0.257);
\draw[postaction={pattern=north east lines},draw=black,fill=viridis4, fill opacity=1.0] (axis cs:-0.12,0) rectangle (axis cs:-0.08,0.572);
\draw[postaction={pattern=north east lines},draw=black,fill=viridis4, fill opacity=1.0] (axis cs:-0.04,0) rectangle (axis cs:0,0.686);
\draw[postaction={pattern=north east lines},draw=black,fill=viridis3] (axis cs:0.16,0) rectangle (axis cs:0.24,0.605904383333333);

% define coordinates at bottom left and top left of rectangle
\coordinate (c1) at (axis cs:-0.5,0.65);
\coordinate (c2) at (axis cs:-0.5,0.70);
\coordinate (c3) at (axis cs:-0.18,0.65);
% draw a rectangle
\draw (c1) rectangle (axis cs:-0.18,0.70);

% \addplot [line width=0.9pt, darkslategray66, forget plot]
% table {%
% -0.36 0.564530176814455
% -0.36 0.56975396959421
% };
% \addplot [line width=0.9pt, darkslategray66, forget plot]
% table {%
% 0.64 0.59153763790257
% 0.64 0.595511039887238
% };
% \addplot [line width=0.9pt, darkslategray66, forget plot]
% table {%
% -0.20 0.603215667488685
% -0.20 0.606723860231615
% };
% \addplot [line width=0.9pt, darkslategray66, forget plot]
% table {%
% -0.04 0.228725206063478
% -0.04 0.230499560890395
% };
% \addplot [line width=0.9pt, darkslategray66, forget plot]
% table {%
% 0.12 0.530760004829933
% 0.12 0.533011537248994
% };
% \addplot [line width=0.9pt, darkslategray66, forget plot]
% table {%
% 0.28 0.601263766599824
% 0.28 0.605041629595162
% };

\node [black] at (-0.16, -0.04) {\tiny$C_1$};
\node [black] at (-0.08, -0.04) {\tiny$C_2$};
\node [black] at (0.0, -0.04) {\tiny$C_3$};

\end{axis}

\begin{axis}[
   height=0.3\columnwidth,
   width=0.6\columnwidth,
   name=ax2,
   scaled ticks=false,
   tick align=outside,
   tick pos=left,
   xmin=-0.5,xmax=-0.18,
   ymin=0.665,ymax=0.69,
   ytick = {0.67,0.68},
   at={($(ax1.south east)+(-4cm,3cm)$)},
   xticklabels={draw=none},
   xtick style={draw=none},
   clip=true,
   ytick style={color=black}
 ]
    \draw[draw=black,fill=viridis1] (axis cs:-0.48,0) rectangle (axis cs:-0.40,0.684695225);
\draw[draw=black,fill=viridis2] (axis cs:-0.4,0) rectangle (axis cs:-0.32,0.684410225);
\draw[draw=black,fill=viridis3] (axis cs:-0.32,0) rectangle (axis cs:-0.24,0.682130031364349);
\draw[draw=black,fill=viridis4, fill opacity=1.0] (axis cs:-0.24,0) rectangle (axis cs:-0.16,0.257);
\draw[draw=black,fill=viridis4, fill opacity=1.0] (axis cs:-0.16,0) rectangle (axis cs:-0.08,0.572);
\draw[draw=black,fill=viridis4, fill opacity=1.0] (axis cs:-0.08,0) rectangle (axis cs:0,0.686);
\draw[draw=black,fill=viridis3] (axis cs:0.12,0) rectangle (axis cs:0.20,0.6849275);

\draw[postaction={pattern=north east lines},draw=black,fill=viridis1] (axis cs:-0.44,0) rectangle (axis cs:-0.36,0.673500916666667);
\draw[postaction={pattern=north east lines},draw=black,fill=viridis2] (axis cs:-0.36,0) rectangle (axis cs:-0.28,0.672573489820892);
\draw[postaction={pattern=north east lines},draw=black,fill=viridis3] (axis cs:-0.28,0) rectangle (axis cs:-0.2,0.667409068053749);
\draw[postaction={pattern=north east lines},draw=black,fill=viridis4, fill opacity=1.0] (axis cs:-0.18,0) rectangle (axis cs:-0.12,0.257);
\draw[postaction={pattern=north east lines},draw=black,fill=viridis4, fill opacity=1.0] (axis cs:-0.12,0) rectangle (axis cs:-0.04,0.572);
\draw[postaction={pattern=north east lines},draw=black,fill=viridis4, fill opacity=1.0] (axis cs:-0.04,0) rectangle (axis cs:0.04,0.686);
\draw[postaction={pattern=north east lines},draw=black,fill=viridis3] (axis cs:0.16,0) rectangle (axis cs:0.24,0.605904383333333);

\end{axis}

% draw dashed lines from rectangle in first axis to corners of second
\draw [dashed] (c3) -- (ax2.south east);
\draw [dashed] (c2) -- (ax2.north west);

\end{tikzpicture}

%% file: G_50_d.tex
% This file was created with tikzplotlib v0.10.1.
\pgfplotsset{
tick label style={font=\footnotesize},
label style={font=\footnotesize},
legend  style={font=\footnotesize}
}
\begin{tikzpicture}

\begin{groupplot}[
    group style={
        group size=2 by 1, % One column, two rows
        horizontal sep=1.5cm, % Vertical spacing between the plots
    },
    width=0.5\columnwidth, % Adjust the width of the plots
    height=0.7\columnwidth % Adjust the height of the plots
]

\definecolor{viridis1}{RGB}{53, 95, 141}
\definecolor{viridis2}{RGB}{33, 145, 140}
\definecolor{viridis3}{RGB}{94, 201, 98}
\definecolor{viridis4}{RGB}{253, 231, 37}

\definecolor{darkgray176}{RGB}{176,176,176}
\definecolor{black}{RGB}{61,61,61}
\definecolor{dimgray1319183}{RGB}{131,91,83}
\definecolor{lightgray204}{RGB}{204,204,204}

\nextgroupplot[
legend cell align={left},
legend style={fill opacity=0.8, draw opacity=1, text opacity=1, draw=lightgray204},
tick align=outside,
tick pos=left,
x grid style={darkgray176},
xlabel={},
xmin=-0.52, xmax=0.786,
xtick style={color=black},
xtick={0,0.5999},
xtick style={draw=none},
xticklabel style={align=center},
xticklabels = {\hspace{-0.5cm}Centralized, Federated},
y grid style={darkgray176},
ymin=14, ymax=62,
ylabel={E2E delay [ms]},
ymajorgrids,
clip=false,
xticklabel shift={3pt},
ytick style={color=black},
]

\path [draw=black, fill=viridis1]
(axis cs:-0.4,20.83814375)
--(axis cs:-0.266666666666667,20.83814375)
--(axis cs:-0.266666666666667,32.1942875)
--(axis cs:-0.4,32.1942875)
--(axis cs:-0.4,20.83814375)
--cycle;
\addplot [black, forget plot]
table {%
-0.333333333333333 20.83814375
-0.333333333333333 18.369142
};
\addplot [black, forget plot]
table {%
-0.333333333333333 32.1942875
-0.333333333333333 49.223029
};
\addplot [black, forget plot]
table {%
-0.366666666666667 18.369142
-0.3 18.369142
};
\addplot [black, forget plot]
table {%
-0.366666666666667 49.223029
-0.3 49.223029
};
\path [draw=black, fill=viridis2]
(axis cs:-0.266666666666667,20.82013125)
--(axis cs:-0.133333333333333,20.82013125)
--(axis cs:-0.133333333333333,31.645687)
--(axis cs:-0.266666666666667,31.645687)
--(axis cs:-0.266666666666667,20.82013125)
--cycle;
\addplot [black, forget plot]
table {%
-0.2 20.82013125
-0.2 18.251328
};
\addplot [black, forget plot]
table {%
-0.2 31.645687
-0.2 47.875088
};
\addplot [black, forget plot]
table {%
-0.233333333333333 18.251328
-0.166666666666667 18.251328
};
\addplot [black, forget plot]
table {%
-0.233333333333333 47.875088
-0.166666666666667 47.875088
};
\path [draw=black, fill=viridis3]
(axis cs:-0.133333333333333,22.28546425)
--(axis cs:-5.55111512312578e-17,22.28546425)
--(axis cs:-5.55111512312578e-17,36.988728)
--(axis cs:-0.133333333333333,36.988728)
--(axis cs:-0.133333333333333,22.28546425)
--cycle;
\addplot [black, forget plot]
table {%
-0.0666666666666667 22.28546425
-0.0666666666666667 18.368806
};
\addplot [black, forget plot]
table {%
-0.0666666666666667 36.988728
-0.0666666666666667 59.034954
};
\addplot [black, forget plot]
table {%
-0.1 18.368806
-0.0333333333333334 18.368806
};
\addplot [black, forget plot]
table {%
-0.1 59.034954
-0.0333333333333334 59.034954
};
\path [draw=black, fill=viridis3]
(axis cs:0.5333,22.1427845)
--(axis cs:0.6666,22.1427845)
--(axis cs:0.6666,36.9642725)
--(axis cs:0.5333,36.9642725)
--(axis cs:0.5333,22.1427845)
--cycle;
\addplot [black, forget plot]
table {%
0.5999 22.1427845
0.5999 18.232166
};
\addplot [black, forget plot]
table {%
0.5999 36.9642725
0.5999 59.169496
};
\addplot [black, forget plot]
table {%
0.5666 18.232166
0.6333 18.232166
};
\addplot [black, forget plot]
table {%
0.5666 59.169496
0.6333 59.169496
};
\path [draw=black, fill=viridis4]
(axis cs:0,22.02796175)
--(axis cs:0.133333333333333,22.02796175)
--(axis cs:0.133333333333333,27.455026)
--(axis cs:0,27.455026)
--(axis cs:0,22.02796175)
--cycle;
\addplot [black, forget plot]
table {%
0.0666666666666667 22.02796175
0.0666666666666667 20.321467
};
\addplot [black, forget plot]
table {%
0.0666666666666667 27.455026
0.0666666666666667 35.577919
};
\addplot [black, forget plot]
table {%
0.0333333333333333 20.321467
0.1 20.321467
};
\addplot [black, forget plot]
table {%
0.0333333333333333 35.577919
0.1 35.577919
};
\path [draw=black, fill=viridis4]
(axis cs:0.133333333333333,21.52289575)
--(axis cs:0.266666666666667,21.52289575)
--(axis cs:0.266666666666667,29.00353925)
--(axis cs:0.133333333333333,29.00353925)
--(axis cs:0.133333333333333,21.52289575)
--cycle;
\addplot [black, forget plot]
table {%
0.2 21.52289575
0.2 19.43715
};
\addplot [black, forget plot]
table {%
0.2 29.00353925
0.2 40.224189
};
\addplot [black, forget plot]
table {%
0.166666666666667 19.43715
0.233333333333333 19.43715
};
\addplot [black, forget plot]
table {%
0.166666666666667 40.224189
0.233333333333333 40.224189
};
\path [draw=black, fill=viridis4]
(axis cs:0.266666666666667,20.83657175)
--(axis cs:0.4,20.83657175)
--(axis cs:0.4,32.000131)
--(axis cs:0.266666666666667,32.000131)
--(axis cs:0.266666666666667,20.83657175)
--cycle;
\addplot [black, forget plot]
table {%
0.333333333333333 20.83657175
0.333333333333333 18.369204
};
\addplot [black, forget plot]
table {%
0.333333333333333 32.000131
0.333333333333333 48.741156
};
\addplot [black, forget plot]
table {%
0.3 18.369204
0.366666666666667 18.369204
};
\addplot [black, forget plot]
table {%
0.3 48.741156
0.366666666666667 48.741156
};
\addplot [black, forget plot]
table {%
-0.4 24.5889435
-0.266666666666667 24.5889435
};
\addplot [black, forget plot]
table {%
-0.266666666666667 24.2848475
-0.133333333333333 24.2848475
};
\addplot [black, forget plot]
table {%
-0.133333333333333 27.896254
-5.55111512312578e-17 27.896254
};
\addplot [black, forget plot]
table {%
0.5333 26.7206175
0.6666 26.7206175
};
\addplot [black, forget plot]
table {%
0 24.151891
0.133333333333333 24.151891
};
\addplot [black, forget plot]
table {%
0.133333333333333 24.2499995
0.266666666666667 24.2499995
};
\addplot [black, forget plot]
table {%
0.266666666666667 24.503239
0.4 24.503239
};
\draw[dashed, thick, red] (axis cs: -0.52,50) -- (axis cs: 0.786,50);
\node [red] at (-0.45, 52) {\footnotesize $\tau$};

\addplot [
draw=black, mark=*, only marks, mark options={fill=white}]
coordinates {
    (-0.333333333333333,29.77) 
};

\addplot [
draw=black, mark=*, only marks, mark options={fill=white}]
coordinates {
    (-0.2,29.50) 
};

\addplot [
draw=black, mark=*, only marks, mark options={fill=white}]
coordinates {
    (-0.0666666666666667,34.45) 
};

\addplot [
draw=black, mark=*, only marks, mark options={fill=white}]
coordinates {
    (0.0666666666666667,28.09) 
};

\addplot [
draw=black, mark=*, only marks, mark options={fill=white}]
coordinates {
    (0.2,28.35) 
};

\addplot [
draw=black, mark=*, only marks, mark options={fill=white}]
coordinates {
    (0.333333333333333,29.67) 
};

\addplot [
draw=black, mark=*, only marks, mark options={fill=white}]
coordinates {
    (0.5999,33.46) 
};

\node [black] at (0.0666666666666667, 12) {\tiny$C_1$};
\node [black] at (0.2, 12) {\tiny$C_2$};
\node [black] at (0.333333333333333 , 12) {\tiny$C_3$};

\nextgroupplot[
legend style={legend cell align=left,
              align=center,
              draw=white!15!black,
              at={(0.5,1.45)},
              anchor=center,
              /tikz/every even column/.append style={column sep=1em},
legend columns=1,
anchor=north,legend rows=2},
tick align=outside,
tick pos=left,
x grid style={darkgray176},
xmin=-0.56, xmax=0.28,
xtick style={color=black},
xtick={-0.24,0.16},
xticklabel style={align=center},
xticklabels = {\hspace{-0.5cm}Centralized, Federated},
clip=false,
y grid style={darkgray176},
xticklabel shift={3pt},
ylabel={$P[{\ell} \leq \tau]$},
ymajorgrids,
ymin=0, ymax=1,
xtick style={draw=none},
ytick style={color=black},
]
\draw[draw=black,fill=viridis1] (axis cs:-0.48,0) rectangle (axis cs:-0.40,0.937738);
\draw[draw=black,fill=viridis2] (axis cs:-0.4,0) rectangle (axis cs:-0.32,0.937925);
\draw[draw=black,fill=viridis3] (axis cs:-0.32,0) rectangle (axis cs:-0.24,0.926686);
\draw[draw=black,fill=viridis4, fill opacity=1.0] (axis cs:-0.24,0) rectangle (axis cs:-0.16,0.954625);
\draw[draw=black,fill=viridis4, fill opacity=1.0] (axis cs:-0.16,0) rectangle (axis cs:-0.08,0.955738);
\draw[draw=black,fill=viridis4, fill opacity=1.0] (axis cs:-0.08,0) rectangle (axis cs:0,0.938875);
\draw[draw=black,fill=viridis3] (axis cs:0.12,0) rectangle (axis cs:0.20,0.933900);

\node [black] at (-0.21, -0.04) {\tiny$C_1$};
\node [black] at (-0.12, -0.04) {\tiny$C_2$};
\node [black] at (-0.03 , -0.04) {\tiny$C_3$};
\end{groupplot}

\end{tikzpicture}

%% file: G_30_d.tex
% This file was created with tikzplotlib v0.10.1.
\pgfplotsset{
tick label style={font=\footnotesize},
label style={font=\footnotesize},
legend  style={font=\footnotesize}
}
\begin{tikzpicture}

\definecolor{viridis1}{RGB}{53, 95, 141}
\definecolor{viridis2}{RGB}{33, 145, 140}
\definecolor{viridis3}{RGB}{94, 201, 98}
\definecolor{viridis4}{RGB}{253, 231, 37}

\definecolor{darkgray176}{RGB}{176,176,176}
\definecolor{black}{RGB}{61,61,61}
\definecolor{dimgray1319183}{RGB}{131,91,83}
\definecolor{lightgray204}{RGB}{204,204,204}

\begin{groupplot}[
    group style={
        group size=2 by 1, % One column, two rows
        horizontal sep=1.5cm, % Vertical spacing between the plots
    },
    width=0.5\columnwidth, % Adjust the width of the plots
    height=0.7\columnwidth % Adjust the height of the plots
]

\nextgroupplot[
legend cell align={left},
legend style={fill opacity=0.8, draw opacity=1, text opacity=1, draw=lightgray204},
tick align=outside,
tick pos=left,
x grid style={darkgray176},
xlabel={},
xmin=-0.52, xmax=0.786,
xtick style={color=black},
xtick={0,0.5999},
xtick style={draw=none},
xticklabel style={align=center},
xticklabels = {\hspace{-0.5cm} Centralized, Federated},
y grid style={darkgray176},
ylabel={E2E delay [ms]},
ymin=14, ymax=62,
ytick style={color=black},
ymajorgrids,
clip=false,
xticklabel shift={3pt},
ytick style={color=black},
]

\path [postaction={pattern=north east lines}, draw=black, fill=viridis1]
(axis cs:-0.4,20.7061675)
--(axis cs:-0.266666666666667,20.7061675)
--(axis cs:-0.266666666666667,29.586452)
--(axis cs:-0.4,29.586452)
--(axis cs:-0.4,20.7061675)
--cycle;
\addplot [black, forget plot]
table {%
-0.333333333333333 20.7061675
-0.333333333333333 18.369142
};
\addplot [black, forget plot]
table {%
-0.333333333333333 29.586452
-0.333333333333333 42.901577
};
\addplot [black, forget plot]
table {%
-0.366666666666667 18.369142
-0.3 18.369142
};
\addplot [black, forget plot]
table {%
-0.366666666666667 42.901577
-0.3 42.901577
};
\path [postaction={pattern=north east lines}, draw=black, fill=viridis2]
(axis cs:-0.266666666666667,20.77573525)
--(axis cs:-0.133333333333333,20.77573525)
--(axis cs:-0.133333333333333,29.5572255)
--(axis cs:-0.266666666666667,29.5572255)
--(axis cs:-0.266666666666667,20.77573525)
--cycle;
\addplot [black, forget plot]
table {%
-0.2 20.77573525
-0.2 18.369142
};
\addplot [black, forget plot]
table {%
-0.2 29.5572255
-0.2 42.729173
};
\addplot [black, forget plot]
table {%
-0.233333333333333 18.369142
-0.166666666666667 18.369142
};
\addplot [black, forget plot]
table {%
-0.233333333333333 42.729173
-0.166666666666667 42.729173
};
\path [postaction={pattern=north east lines}, draw=black, fill=viridis3]
(axis cs:-0.133333333333333,22.13930475)
--(axis cs:-5.55111512312578e-17,22.13930475)
--(axis cs:-5.55111512312578e-17,37.000079)
--(axis cs:-0.133333333333333,37.000079)
--(axis cs:-0.133333333333333,22.13930475)
--cycle;
\addplot [black, forget plot]
table {%
-0.0666666666666667 22.13930475
-0.0666666666666667 18.002614
};
\addplot [black, forget plot]
table {%
-0.0666666666666667 37.000079
-0.0666666666666667 59.276773
};
\addplot [black, forget plot]
table {%
-0.1 18.002614
-0.0333333333333334 18.002614
};
\addplot [black, forget plot]
table {%
-0.1 59.276773
-0.0333333333333334 59.276773
};
\path [postaction={pattern=north east lines}, draw=black, fill=viridis3]
(axis cs:0.5333,21.581434)
--(axis cs:0.6666,21.581434)
--(axis cs:0.6666,37.26711675)
--(axis cs:0.5333,37.26711675)
--(axis cs:0.5333,21.581434)
--cycle;
\addplot [black, forget plot]
table {%
0.5999 21.581434
0.5999 18.369204
};
\addplot [black, forget plot]
table {%
0.5999 37.26711675
0.5999 60.758934
};
\addplot [black, forget plot]
table {%
0.5666 18.369204
0.6333 18.369204
};
\addplot [black, forget plot]
table {%
0.5666 60.758934
0.6333 60.758934
};
\path [postaction={pattern=north east lines}, draw=black, fill=viridis4]
(axis cs:0,21.8189405)
--(axis cs:0.133333333333333,21.8189405)
--(axis cs:0.133333333333333,26.65526125)
--(axis cs:0,26.65526125)
--(axis cs:0,21.8189405)
--cycle;
\addplot [black, forget plot]
table {%
0.0666666666666667 21.8189405
0.0666666666666667 20.546685
};
\addplot [black, forget plot]
table {%
0.0666666666666667 26.65526125
0.0666666666666667 33.907532
};
\addplot [black, forget plot]
table {%
0.0333333333333333 20.546685
0.1 20.546685
};
\addplot [black, forget plot]
table {%
0.0333333333333333 33.907532
0.1 33.907532
};
\path [postaction={pattern=north east lines}, draw=black, fill=viridis4]
(axis cs:0.133333333333333,21.318849)
--(axis cs:0.266666666666667,21.318849)
--(axis cs:0.266666666666667,27.48676625)
--(axis cs:0.133333333333333,27.48676625)
--(axis cs:0.133333333333333,21.318849)
--cycle;
\addplot [black, forget plot]
table {%
0.2 21.318849
0.2 19.625115
};
\addplot [black, forget plot]
table {%
0.2 27.48676625
0.2 36.69645
};
\addplot [black, forget plot]
table {%
0.166666666666667 19.625115
0.233333333333333 19.625115
};
\addplot [black, forget plot]
table {%
0.166666666666667 36.69645
0.233333333333333 36.69645
};
\path [postaction={pattern=north east lines}, draw=black, fill=viridis4]
(axis cs:0.266666666666667,20.770125)
--(axis cs:0.4,20.770125)
--(axis cs:0.4,29.89289075)
--(axis cs:0.266666666666667,29.89289075)
--(axis cs:0.266666666666667,20.770125)
--cycle;
\addplot [black, forget plot]
table {%
0.333333333333333 20.770125
0.333333333333333 18.250191
};
\addplot [black, forget plot]
table {%
0.333333333333333 29.89289075
0.333333333333333 43.572383
};
\addplot [black, forget plot]
table {%
0.3 18.250191
0.366666666666667 18.250191
};
\addplot [black, forget plot]
table {%
0.3 43.572383
0.366666666666667 43.572383
};
\addplot [black, forget plot]
table {%
-0.4 23.483942
-0.266666666666667 23.483942
};
\addplot [black, forget plot]
table {%
-0.266666666666667 23.633411
-0.133333333333333 23.633411
};
\addplot [black, forget plot]
table {%
-0.133333333333333 26.5746885
-5.55111512312578e-17 26.5746885
};
\addplot [black, forget plot]
table {%
0.5333 24.018015
0.6666 24.018015
};
\addplot [black, forget plot]
table {%
0 23.652994
0.133333333333333 23.652994
};
\addplot [black, forget plot]
table {%
0.133333333333333 23.449721
0.266666666666667 23.449721
};
\addplot [black, forget plot]
table {%
0.266666666666667 23.643241
0.4 23.643241
};

\draw[dashed, thick, red] (axis cs: -0.52,30) -- (axis cs: 0.786,30);
\node [red] at (-0.45, 32) {\footnotesize $\tau$};

\addplot [
draw=black, mark=*, only marks, mark options={fill=white}]
coordinates {
    (-0.333333333333333,28.14) 
};

\addplot [
draw=black, mark=*, only marks, mark options={fill=white}]
coordinates {
    (-0.2,28.31) 
};

\addplot [
draw=black, mark=*, only marks, mark options={fill=white}]
coordinates {
    (-0.0666666666666667,33.97) 
};

\addplot [
draw=black, mark=*, only marks, mark options={fill=white}]
coordinates {
    (0.0666666666666667,27.64) 
};

\addplot [
draw=black, mark=*, only marks, mark options={fill=white}]
coordinates {
    (0.2,27.39) 
};

\addplot [
draw=black, mark=*, only marks, mark options={fill=white}]
coordinates {
    (0.333333333333333,28.82) 
};

\addplot [
draw=black, mark=*, only marks, mark options={fill=white}]
coordinates {
    (0.5999,30.56) 
};

\node [black] at (0.0666666666666667, 12) {\tiny$C_1$};
\node [black] at (0.2, 12) {\tiny$C_2$};
\node [black] at (0.333333333333333 , 12) {\tiny$C_3$};

\nextgroupplot[
legend style={legend cell align=left,
              align=center,
              draw=white!15!black,
              at={(0.5,1.45)},
              anchor=center,
              /tikz/every even column/.append style={column sep=1em},
legend columns=1,
anchor=north,legend rows=2},
tick align=outside,
tick pos=left,
x grid style={darkgray176},
xmin=-0.56, xmax=0.28,
xtick style={color=black},
xtick={-0.24,0.16},
xticklabel style={align=center},
xticklabels = {\hspace{-0.5cm} Centralized, Federated},
clip=false,
y grid style={darkgray176},
xticklabel shift={3pt},
ylabel={$P[{\ell} \leq \tau]$},
ymajorgrids,
ymin=0, ymax=1,
xtick style={draw=none},
ytick style={color=black},
]
\draw[draw=black, postaction={pattern=north east lines}, fill=viridis1] (axis cs:-0.48,0) rectangle (axis cs:-0.40,0.755450);
\draw[draw=black, postaction={pattern=north east lines}, fill=viridis2] (axis cs:-0.4,0) rectangle (axis cs:-0.32,0.756341);
\draw[draw=black, postaction={pattern=north east lines}, fill=viridis3] (axis cs:-0.32,0) rectangle (axis cs:-0.24,0.529658);
\draw[draw=black, postaction={pattern=north east lines}, fill=viridis4, fill opacity=1.0] (axis cs:-0.24,0) rectangle (axis cs:-0.16,0.833961);
\draw[draw=black, postaction={pattern=north east lines}, fill=viridis4, fill opacity=1.0] (axis cs:-0.16,0) rectangle (axis cs:-0.08,0.809007);
\draw[draw=black, postaction={pattern=north east lines}, fill=viridis4, fill opacity=1.0] (axis cs:-0.08,0) rectangle (axis cs:0,0.748315);
\draw[draw=black, postaction={pattern=north east lines}, fill=viridis3] (axis cs:0.12,0) rectangle (axis cs:0.20,0.619006);

\node [black] at (-0.21, -0.04) {\tiny$C_1$};
\node [black] at (-0.12, -0.04) {\tiny$C_2$};
\node [black] at (-0.03 , -0.04) {\tiny$C_3$};

\end{groupplot}

\end{tikzpicture}

%% file: legend_B.tex
\definecolor{brown1926061}{RGB}{192,60,61}
\definecolor{darkgray176}{RGB}{176,176,176}
\definecolor{darkslategray66}{RGB}{66,66,66}
\definecolor{lightgray204}{RGB}{204,204,204}
\definecolor{mediumpurple147113178}{RGB}{147,113,178}
\definecolor{peru22412844}{RGB}{224,128,44}
\definecolor{seagreen5814558}{RGB}{58,145,58}
\definecolor{steelblue49115161}{RGB}{49,115,161}

\definecolor{viridis1}{RGB}{53, 95, 141}
\definecolor{viridis2}{RGB}{33, 145, 140}
\definecolor{viridis3}{RGB}{94, 201, 98}
\definecolor{viridis4}{RGB}{253, 231, 37}

\pgfplotsset{
tick label style={font=\footnotesize},
label style={font=\footnotesize},
legend  style={font=\footnotesize}
}
\begin{tikzpicture}
\pgfplotsset{every tick label/.append style={font=\scriptsize}}

\pgfplotsset{compat=1.11,
  /pgfplots/ybar legend/.style={
    /pgfplots/legend image code/.code={%
      \draw[##1,/tikz/.cd,yshift=-0.25em]
      (0cm,0cm) rectangle (10pt,0.6em);},
  },
}

\begin{axis}[%
width=0,
height=0,
at={(0,0)},
scale only axis,
xmin=0,
xmax=0,
xtick={},
ymin=0,
ymax=0,
ytick={},
axis background/.style={fill=white},
legend style={legend cell align=left,
              align=center,
              draw=white!15!black,
              at={(0.5, 1.3)},
              anchor=center,
              /tikz/every even column/.append style={column sep=1em}},
legend columns=7,
]
\addplot[ybar,ybar legend,draw=black,fill=viridis1,line width=0.08pt]
table[row sep=crcr]{%
  0  0\\
};
\addlegendentry{CCS}

\addplot[ybar legend,ybar,draw=black,fill=viridis2,fill opacity=1.0,line width=0.08pt]
  table[row sep=crcr]{%
  0  0\\
};
\addlegendentry{ICS}

\addplot[ybar legend,ybar,draw=black,fill=viridis3,fill opacity=1.0,line width=0.08pt]
  table[row sep=crcr]{%
  0  0\\
};
\addlegendentry{C}

\addplot[ybar legend,ybar,draw=black,fill=viridis4,fill opacity=1.0,line width=0.08pt]
  table[row sep=crcr]{%
  0  0\\
};
\addlegendentry{S}

\addplot[ybar legend,ybar,draw=black,fill=white,fill opacity=1.0,line width=0.08pt]
  table[row sep=crcr]{%
  0  0\\
};
\addlegendentry{$\tau\!=\!50\,$ms}

\addplot[postaction={pattern=north east lines}, ybar legend,ybar,draw=black,fill=white,fill opacity=1.0,line width=0.08pt]
  table[row sep=crcr]{%
  0  0\\
};
\addlegendentry{$\tau\!=\!30\,$ms}

\end{axis}
\end{tikzpicture}%

%% file: B_r.tex
% This file was created with tikzplotlib v0.10.1.
\pgfplotsset{
tick label style={font=\footnotesize},
label style={font=\footnotesize},
legend  style={font=\footnotesize}
}
\begin{tikzpicture}

\definecolor{viridis1}{RGB}{53, 95, 141}
\definecolor{viridis2}{RGB}{33, 145, 140}
\definecolor{viridis3}{RGB}{94, 201, 98}
\definecolor{viridis4}{RGB}{253, 231, 37}

\definecolor{brown1926061}{RGB}{192,60,61}
\definecolor{darkgray176}{RGB}{176,176,176}
\definecolor{darkslategray66}{RGB}{66,66,66}
\definecolor{lightgray204}{RGB}{204,204,204}
\definecolor{mediumpurple147113178}{RGB}{147,113,178}
\definecolor{peru22412844}{RGB}{224,128,44}
\definecolor{seagreen5814558}{RGB}{58,145,58}
\definecolor{steelblue49115161}{RGB}{49,115,161}
\definecolor{antiquewhite239222208}{RGB}{239,222,208}
\definecolor{antiquewhite240226214}{RGB}{240,226,214}
\definecolor{antiquewhite240229220}{RGB}{240,229,220}
\definecolor{antiquewhite241231223}{RGB}{241,231,223}
\definecolor{bisque239218200}{RGB}{239,218,200}
\definecolor{brown1926061}{RGB}{192,60,61}
\definecolor{burlywood233184141}{RGB}{233,184,141}
\definecolor{burlywood235196161}{RGB}{235,196,161}
\definecolor{cadetblue119161191}{RGB}{119,161,191}
\definecolor{crimson2143940}{RGB}{214,39,40}
\definecolor{darkgray144177201}{RGB}{144,177,201}
\definecolor{darkgray166139191}{RGB}{166,139,191}
\definecolor{darkgray176}{RGB}{176,176,176}
\definecolor{darkorange25512714}{RGB}{255,127,14}
\definecolor{darksalmon231169116}{RGB}{231,169,116}
\definecolor{darkseagreen123181123}{RGB}{123,181,123}
\definecolor{darkseagreen146193146}{RGB}{146,193,146}
\definecolor{darkseagreen165204165}{RGB}{165,204,165}
\definecolor{darkslategray61}{RGB}{61,61,61}
\definecolor{dimgray114}{RGB}{114,114,114}
\definecolor{forestgreen4416044}{RGB}{44,160,44}
\definecolor{gainsboro209221229}{RGB}{209,221,229}
\definecolor{gainsboro213230213}{RGB}{213,230,213}
\definecolor{gainsboro215225232}{RGB}{215,225,232}
\definecolor{gainsboro217208225}{RGB}{217,208,225}
\definecolor{gainsboro218232218}{RGB}{218,232,218}
\definecolor{gainsboro220228234}{RGB}{220,228,234}
\definecolor{gainsboro222215229}{RGB}{222,215,229}
\definecolor{gainsboro222234221}{RGB}{222,234,221}
\definecolor{gainsboro224236224}{RGB}{224,236,224}
\definecolor{gainsboro226220231}{RGB}{226,220,231}
\definecolor{gainsboro227237227}{RGB}{227,237,227}
\definecolor{gainsboro229224233}{RGB}{229,224,233}
\definecolor{gainsboro234212212}{RGB}{234,212,212}
\definecolor{gainsboro235217217}{RGB}{235,217,217}
\definecolor{gainsboro237222222}{RGB}{237,222,222}
\definecolor{indianred2029798}{RGB}{202,97,98}
\definecolor{lavender224230236}{RGB}{224,230,236}
\definecolor{lavender227232237}{RGB}{227,232,237}
\definecolor{lavender230234238}{RGB}{230,234,238}
\definecolor{lavender231227235}{RGB}{231,227,235}
\definecolor{lavender233230236}{RGB}{233,230,236}
\definecolor{lavender235232237}{RGB}{235,232,237}
\definecolor{lavender236234238}{RGB}{236,234,238}
\definecolor{lightgray201223201}{RGB}{201,223,201}
\definecolor{lightgray202216226}{RGB}{202,216,226}
\definecolor{lightgray204}{RGB}{204,204,204}
\definecolor{lightgray208227208}{RGB}{208,227,208}
\definecolor{lightgray232205205}{RGB}{232,205,205}
\definecolor{lightpink226183184}{RGB}{226,183,184}
\definecolor{lightsteelblue164191210}{RGB}{164,191,210}
\definecolor{lightsteelblue179201216}{RGB}{179,201,216}
\definecolor{lightsteelblue191209222}{RGB}{191,209,222}
\definecolor{linen238228228}{RGB}{238,228,228}
\definecolor{linen239231231}{RGB}{239,231,231}
\definecolor{linen241233226}{RGB}{241,233,226}
\definecolor{linen241234229}{RGB}{241,234,229}
\definecolor{mediumpurple147113178}{RGB}{147,113,178}
\definecolor{mediumpurple148103189}{RGB}{148,103,189}
\definecolor{mediumseagreen9416594}{RGB}{94,165,94}
\definecolor{mistyrose238226226}{RGB}{238,226,226}
\definecolor{palevioletred210126127}{RGB}{210,126,127}
\definecolor{peru22412844}{RGB}{224,128,44}
\definecolor{rosybrown217150150}{RGB}{217,150,150}
\definecolor{sandybrown22815183}{RGB}{228,151,83}
\definecolor{seagreen5814558}{RGB}{58,145,58}
\definecolor{silver180212180}{RGB}{180,212,180}
\definecolor{silver181160201}{RGB}{181,160,201}
\definecolor{silver191218191}{RGB}{191,218,191}
\definecolor{silver194176209}{RGB}{194,176,209}
\definecolor{steelblue31119180}{RGB}{31,119,180}
\definecolor{steelblue49115161}{RGB}{49,115,161}
\definecolor{steelblue88141177}{RGB}{88,141,177}
\definecolor{tan222169169}{RGB}{222,169,169}
\definecolor{thistle203190216}{RGB}{203,190,216}
\definecolor{thistle211200221}{RGB}{211,200,221}
\definecolor{thistle229195195}{RGB}{229,195,195}
\definecolor{wheat237205177}{RGB}{237,205,177}
\definecolor{wheat238212190}{RGB}{238,212,190}

\begin{axis}[
height=0.50\columnwidth,
width=\columnwidth,
legend style={legend cell align=left,
              align=center,
              draw=white!15!black,
              at={(0.5,1.45)},
              anchor=center,
              /tikz/every even column/.append style={column sep=1em},
legend columns=1,
anchor=north,legend rows=2},
tick align=outside,
tick pos=left,
x grid style={darkgray176},
xmin=-0.52, xmax=0.16,
xtick style={color=black},
xtick={-0.32,-0.04},
xticklabels = {Centralized, Federated},
clip=false,
y grid style={darkgray176},
xticklabel shift={3pt},
ylabel={Reward},
ymajorgrids,
ymin=-1.1, ymax=0,
xtick style={draw=none},
ytick style={color=black},
]

\draw[postaction={pattern=north east lines},draw=black,fill=viridis2] (axis cs:-0.36,0) rectangle (axis cs:-0.28,-0.707956401700968);
\draw[postaction={pattern=north east lines},draw=black,fill=viridis3] (axis cs:-0.28,0) rectangle (axis cs:-0.2,-0.977977794434576);
\draw[postaction={pattern=north east lines},draw=black,fill=viridis3] (axis cs:-0.04,0) rectangle (axis cs:0.04,-0.6124412989025);

\draw[draw=black,fill=viridis2] (axis cs:-0.4,0) rectangle (axis cs:-0.32,-0.400127626380768);
\draw[draw=black,fill=viridis3] (axis cs:-0.32,0) rectangle (axis cs:-0.24,-0.525991343732245);
\draw[draw=black,fill=viridis3] (axis cs:-0.08,0) rectangle (axis cs:0,-0.0459322426410465);

% \draw[postaction={pattern=north east lines},draw=black,fill=viridis1] (axis cs:-0.44,0) rectangle (axis cs:-0.36,0.424639863169146);

% \addplot [line width=0.9pt, darkslategray66, forget plot]
% table {%
% -0.36 0.564530176814455
% -0.36 0.56975396959421
% };
% \addplot [line width=0.9pt, darkslategray66, forget plot]
% table {%
% 0.64 0.59153763790257
% 0.64 0.595511039887238
% };
% \addplot [line width=0.9pt, darkslategray66, forget plot]
% table {%
% -0.20 0.603215667488685
% -0.20 0.606723860231615
% };
% \addplot [line width=0.9pt, darkslategray66, forget plot]
% table {%
% -0.04 0.228725206063478
% -0.04 0.230499560890395
% };
% \addplot [line width=0.9pt, darkslategray66, forget plot]
% table {%
% 0.12 0.530760004829933
% 0.12 0.533011537248994
% };
% \addplot [line width=0.9pt, darkslategray66, forget plot]
% table {%
% 0.28 0.601263766599824
% 0.28 0.605041629595162
% };

\end{axis}

\end{tikzpicture}

%% file: B_mAP.tex
% This file was created with tikzplotlib v0.10.1.
\pgfplotsset{
tick label style={font=\footnotesize},
label style={font=\footnotesize},
legend  style={font=\footnotesize}
}
\usetikzlibrary{calc}
\begin{tikzpicture}

\definecolor{viridis1}{RGB}{53, 95, 141}
\definecolor{viridis2}{RGB}{33, 145, 140}
\definecolor{viridis3}{RGB}{94, 201, 98}
\definecolor{viridis4}{RGB}{253, 231, 37}

\definecolor{brown1926061}{RGB}{192,60,61}
\definecolor{darkgray176}{RGB}{176,176,176}
\definecolor{darkslategray66}{RGB}{66,66,66}
\definecolor{lightgray204}{RGB}{204,204,204}
\definecolor{mediumpurple147113178}{RGB}{147,113,178}
\definecolor{peru22412844}{RGB}{224,128,44}
\definecolor{seagreen5814558}{RGB}{58,145,58}
\definecolor{steelblue49115161}{RGB}{49,115,161}
\definecolor{antiquewhite239222208}{RGB}{239,222,208}
\definecolor{antiquewhite240226214}{RGB}{240,226,214}
\definecolor{antiquewhite240229220}{RGB}{240,229,220}
\definecolor{antiquewhite241231223}{RGB}{241,231,223}
\definecolor{bisque239218200}{RGB}{239,218,200}
\definecolor{brown1926061}{RGB}{192,60,61}
\definecolor{burlywood233184141}{RGB}{233,184,141}
\definecolor{burlywood235196161}{RGB}{235,196,161}
\definecolor{cadetblue119161191}{RGB}{119,161,191}
\definecolor{crimson2143940}{RGB}{214,39,40}
\definecolor{darkgray144177201}{RGB}{144,177,201}
\definecolor{darkgray166139191}{RGB}{166,139,191}
\definecolor{darkgray176}{RGB}{176,176,176}
\definecolor{darkorange25512714}{RGB}{255,127,14}
\definecolor{darksalmon231169116}{RGB}{231,169,116}
\definecolor{darkseagreen123181123}{RGB}{123,181,123}
\definecolor{darkseagreen146193146}{RGB}{146,193,146}
\definecolor{darkseagreen165204165}{RGB}{165,204,165}
\definecolor{darkslategray61}{RGB}{61,61,61}
\definecolor{dimgray114}{RGB}{114,114,114}
\definecolor{forestgreen4416044}{RGB}{44,160,44}
\definecolor{gainsboro209221229}{RGB}{209,221,229}
\definecolor{gainsboro213230213}{RGB}{213,230,213}
\definecolor{gainsboro215225232}{RGB}{215,225,232}
\definecolor{gainsboro217208225}{RGB}{217,208,225}
\definecolor{gainsboro218232218}{RGB}{218,232,218}
\definecolor{gainsboro220228234}{RGB}{220,228,234}
\definecolor{gainsboro222215229}{RGB}{222,215,229}
\definecolor{gainsboro222234221}{RGB}{222,234,221}
\definecolor{gainsboro224236224}{RGB}{224,236,224}
\definecolor{gainsboro226220231}{RGB}{226,220,231}
\definecolor{gainsboro227237227}{RGB}{227,237,227}
\definecolor{gainsboro229224233}{RGB}{229,224,233}
\definecolor{gainsboro234212212}{RGB}{234,212,212}
\definecolor{gainsboro235217217}{RGB}{235,217,217}
\definecolor{gainsboro237222222}{RGB}{237,222,222}
\definecolor{indianred2029798}{RGB}{202,97,98}
\definecolor{lavender224230236}{RGB}{224,230,236}
\definecolor{lavender227232237}{RGB}{227,232,237}
\definecolor{lavender230234238}{RGB}{230,234,238}
\definecolor{lavender231227235}{RGB}{231,227,235}
\definecolor{lavender233230236}{RGB}{233,230,236}
\definecolor{lavender235232237}{RGB}{235,232,237}
\definecolor{lavender236234238}{RGB}{236,234,238}
\definecolor{lightgray201223201}{RGB}{201,223,201}
\definecolor{lightgray202216226}{RGB}{202,216,226}
\definecolor{lightgray204}{RGB}{204,204,204}
\definecolor{lightgray208227208}{RGB}{208,227,208}
\definecolor{lightgray232205205}{RGB}{232,205,205}
\definecolor{lightpink226183184}{RGB}{226,183,184}
\definecolor{lightsteelblue164191210}{RGB}{164,191,210}
\definecolor{lightsteelblue179201216}{RGB}{179,201,216}
\definecolor{lightsteelblue191209222}{RGB}{191,209,222}
\definecolor{linen238228228}{RGB}{238,228,228}
\definecolor{linen239231231}{RGB}{239,231,231}
\definecolor{linen241233226}{RGB}{241,233,226}
\definecolor{linen241234229}{RGB}{241,234,229}
\definecolor{mediumpurple147113178}{RGB}{147,113,178}
\definecolor{mediumpurple148103189}{RGB}{148,103,189}
\definecolor{mediumseagreen9416594}{RGB}{94,165,94}
\definecolor{mistyrose238226226}{RGB}{238,226,226}
\definecolor{palevioletred210126127}{RGB}{210,126,127}
\definecolor{peru22412844}{RGB}{224,128,44}
\definecolor{rosybrown217150150}{RGB}{217,150,150}
\definecolor{sandybrown22815183}{RGB}{228,151,83}
\definecolor{seagreen5814558}{RGB}{58,145,58}
\definecolor{silver180212180}{RGB}{180,212,180}
\definecolor{silver181160201}{RGB}{181,160,201}
\definecolor{silver191218191}{RGB}{191,218,191}
\definecolor{silver194176209}{RGB}{194,176,209}
\definecolor{steelblue31119180}{RGB}{31,119,180}
\definecolor{steelblue49115161}{RGB}{49,115,161}
\definecolor{steelblue88141177}{RGB}{88,141,177}
\definecolor{tan222169169}{RGB}{222,169,169}
\definecolor{thistle203190216}{RGB}{203,190,216}
\definecolor{thistle211200221}{RGB}{211,200,221}
\definecolor{thistle229195195}{RGB}{229,195,195}
\definecolor{wheat237205177}{RGB}{237,205,177}
\definecolor{wheat238212190}{RGB}{238,212,190}

\begin{axis}[
height=0.50\columnwidth,
width=\columnwidth,
legend style={legend cell align=left,
              align=center,
              draw=white!15!black,
              at={(0.5,1.45)},
              anchor=center,
              /tikz/every even column/.append style={column sep=1em},
legend columns=1,
anchor=north,legend rows=2},
tick align=outside,
tick pos=left,
x grid style={darkgray176},
xmin=-0.6, xmax=0.36,
xtick style={color=black},
xtick={-0.16,0.20},
xticklabels = {Centralized, Federated},
y grid style={darkgray176},
xlabel shift=-4,
ylabel={mAP},
ymajorgrids,
ymin=0, ymax=0.75,
xtick style={draw=none},
clip=false,
xticklabel shift={3pt},
name=ax1,
ytick style={color=black},
]
    \draw[draw=black,fill=viridis1] (axis cs:-0.48,0) rectangle (axis cs:-0.40,0.562687137183013);
\draw[draw=black,fill=viridis2] (axis cs:-0.4,0) rectangle (axis cs:-0.32,0.5051);
\draw[draw=black,fill=viridis3] (axis cs:-0.32,0) rectangle (axis cs:-0.24,0.485139670312741);
\draw[draw=black,fill=viridis4, fill opacity=1.0] (axis cs:-0.24,0) rectangle (axis cs:-0.16,0.257);
\draw[draw=black,fill=viridis4, fill opacity=1.0] (axis cs:-0.16,0) rectangle (axis cs:-0.08,0.572);
\draw[draw=black,fill=viridis4, fill opacity=1.0] (axis cs:-0.08,0) rectangle (axis cs:0,0.686);
\draw[draw=black,fill=viridis3] (axis cs:0.12,0) rectangle (axis cs:0.20,0.414871197674419);

\draw[postaction={pattern=north east lines},draw=black,fill=viridis1] (axis cs:-0.44,0) rectangle (axis cs:-0.36,0.506154877014419);
\draw[postaction={pattern=north east lines},draw=black,fill=viridis2] (axis cs:-0.36,0) rectangle (axis cs:-0.28,0.430539790940767);
\draw[postaction={pattern=north east lines},draw=black,fill=viridis3] (axis cs:-0.28,0) rectangle (axis cs:-0.2,0.410320897074113);
\draw[postaction={pattern=north east lines},draw=black,fill=viridis4, fill opacity=1.0] (axis cs:-0.18,0) rectangle (axis cs:-0.16,0.257);
\draw[postaction={pattern=north east lines},draw=black,fill=viridis4, fill opacity=1.0] (axis cs:-0.12,0) rectangle (axis cs:-0.08,0.572);
\draw[postaction={pattern=north east lines},draw=black,fill=viridis4, fill opacity=1.0] (axis cs:-0.04,0) rectangle (axis cs:0,0.686);
\draw[postaction={pattern=north east lines},draw=black,fill=viridis3] (axis cs:0.16,0) rectangle (axis cs:0.24,0.3614486375);

% % define coordinates at bottom left and top left of rectangle
% \coordinate (c1) at (axis cs:-0.5,0.63);
% \coordinate (c2) at (axis cs:-0.5,0.70);
% \coordinate (c3) at (axis cs:-0.18,0.63);
% % draw a rectangle
% \draw (c1) rectangle (axis cs:-0.18,0.70);

% \addplot [line width=0.9pt, darkslategray66, forget plot]
% table {%
% -0.36 0.564530176814455
% -0.36 0.56975396959421
% };
% \addplot [line width=0.9pt, darkslategray66, forget plot]
% table {%
% 0.64 0.59153763790257
% 0.64 0.595511039887238
% };
% \addplot [line width=0.9pt, darkslategray66, forget plot]
% table {%
% -0.20 0.603215667488685
% -0.20 0.606723860231615
% };
% \addplot [line width=0.9pt, darkslategray66, forget plot]
% table {%
% -0.04 0.228725206063478
% -0.04 0.230499560890395
% };
% \addplot [line width=0.9pt, darkslategray66, forget plot]
% table {%
% 0.12 0.530760004829933
% 0.12 0.533011537248994
% };
% \addplot [line width=0.9pt, darkslategray66, forget plot]
% table {%
% 0.28 0.601263766599824
% 0.28 0.605041629595162
% };

\node [black] at (-0.16, -0.04) {\tiny$C_1$};
\node [black] at (-0.08, -0.04) {\tiny$C_2$};
\node [black] at (0.0, -0.04) {\tiny$C_3$};

\end{axis}

\end{tikzpicture}

%% file: B_50_d.tex
% This file was created with tikzplotlib v0.10.1.
\pgfplotsset{
tick label style={font=\footnotesize},
label style={font=\footnotesize},
legend  style={font=\footnotesize}
}
\begin{tikzpicture}

\begin{groupplot}[
    group style={
        group size=2 by 1, % One column, two rows
        horizontal sep=1.5cm, % Vertical spacing between the plots
    },
    width=0.5\columnwidth, % Adjust the width of the plots
    height=0.7\columnwidth % Adjust the height of the plots
]

\definecolor{viridis1}{RGB}{53, 95, 141}
\definecolor{viridis2}{RGB}{33, 145, 140}
\definecolor{viridis3}{RGB}{94, 201, 98}
\definecolor{viridis4}{RGB}{253, 231, 37}

\definecolor{darkgray176}{RGB}{176,176,176}
\definecolor{black}{RGB}{61,61,61}
\definecolor{dimgray1319183}{RGB}{131,91,83}
\definecolor{lightgray204}{RGB}{204,204,204}
\definecolor{darkslategray61}{RGB}{61,61,61}

\nextgroupplot[
legend cell align={left},
legend style={fill opacity=0.8, draw opacity=1, text opacity=1, draw=lightgray204},
tick align=outside,
tick pos=left,
x grid style={darkgray176},
xlabel={},
xmin=-0.52, xmax=0.786,
xtick style={color=black},
xtick={0,0.5999},
xtick style={draw=none},
xticklabel style={align=center},
xticklabels = {\hspace{-0.5cm} Centralized, Federated},
y grid style={darkgray176},
ymin=10, ymax=450,
ylabel={E2E latency [ms]},
ymajorgrids,
clip=false,
xticklabel shift={3pt},
ytick style={color=black},
]

\path [draw=darkslategray61, fill=viridis1]
(axis cs:-0.4,36.9195815)
--(axis cs:-0.266666666666667,36.9195815)
--(axis cs:-0.266666666666667,154.9910585)
--(axis cs:-0.4,154.9910585)
--(axis cs:-0.4,36.9195815)
--cycle;
\addplot [darkslategray61, forget plot]
table {%
-0.333333333333333 36.9195815
-0.333333333333333 18.374336
};
\addplot [darkslategray61, forget plot]
table {%
-0.333333333333333 154.9910585
-0.333333333333333 332.097152
};
\addplot [darkslategray61, forget plot]
table {%
-0.366666666666667 18.374336
-0.3 18.374336
};
\addplot [darkslategray61, forget plot]
table {%
-0.366666666666667 332.097152
-0.3 332.097152
};
\path [draw=darkslategray61, fill=viridis2]
(axis cs:-0.266666666666667,32.2493395)
--(axis cs:-0.133333333333333,32.2493395)
--(axis cs:-0.133333333333333,100.51978375)
--(axis cs:-0.266666666666667,100.51978375)
--(axis cs:-0.266666666666667,32.2493395)
--cycle;
\addplot [darkslategray61, forget plot]
table {%
-0.2 32.2493395
-0.2 18.301659
};
\addplot [darkslategray61, forget plot]
table {%
-0.2 100.51978375
-0.2 202.910948
};
\addplot [darkslategray61, forget plot]
table {%
-0.233333333333333 18.301659
-0.166666666666667 18.301659
};
\addplot [darkslategray61, forget plot]
table {%
-0.233333333333333 202.910948
-0.166666666666667 202.910948
};
\path [draw=darkslategray61, fill=viridis3]
(axis cs:-0.133333333333333,38.7588625)
--(axis cs:-5.55111512312578e-17,38.7588625)
--(axis cs:-5.55111512312578e-17,118.607093)
--(axis cs:-0.133333333333333,118.607093)
--(axis cs:-0.133333333333333,38.7588625)
--cycle;
\addplot [darkslategray61, forget plot]
table {%
-0.0666666666666667 38.7588625
-0.0666666666666667 18.568997
};
\addplot [darkslategray61, forget plot]
table {%
-0.0666666666666667 118.607093
-0.0666666666666667 238.259743
};
\addplot [darkslategray61, forget plot]
table {%
-0.1 18.568997
-0.0333333333333334 18.568997
};
\addplot [darkslategray61, forget plot]
table {%
-0.1 238.259743
-0.0333333333333334 238.259743
};
\path [draw=darkslategray61, fill=viridis3]
(axis cs:0.5333,38.65234975)
--(axis cs:0.6666,38.65234975)
--(axis cs:0.6666,60.12496775)
--(axis cs:0.5333,60.12496775)
--(axis cs:0.5333,38.65234975)
--cycle;
\addplot [darkslategray61, forget plot]
table {%
0.5999 38.65234975
0.5999 18.769472
};
\addplot [darkslategray61, forget plot]
table {%
0.5999 60.12496775
0.5999 92.267881
};
\addplot [darkslategray61, forget plot]
table {%
0.5666 18.769472
0.6333 18.769472
};
\addplot [darkslategray61, forget plot]
table {%
0.5666 92.267881
0.6333 92.267881
};
\path [draw=darkslategray61, fill=viridis4]
(axis cs:0,39.839296)
--(axis cs:0.133333333333333,39.839296)
--(axis cs:0.133333333333333,60.9285595)
--(axis cs:0,60.9285595)
--(axis cs:0,39.839296)
--cycle;
\addplot [darkslategray61, forget plot]
table {%
0.0666666666666667 39.839296
0.0666666666666667 20.375905
};
\addplot [darkslategray61, forget plot]
table {%
0.0666666666666667 60.9285595
0.0666666666666667 92.51792
};
\addplot [darkslategray61, forget plot]
table {%
0.0333333333333333 20.375905
0.1 20.375905
};
\addplot [darkslategray61, forget plot]
table {%
0.0333333333333333 92.51792
0.1 92.51792
};
\path [draw=darkslategray61, fill=viridis4]
(axis cs:0.133333333333333,38.749791)
--(axis cs:0.266666666666667,38.749791)
--(axis cs:0.266666666666667,131.945618)
--(axis cs:0.133333333333333,131.945618)
--(axis cs:0.133333333333333,38.749791)
--cycle;
\addplot [darkslategray61, forget plot]
table {%
0.2 38.749791
0.2 19.8179
};
\addplot [darkslategray61, forget plot]
table {%
0.2 131.945618
0.2 271.391127
};
\addplot [darkslategray61, forget plot]
table {%
0.166666666666667 19.8179
0.233333333333333 19.8179
};
\addplot [darkslategray61, forget plot]
table {%
0.166666666666667 271.391127
0.233333333333333 271.391127
};
\path [draw=darkslategray61, fill=viridis4]
(axis cs:0.266666666666667,39.2333455)
--(axis cs:0.4,39.2333455)
--(axis cs:0.4,197.0003185)
--(axis cs:0.266666666666667,197.0003185)
--(axis cs:0.266666666666667,39.2333455)
--cycle;
\addplot [darkslategray61, forget plot]
table {%
0.333333333333333 39.2333455
0.333333333333333 18.537708
};
\addplot [darkslategray61, forget plot]
table {%
0.333333333333333 197.0003185
0.333333333333333 433.555602
};
\addplot [darkslategray61, forget plot]
table {%
0.3 18.537708
0.366666666666667 18.537708
};
\addplot [darkslategray61, forget plot]
table {%
0.3 433.555602
0.366666666666667 433.555602
};
\addplot [darkslategray61, forget plot]
table {%
-0.4 56.892957
-0.266666666666667 56.892957
};
\addplot [darkslategray61, forget plot]
table {%
-0.266666666666667 41.4764315
-0.133333333333333 41.4764315
};
\addplot [darkslategray61, forget plot]
table {%
-0.133333333333333 57.250231
-5.55111512312578e-17 57.250231
};
\addplot [darkslategray61, forget plot]
table {%
0.5333 40.089247
0.6666 40.089247
};
\addplot [darkslategray61, forget plot]
table {%
0 40.1876565
0.133333333333333 40.1876565
};
\addplot [darkslategray61, forget plot]
table {%
0.133333333333333 58.668787
0.266666666666667 58.668787
};
\addplot [darkslategray61, forget plot]
table {%
0.266666666666667 77.357387
0.4 77.357387
};

\draw[dashed, thick, red] (axis cs: -0.52,50) -- (axis cs: 0.786,50);
\node [red] at (-0.45, 52) {\footnotesize $\tau$};

\addplot [
draw=black, mark=*, only marks, mark options={fill=white}]
coordinates {
    (-0.333333333333333,143.474593) 
};

\addplot [
draw=black, mark=*, only marks, mark options={fill=white}]
coordinates {
    (-0.2,108.354621) 
};

\addplot [
draw=black, mark=*, only marks, mark options={fill=white}]
coordinates {
    (-0.0666666666666667,117.815143) 
};

\addplot [
draw=black, mark=*, only marks, mark options={fill=white}]
coordinates {
    (0.0666666666666667,83.587268) 
};

\addplot [
draw=black, mark=*, only marks, mark options={fill=white}]
coordinates {
    (0.2,123.399945) 
};

\addplot [
draw=black, mark=*, only marks, mark options={fill=white}]
coordinates {
    (0.333333333333333,177.389481) 
};

\addplot [
draw=black, mark=*, only marks, mark options={fill=white}]
coordinates {
    (0.5999,71.908685) 
};

\node [black] at (0.0666666666666667, -5) {\tiny$C_1$};
\node [black] at (0.2, -5) {\tiny$C_2$};
\node [black] at (0.333333333333333 , -5) {\tiny$C_3$};

\nextgroupplot[
legend style={legend cell align=left,
              align=center,
              draw=white!15!black,
              at={(0.5,1.45)},
              anchor=center,
              /tikz/every even column/.append style={column sep=1em},
legend columns=1,
anchor=north,legend rows=2},
tick align=outside,
tick pos=left,
x grid style={darkgray176},
xmin=-0.56, xmax=0.28,
xtick style={color=black},
xtick={-0.24,0.16},
xticklabel style={align=center},
xticklabels = {\hspace{-0.5cm} Centralized, Federated},
clip=false,
y grid style={darkgray176},
xticklabel shift={3pt},
ylabel={$P[{\ell} \leq \tau]$},
ymajorgrids,
ymin=0, ymax=1,
xtick style={draw=none},
ytick style={color=black},
]
\draw[draw=black,fill=viridis1] (axis cs:-0.48,0) rectangle (axis cs:-0.40,0.475405);
\draw[draw=black,fill=viridis2] (axis cs:-0.4,0) rectangle (axis cs:-0.32,0.551517);
\draw[draw=black,fill=viridis3] (axis cs:-0.32,0) rectangle (axis cs:-0.24,0.477923);
\draw[draw=black,fill=viridis4, fill opacity=1.0] (axis cs:-0.24,0) rectangle (axis cs:-0.16,0.615085);
\draw[draw=black,fill=viridis4, fill opacity=1.0] (axis cs:-0.16,0) rectangle (axis cs:-0.08,0.440850);
\draw[draw=black,fill=viridis4, fill opacity=1.0] (axis cs:-0.08,0) rectangle (axis cs:0,0.264069);
\draw[draw=black,fill=viridis3] (axis cs:0.12,0) rectangle (axis cs:0.20,0.628105);

\node [black] at (-0.21, -0.04) {\tiny$C_1$};
\node [black] at (-0.12, -0.04) {\tiny$C_2$};
\node [black] at (-0.03 , -0.04) {\tiny$C_3$};
\end{groupplot}

\end{tikzpicture}

%% file: B_30_d.tex
% This file was created with tikzplotlib v0.10.1.
\pgfplotsset{
tick label style={font=\footnotesize},
label style={font=\footnotesize},
legend  style={font=\footnotesize}
}
\begin{tikzpicture}

\begin{groupplot}[
    group style={
        group size=2 by 1, % One column, two rows
        horizontal sep=1.5cm, % Vertical spacing between the plots
    },
    width=0.5\columnwidth, % Adjust the width of the plots
    height=0.7\columnwidth % Adjust the height of the plots
]

\definecolor{viridis1}{RGB}{53, 95, 141}
\definecolor{viridis2}{RGB}{33, 145, 140}
\definecolor{viridis3}{RGB}{94, 201, 98}
\definecolor{viridis4}{RGB}{253, 231, 37}

\definecolor{darkgray176}{RGB}{176,176,176}
\definecolor{black}{RGB}{61,61,61}
\definecolor{dimgray1319183}{RGB}{131,91,83}
\definecolor{lightgray204}{RGB}{204,204,204}
\definecolor{darkslategray61}{RGB}{61,61,61}

\nextgroupplot[
legend cell align={left},
legend style={fill opacity=0.8, draw opacity=1, text opacity=1, draw=lightgray204},
tick align=outside,
tick pos=left,
x grid style={darkgray176},
xlabel={},
xmin=-0.52, xmax=0.786,
xtick style={color=black},
xtick={0,0.5999},
xtick style={draw=none},
xticklabel style={align=center},
xticklabels = {\hspace{-0.5cm} Centralized, Federated},
y grid style={darkgray176},
ymin=10, ymax=450,
ylabel={E2E latency [ms]},
ymajorgrids,
clip=false,
xticklabel shift={3pt},
ytick style={color=black},
]

\path [draw=darkslategray61, postaction={pattern=north east lines}, fill=viridis1]
(axis cs:-0.4,34.7501235)
--(axis cs:-0.266666666666667,34.7501235)
--(axis cs:-0.266666666666667,104.356669)
--(axis cs:-0.4,104.356669)
--(axis cs:-0.4,34.7501235)
--cycle;
\addplot [darkslategray61, forget plot]
table {%
-0.333333333333333 34.7501235
-0.333333333333333 18.522695
};
\addplot [darkslategray61, forget plot]
table {%
-0.333333333333333 104.356669
-0.333333333333333 208.705203
};
\addplot [darkslategray61, forget plot]
table {%
-0.366666666666667 18.522695
-0.3 18.522695
};
\addplot [darkslategray61, forget plot]
table {%
-0.366666666666667 208.705203
-0.3 208.705203
};
\path [draw=darkslategray61, postaction={pattern=north east lines}, fill=viridis2]
(axis cs:-0.266666666666667,28.054681)
--(axis cs:-0.133333333333333,28.054681)
--(axis cs:-0.133333333333333,79.624669)
--(axis cs:-0.266666666666667,79.624669)
--(axis cs:-0.266666666666667,28.054681)
--cycle;
\addplot [darkslategray61, forget plot]
table {%
-0.2 28.054681
-0.2 18.312126
};
\addplot [darkslategray61, forget plot]
table {%
-0.2 79.624669
-0.2 156.973223
};
\addplot [darkslategray61, forget plot]
table {%
-0.233333333333333 18.312126
-0.166666666666667 18.312126
};
\addplot [darkslategray61, forget plot]
table {%
-0.233333333333333 156.973223
-0.166666666666667 156.973223
};
\path [draw=darkslategray61, postaction={pattern=north east lines}, fill=viridis3]
(axis cs:-0.133333333333333,38.964332)
--(axis cs:-5.55111512312578e-17,38.964332)
--(axis cs:-5.55111512312578e-17,99.83036)
--(axis cs:-0.133333333333333,99.83036)
--(axis cs:-0.133333333333333,38.964332)
--cycle;
\addplot [darkslategray61, forget plot]
table {%
-0.0666666666666667 38.964332
-0.0666666666666667 18.529606
};
\addplot [darkslategray61, forget plot]
table {%
-0.0666666666666667 99.83036
-0.0666666666666667 191.112658
};
\addplot [darkslategray61, forget plot]
table {%
-0.1 18.529606
-0.0333333333333334 18.529606
};
\addplot [darkslategray61, forget plot]
table {%
-0.1 191.112658
-0.0333333333333334 191.112658
};
\path [draw=darkslategray61, postaction={pattern=north east lines}, fill=viridis3]
(axis cs:0.5333,38.875023)
--(axis cs:0.6666,38.875023)
--(axis cs:0.6666,60.116081)
--(axis cs:0.5333,60.116081)
--(axis cs:0.5333,38.875023)
--cycle;
\addplot [darkslategray61, forget plot]
table {%
0.5999 38.875023
0.5999 18.184975
};
\addplot [darkslategray61, forget plot]
table {%
0.5999 60.116081
0.5999 91.926409
};
\addplot [darkslategray61, forget plot]
table {%
0.5666 18.184975
0.6333 18.184975
};
\addplot [darkslategray61, forget plot]
table {%
0.5666 91.926409
0.6333 91.926409
};
\path [draw=darkslategray61, postaction={pattern=north east lines}, fill=viridis4]
(axis cs:0,39.83929725)
--(axis cs:0.133333333333333,39.83929725)
--(axis cs:0.133333333333333,61.00675025)
--(axis cs:0,61.00675025)
--(axis cs:0,39.83929725)
--cycle;
\addplot [darkslategray61, forget plot]
table {%
0.0666666666666667 39.83929725
0.0666666666666667 20.375905
};
\addplot [darkslategray61, forget plot]
table {%
0.0666666666666667 61.00675025
0.0666666666666667 92.749885
};
\addplot [darkslategray61, forget plot]
table {%
0.0333333333333333 20.375905
0.1 20.375905
};
\addplot [darkslategray61, forget plot]
table {%
0.0333333333333333 92.749885
0.1 92.749885
};
\path [draw=darkslategray61, postaction={pattern=north east lines}, fill=viridis4]
(axis cs:0.133333333333333,38.7497765)
--(axis cs:0.266666666666667,38.7497765)
--(axis cs:0.266666666666667,125.72897375)
--(axis cs:0.133333333333333,125.72897375)
--(axis cs:0.133333333333333,38.7497765)
--cycle;
\addplot [darkslategray61, forget plot]
table {%
0.2 38.7497765
0.2 19.8179
};
\addplot [darkslategray61, forget plot]
table {%
0.2 125.72897375
0.2 255.499947
};
\addplot [darkslategray61, forget plot]
table {%
0.166666666666667 19.8179
0.233333333333333 19.8179
};
\addplot [darkslategray61, forget plot]
table {%
0.166666666666667 255.499947
0.233333333333333 255.499947
};
\path [draw=darkslategray61,postaction={pattern=north east lines}, fill=viridis4]
(axis cs:0.266666666666667,38.906027)
--(axis cs:0.4,38.906027)
--(axis cs:0.4,196.946148)
--(axis cs:0.266666666666667,196.946148)
--(axis cs:0.266666666666667,38.906027)
--cycle;
\addplot [darkslategray61, forget plot]
table {%
0.333333333333333 38.906027
0.333333333333333 18.537708
};
\addplot [darkslategray61, forget plot]
table {%
0.333333333333333 196.946148
0.333333333333333 433.79206
};
\addplot [darkslategray61, forget plot]
table {%
0.3 18.537708
0.366666666666667 18.537708
};
\addplot [darkslategray61, forget plot]
table {%
0.3 433.79206
0.366666666666667 433.79206
};
\addplot [darkslategray61, forget plot]
table {%
-0.4 42.24998
-0.266666666666667 42.24998
};
\addplot [darkslategray61, forget plot]
table {%
-0.266666666666667 40.027001
-0.133333333333333 40.027001
};
\addplot [darkslategray61, forget plot]
table {%
-0.133333333333333 40.570635
-5.55111512312578e-17 40.570635
};
\addplot [darkslategray61, forget plot]
table {%
0.5333 40.1240855
0.6666 40.1240855
};
\addplot [darkslategray61, forget plot]
table {%
0 40.1954145
0.133333333333333 40.1954145
};
\addplot [darkslategray61, forget plot]
table {%
0.133333333333333 58.660721
0.266666666666667 58.660721
};
\addplot [darkslategray61, forget plot]
table {%
0.266666666666667 77.339295
0.4 77.339295
};

\draw[dashed, thick, red] (axis cs: -0.52,30) -- (axis cs: 0.786,30);
\node [red] at (-0.45, 32) {\footnotesize $\tau$};

\addplot [
draw=black, mark=*, only marks, mark options={fill=white}]
coordinates {
    (-0.333333333333333,108.536766) 
};

\addplot [
draw=black, mark=*, only marks, mark options={fill=white}]
coordinates {
    (-0.2, 90.660583) 
};

\addplot [
draw=black, mark=*, only marks, mark options={fill=white}]
coordinates {
    (-0.0666666666666667,108.645856) 
};

\addplot [
draw=black, mark=*, only marks, mark options={fill=white}]
coordinates {
    (0.0666666666666667,83.437447) 
};

\addplot [
draw=black, mark=*, only marks, mark options={fill=white}]
coordinates {
    (0.2,122.988198) 
};

\addplot [
draw=black, mark=*, only marks, mark options={fill=white}]
coordinates {
    (0.333333333333333,175.965340) 
};

\addplot [
draw=black, mark=*, only marks, mark options={fill=white}]
coordinates {
    (0.5999,71.508381) 
};

\node [black] at (0.0666666666666667, -5) {\tiny$C_1$};
\node [black] at (0.2, -5) {\tiny$C_2$};
\node [black] at (0.333333333333333 , -5) {\tiny$C_3$};

\nextgroupplot[
legend style={legend cell align=left,
              align=center,
              draw=white!15!black,
              at={(0.5,1.45)},
              anchor=center,
              /tikz/every even column/.append style={column sep=1em},
legend columns=1,
anchor=north,legend rows=2},
tick align=outside,
tick pos=left,
x grid style={darkgray176},
xmin=-0.56, xmax=0.28,
xtick style={color=black},
xtick={-0.24,0.16},
xticklabel style={align=center},
xticklabels = {\hspace{-0.5cm} Centralized, Federated},
clip=false,
y grid style={darkgray176},
xticklabel shift={3pt},
ylabel={$P[{\ell} \leq \tau]$},
ymajorgrids,
ymin=0, ymax=1,
xtick style={draw=none},
ytick style={color=black},
]
\draw[draw=black,postaction={pattern=north east lines}, fill=viridis1] (axis cs:-0.48,0) rectangle (axis cs:-0.40,0.220337);
\draw[draw=black,postaction={pattern=north east lines}, fill=viridis2] (axis cs:-0.4,0) rectangle (axis cs:-0.32,0.265861);
\draw[draw=black,postaction={pattern=north east lines}, fill=viridis3] (axis cs:-0.32,0) rectangle (axis cs:-0.24,0.136759);
\draw[draw=black,postaction={pattern=north east lines}, fill=viridis4, fill opacity=1.0] (axis cs:-0.24,0) rectangle (axis cs:-0.16,0.176026);
\draw[draw=black,postaction={pattern=north east lines}, fill=viridis4, fill opacity=1.0] (axis cs:-0.16,0) rectangle (axis cs:-0.08,0.093114);
\draw[draw=black,postaction={pattern=north east lines}, fill=viridis4, fill opacity=1.0] (axis cs:-0.08,0) rectangle (axis cs:0,0.048383);
\draw[draw=black,postaction={pattern=north east lines}, fill=viridis3] (axis cs:0.12,0) rectangle (axis cs:0.20,0.170700);

\node [black] at (-0.21, -0.04) {\tiny$C_1$};
\node [black] at (-0.12, -0.04) {\tiny$C_2$};
\node [black] at (-0.03 , -0.04) {\tiny$C_3$};
\end{groupplot}

\end{tikzpicture}

%% file: G_MA_r.tex
% This file was created with tikzplotlib v0.10.1.

\pgfplotsset{
tick label style={font=\footnotesize},
label style={font=\footnotesize},
legend  style={font=\footnotesize}
}

\begin{tikzpicture}

\begin{groupplot}[
    group style={
        group size=1 by 2, % One column, two rows
        vertical sep=0.25cm, % Vertical spacing between the plots
    },
    width=\columnwidth, % Adjust the width of the plots
    height=0.4\columnwidth % Adjust the height of the plots
]

\definecolor{magma1}{RGB}{254, 159, 109}
\definecolor{magma2}{RGB}{222, 73, 104}
\definecolor{magma3}{RGB}{140, 41, 129}
\definecolor{magma4}{RGB}{59, 15, 112}
\definecolor{darkgray176}{RGB}{176,176,176}

\nextgroupplot[
legend style={legend cell align=left,
              align=center,
              draw=white!15!black,
              at={(0.5,1.45)},
              anchor=center,
              /tikz/every even column/.append style={column sep=1em},
legend columns=1,
anchor=north,legend rows=2},
tick align=outside,
tick pos=left,
x grid style={darkgray176},
xtick style={color=black},
xtick={0,1,2,3,4,5,6,7},
xmin=-0.6, xmax=6.6,
y grid style={darkgray176},
ytick style={color=black},
ytick={0.45,0.50,0.55},
ylabel={Reward},
ymajorgrids,
ymin=0.45, ymax=0.55,
xticklabel=\empty 
]
\draw[draw=black,fill=magma1] (axis cs:-0.4,0) rectangle (axis cs:0.4,0.4632574544469);
\draw[draw=black,fill=magma1] (axis cs:0.6,0) rectangle (axis cs:1.4, 0.49745388121639);
\draw[draw=black,fill=magma2] (axis cs:1.6,0) rectangle (axis cs:2.4,0.503323533189428);
\draw[draw=black,fill=magma2] (axis cs:2.6,0) rectangle (axis cs:3.4,0.523250693021421);
\draw[draw=black,fill=magma3] (axis cs:3.6,0) rectangle (axis cs:4.4,0.497211963551661);
\draw[draw=black,fill=magma3] (axis cs:4.6,0) rectangle (axis cs:5.4,0.517022248289555);
\draw[draw=black,fill=magma4] (axis cs:5.6,0) rectangle (axis cs:6.4,0.529296904155714);

\draw[black] (axis cs: -0.6,0) -- (axis cs: 8,0);

\nextgroupplot[
name=bottomaxis,
after end axis/.append code={
  % ---- square bracket under the axis (overlay, no bbox impact) ----
\draw[overlay,thick] (axis cs:2,-0.14) -- (axis cs:6,-0.14);  % horizontal line
  \draw[overlay,thick] (axis cs:2,-0.14) -- (axis cs:2,-0.11); % left short tick
  \draw[overlay,thick] (axis cs:6,-0.14) -- (axis cs:6,-0.11); % right short tick
  \node[overlay,font=\footnotesize] at (axis cs:4,-0.17) {Meta-learning};
},
legend style={legend cell align=left,
              align=center,
              draw=white!15!black,
              at={(0.5,1.45)},
              anchor=center,
              /tikz/every even column/.append style={column sep=1em},
legend columns=1,
anchor=north,legend rows=2},
tick align=outside,
tick pos=left,
x grid style={darkgray176},
xtick style={color=black},
xtick={0,1,2,3,4,5,6,7},
xticklabels={ICS,C(F),R,EG,LTS,NLTS,DDQL},
xmin=-0.6, xmax=6.6,
y grid style={darkgray176},
ytick style={color=black},
ylabel={Reward},
ymajorgrids,
ymin=0, ymax=0.3,
]

\draw[draw=black,fill=magma1] (axis cs:-0.4,0) rectangle (axis cs:0.4,0.00842236536220625);
\draw[draw=black,fill=magma1] (axis cs:0.6,0) rectangle (axis cs:1.4, 0.1969451721679);
\draw[draw=black,fill=magma2] (axis cs:1.6,0) rectangle (axis cs:2.4,0.18783596500883);
\draw[draw=black,fill=magma2] (axis cs:2.6,0) rectangle (axis cs:3.4,0.172560713991722);
\draw[draw=black,fill=magma3] (axis cs:3.6,0) rectangle (axis cs:4.4,0.196331066880203);
\draw[draw=black,fill=magma3] (axis cs:4.6,0) rectangle (axis cs:5.4,0.239161884989458);
\draw[draw=black,fill=magma4] (axis cs:5.6,0) rectangle (axis cs:6.4,0.223576654108235);

\draw[black] (axis cs: -0.6,0) -- (axis cs: 8,0);

\draw[
    decorate,
    decoration={brace, mirror, amplitude=5pt},
    thick
]
(axis cs:1.6,-0.1) -- (axis cs:6.4,-0.1)
node[midway, yshift=-10pt, font=\footnotesize] {Meta-learning agents};

\end{groupplot}

\end{tikzpicture}

%% file: G_MA_mAP.tex
% This file was created with tikzplotlib v0.10.1.

\pgfplotsset{
tick label style={font=\footnotesize},
label style={font=\footnotesize},
legend  style={font=\footnotesize}
}

\begin{tikzpicture}

\begin{groupplot}[
    group style={
        group size=1 by 2, % One column, two rows
        vertical sep=0.25cm, % Vertical spacing between the plots
    },
    width=\columnwidth, % Adjust the width of the plots
    height=0.4\columnwidth % Adjust the height of the plots
]

\definecolor{magma1}{RGB}{254, 159, 109}
\definecolor{magma2}{RGB}{222, 73, 104}
\definecolor{magma3}{RGB}{140, 41, 129}
\definecolor{magma4}{RGB}{59, 15, 112}
\definecolor{darkgray176}{RGB}{176,176,176}

\nextgroupplot[
legend style={legend cell align=left,
              align=center,
              draw=white!15!black,
              at={(0.5,1.45)},
              anchor=center,
              /tikz/every even column/.append style={column sep=1em},
legend columns=1,
anchor=north,legend rows=2},
tick align=outside,
tick pos=left,
x grid style={darkgray176},
xtick style={color=black},
xtick={0,1,2,3,4,5,6,7},
xmin=-0.6, xmax=6.6,
y grid style={darkgray176},
ytick style={color=black},
ylabel={mAP},
ymajorgrids,
ymin=0.62, ymax=0.68,
xticklabel=\empty 
]
\draw[draw=black,fill=magma1] (axis cs:-0.4,0) rectangle (axis cs:0.4,0.667060073641876);
\draw[draw=black,fill=magma1] (axis cs:0.6,0) rectangle (axis cs:1.4, 0.663747902657744);
\draw[draw=black,fill=magma2] (axis cs:1.6,0) rectangle (axis cs:2.4,0.648024954863799);
\draw[draw=black,fill=magma2] (axis cs:2.6,0) rectangle (axis cs:3.4,0.643297366141825);
\draw[draw=black,fill=magma3] (axis cs:3.6,0) rectangle (axis cs:4.4,0.671331915580321);
\draw[draw=black,fill=magma3] (axis cs:4.6,0) rectangle (axis cs:5.4,0.64078370707768);
\draw[draw=black,fill=magma4] (axis cs:5.6,0) rectangle (axis cs:6.4,0.637144494065403);

\draw[black] (axis cs: -0.6,0) -- (axis cs: 8,0);

\nextgroupplot[
legend style={legend cell align=left,
              align=center,
              draw=white!15!black,
              at={(0.5,1.45)},
              anchor=center,
              /tikz/every even column/.append style={column sep=1em},
legend columns=1,
anchor=north,legend rows=2},
tick align=outside,
tick pos=left,
x grid style={darkgray176},
xtick style={color=black},
xtick={0,1,2,3,4,5,6,7},
xticklabels={ICS,C(F),R,EG,LTS,NLTS,DDQL},
xmin=-0.6, xmax=6.6,
y grid style={darkgray176},
ytick style={color=black},
ylabel={mAP},
ymajorgrids,
ymin=0.45, ymax=0.65,
]

\draw[draw=black,fill=magma1] (axis cs:-0.4,0) rectangle (axis cs:0.4,0.611142797343976);
\draw[draw=black,fill=magma1] (axis cs:0.6,0) rectangle (axis cs:1.4, 0.54555017);
\draw[draw=black,fill=magma2] (axis cs:1.6,0) rectangle (axis cs:2.4,0.555952284465014);
\draw[draw=black,fill=magma2] (axis cs:2.6,0) rectangle (axis cs:3.4,0.617408520789993);
\draw[draw=black,fill=magma3] (axis cs:3.6,0) rectangle (axis cs:4.4,0.577260654622145);
\draw[draw=black,fill=magma3] (axis cs:4.6,0) rectangle (axis cs:5.4,0.479362345668414);
\draw[draw=black,fill=magma4] (axis cs:5.6,0) rectangle (axis cs:6.4,0.490499726131601);

\draw[black] (axis cs: -0.6,0) -- (axis cs: 8,0);

\end{groupplot}

\end{tikzpicture}

%% file: G_MA_d.tex
% This file was created with tikzplotlib v0.10.1.
\pgfplotsset{
tick label style={font=\footnotesize},
label style={font=\footnotesize},
legend  style={font=\footnotesize}
}

\begin{tikzpicture}

\definecolor{magma1}{RGB}{254, 159, 109}
\definecolor{magma2}{RGB}{222, 73, 104}
\definecolor{magma3}{RGB}{140, 41, 129}
\definecolor{magma4}{RGB}{59, 15, 112}
\definecolor{darkgray176}{RGB}{176,176,176}
\definecolor{darkgray176}{RGB}{176,176,176}
\definecolor{darkslategray63}{RGB}{63,63,63}
\definecolor{steelblue49115161}{RGB}{49,115,161}

\begin{groupplot}[
    group style={
        group size=2 by 1, % One column, two rows
        horizontal sep=0.25cm, % Vertical spacing between the plots
    },
    width=0.55\columnwidth, % Adjust the width of the plots
    height=0.7\columnwidth % Adjust the height of the plots
]

\nextgroupplot[
log basis y={10},
minor ytick={2,3,4,5,6,7,8,9,20,30,40,50,60,70,80,90,200,300,400,500,600,700,800,900,2000,3000,4000,5000,6000,7000,8000,9000,20000,30000,40000,50000,60000,70000,80000,90000},
tick align=outside,
tick pos=left,
x grid style={darkgray176},
xmin=-0.75, xmax=6.75,
xtick style={color=black},
xtick={0,1,2,3,4,5,6},
xticklabels={ICS,C(F),R,EG,LTS,NLTS,DDQL},
xticklabel style={rotate=90},
y grid style={darkgray176},
ylabel={E2E latency [ms]},
ymin=20, ymax=200,
ymode=log,
ytick style={color=black},
ytick={1,10,100,1000,10000},
ymajorgrids,
yticklabels={
  \(\displaystyle {10^{0}}\),
  \(\displaystyle {10^{1}}\),
  \(\displaystyle {10^{2}}\),
  \(\displaystyle {10^{3}}\),
  \(\displaystyle {10^{4}}\)
}
]
\path [draw=darkslategray63, fill=magma1]
(axis cs:-0.4,23.5436618633127)
--(axis cs:0.4,23.5436618633127)
--(axis cs:0.4,23.9501498207175)
--(axis cs:-0.4,23.9501498207175)
--(axis cs:-0.4,23.5436618633127)
--cycle;
\addplot [darkslategray63]
table {%
0 23.5436618633127
0 23.136739114
};
\addplot [darkslategray63]
table {%
0 23.9501498207175
0 24.0524975745
};
\addplot [darkslategray63]
table {%
-0.2 23.136739114
0.2 23.136739114
};
\addplot [darkslategray63]
table {%
-0.2 24.0524975745
0.2 24.0524975745
};
\addplot [black, mark=o, mark size=2, mark options={solid,fill opacity=0,draw=darkslategray63}, only marks]
table {%
0 22.706275231
0 22.794673463
0 22.6540059825
0 116.373047381538
0 156.419234255958
0 153.196467238542
0 120.44647808766
0 25.095609399
0 106.148341349813
0 25.13725092
0 114.109305112258
0 81.7694493400652
0 101.810414852174
};
\path [draw=darkslategray63, fill=magma1]
(axis cs:0.6,28.5886582843698)
--(axis cs:1.4,28.5886582843698)
--(axis cs:1.4,29.5648102610597)
--(axis cs:0.6,29.5648102610597)
--(axis cs:0.6,28.5886582843698)
--cycle;
\addplot [darkslategray63]
table {%
1 28.5886582843698
1 27.2461073925
};
\addplot [darkslategray63]
table {%
1 29.5648102610597
1 30.7294171825
};
\addplot [darkslategray63]
table {%
0.8 27.2461073925
1.2 27.2461073925
};
\addplot [darkslategray63]
table {%
0.8 30.7294171825
1.2 30.7294171825
};
\addplot [black, mark=o, mark size=2, mark options={solid,fill opacity=0,draw=darkslategray63}, only marks]
table {%
1 26.840777752
1 26.7238635885
1 26.973745367
1 77.1132085055
1 102.841326038
1 102.281000958572
1 77.933675218
1 72.9052078145
1 91.6599734095
1 75.47467097
1 66.4231530095
};
\path [draw=darkslategray63, fill=magma2]
(axis cs:1.6,24.3853768615966)
--(axis cs:2.4,24.3853768615966)
--(axis cs:2.4,25.3405843194635)
--(axis cs:1.6,25.3405843194635)
--(axis cs:1.6,24.3853768615966)
--cycle;
\addplot [darkslategray63]
table {%
2 24.3853768615966
2 23.4862738055972
};
\addplot [darkslategray63]
table {%
2 25.3405843194635
2 25.7724851812282
};
\addplot [darkslategray63]
table {%
1.8 23.4862738055972
2.2 23.4862738055972
};
\addplot [darkslategray63]
table {%
1.8 25.7724851812282
2.2 25.7724851812282
};
\addplot [black, mark=o, mark size=2, mark options={solid,fill opacity=0,draw=darkslategray63}, only marks]
table {%
2 93.0232710660475
2 82.5177581115
2 78.9708739795
2 77.182323436
2 68.6308648905
2 82.5832736205
2 75.1936636365
2 73.6201998575
};
\path [draw=darkslategray63, fill=magma2]
(axis cs:2.6,23.7459019443954)
--(axis cs:3.4,23.7459019443954)
--(axis cs:3.4,25.9319817215513)
--(axis cs:2.6,25.9319817215513)
--(axis cs:2.6,23.7459019443954)
--cycle;
\addplot [darkslategray63]
table {%
3 23.7459019443954
3 22.7384163954069
};
\addplot [darkslategray63]
table {%
3 25.9319817215513
3 27.3951778132801
};
\addplot [darkslategray63]
table {%
2.8 22.7384163954069
3.2 22.7384163954069
};
\addplot [darkslategray63]
table {%
2.8 27.3951778132801
3.2 27.3951778132801
};
\addplot [black, mark=o, mark size=2, mark options={solid,fill opacity=0,draw=darkslategray63}, only marks]
table {%
3 94.639377899
3 78.6527657695
3 73.0585680095
3 73.066103089
3 64.0753156485
3 73.7698770515
3 64.2068027015
3 68.868125662
};
\path [draw=darkslategray63, fill=magma3]
(axis cs:3.6,26.1145585349172)
--(axis cs:4.4,26.1145585349172)
--(axis cs:4.4,27.257350878872)
--(axis cs:3.6,27.257350878872)
--(axis cs:3.6,26.1145585349172)
--cycle;
\addplot [darkslategray63]
table {%
4 26.1145585349172
4 24.5334840024975
};
\addplot [darkslategray63]
table {%
4 27.257350878872
4 28.130057416
};
\addplot [darkslategray63]
table {%
3.8 24.5334840024975
4.2 24.5334840024975
};
\addplot [darkslategray63]
table {%
3.8 28.130057416
4.2 28.130057416
};
\addplot [black, mark=o, mark size=2, mark options={solid,fill opacity=0,draw=darkslategray63}, only marks]
table {%
4 92.847042022402
4 95.202306190078
4 100.9946816465
4 97.8249244175
4 71.1289272965
4 91.3109513965
4 76.212793251
4 84.9194696485
};
\path [draw=darkslategray63, fill=magma3]
(axis cs:4.6,26.2744316463802)
--(axis cs:5.4,26.2744316463802)
--(axis cs:5.4,27.2250441912517)
--(axis cs:4.6,27.2250441912517)
--(axis cs:4.6,26.2744316463802)
--cycle;
\addplot [darkslategray63]
table {%
5 26.2744316463802
5 25.1563296185
};
\addplot [darkslategray63]
table {%
5 27.2250441912517
5 28.230921025
};
\addplot [darkslategray63]
table {%
4.8 25.1563296185
5.2 25.1563296185
};
\addplot [darkslategray63]
table {%
4.8 28.230921025
5.2 28.230921025
};
\addplot [black, mark=o, mark size=2, mark options={solid,fill opacity=0,draw=darkslategray63}, only marks]
table {%
5 24.612164907
5 24.5681465925
5 24.686149538961
5 66.4337433275
5 67.7816015545
5 66.865217134
5 71.9126032835
5 67.673667714
5 86.5955606363636
5 81.232872994
5 79.6303096665
};
\path [draw=darkslategray63, fill=magma4]
(axis cs:5.6,23.7304508095164)
--(axis cs:6.4,23.7304508095164)
--(axis cs:6.4,24.9194425892054)
--(axis cs:5.6,24.9194425892054)
--(axis cs:5.6,23.7304508095164)
--cycle;
\addplot [darkslategray63]
table {%
6 23.7304508095164
6 22.8665813368316
};
\addplot [darkslategray63]
table {%
6 24.9194425892054
6 25.6704605555
};
\addplot [darkslategray63]
table {%
5.8 22.8665813368316
6.2 22.8665813368316
};
\addplot [darkslategray63]
table {%
5.8 25.6704605555
6.2 25.6704605555
};
\addplot [black, mark=o, mark size=2, mark options={solid,fill opacity=0,draw=darkslategray63}, only marks]
table {%
6 68.3274768764616
6 76.9918256295
6 69.288412625
6 71.172322214
6 62.8327490895
6 72.190574636
6 65.8705782385
6 68.5002643245
};
\addplot [darkslategray63]
table {%
-0.4 23.6621204306044
0.4 23.6621204306044
};
\addplot [darkslategray63]
table {%
0.6 28.8100232074444
1.4 28.8100232074444
};
\addplot [darkslategray63]
table {%
1.6 24.6582614367026
2.4 24.6582614367026
};
\addplot [darkslategray63]
table {%
2.6 24.0519227674032
3.4 24.0519227674032
};
\addplot [darkslategray63]
table {%
3.6 26.3096177436113
4.4 26.3096177436113
};
\addplot [darkslategray63]
table {%
4.6 26.4915850578566
5.4 26.4915850578566
};
\addplot [darkslategray63]
table {%
5.6 24.0363922124147
6.4 24.0363922124147
};

\nextgroupplot[
log basis y={10},
minor ytick={2,3,4,5,6,7,8,9,20,30,40,50,60,70,80,90,200,300,400,500,600,700,800,900,2000,3000,4000,5000,6000,7000,8000,9000,20000,30000,40000,50000,60000,70000,80000,90000},
tick align=outside,
tick pos=left,
x grid style={darkgray176},
xmin=-0.75, xmax=6.75,
xtick style={color=black},
xtick={0,1,2,3,4,5,6},
xticklabels={ICS,C(F),R,EG,LTS,NLTS,DDQL},
xticklabel style={rotate=90},
y grid style={darkgray176},
ylabel={},
ymin=20, ymax=200,
ymode=log,
ytick style={color=black},
ytick={1,10,100,1000,10000},
ymajorgrids,
yticklabels={}
]
\path [draw=darkslategray63, fill=magma1]
(axis cs:-0.4,23.5576064676673)
--(axis cs:0.4,23.5576064676673)
--(axis cs:0.4,109.798036294139)
--(axis cs:-0.4,109.798036294139)
--(axis cs:-0.4,23.5576064676673)
--cycle;
\addplot [darkslategray63]
table {%
0 23.5576064676673
0 22.731921177
};
\addplot [darkslategray63]
table {%
0 109.798036294139
0 155.17701154023
};
\addplot [darkslategray63]
table {%
-0.2 22.731921177
0.2 22.731921177
};
\addplot [darkslategray63]
table {%
-0.2 155.17701154023
0.2 155.17701154023
};
\path [draw=darkslategray63, fill=magma1]
(axis cs:0.6,28.8191897056289)
--(axis cs:1.4,28.8191897056289)
--(axis cs:1.4,68.3307346142641)
--(axis cs:0.6,68.3307346142641)
--(axis cs:0.6,28.8191897056289)
--cycle;
\addplot [darkslategray63]
table {%
1 28.8191897056289
1 26.6372136115
};
\addplot [darkslategray63]
table {%
1 68.3307346142641
1 103.8588044225
};
\addplot [darkslategray63]
table {%
0.8 26.6372136115
1.2 26.6372136115
};
\addplot [darkslategray63]
table {%
0.8 103.8588044225
1.2 103.8588044225
};
\path [draw=darkslategray63, fill=magma2]
(axis cs:1.6,24.61413993989)
--(axis cs:2.4,24.61413993989)
--(axis cs:2.4,73.7112495651719)
--(axis cs:1.6,73.7112495651719)
--(axis cs:1.6,24.61413993989)
--cycle;
\addplot [darkslategray63]
table {%
2 24.61413993989
2 23.3481596711577
};
\addplot [darkslategray63]
table {%
2 73.7112495651719
2 99.8048204217578
};
\addplot [darkslategray63]
table {%
1.8 23.3481596711577
2.2 23.3481596711577
};
\addplot [darkslategray63]
table {%
1.8 99.8048204217578
2.2 99.8048204217578
};
\path [draw=darkslategray63, fill=magma2]
(axis cs:2.6,23.8551385307475)
--(axis cs:3.4,23.8551385307475)
--(axis cs:3.4,79.6977904927435)
--(axis cs:2.6,79.6977904927435)
--(axis cs:2.6,23.8551385307475)
--cycle;
\addplot [darkslategray63]
table {%
3 23.8551385307475
3 22.5698807311344
};
\addplot [darkslategray63]
table {%
3 79.6977904927435
3 104.467441068868
};
\addplot [darkslategray63]
table {%
2.8 22.5698807311344
3.2 22.5698807311344
};
\addplot [darkslategray63]
table {%
2.8 104.467441068868
3.2 104.467441068868
};
\path [draw=darkslategray63, fill=magma3]
(axis cs:3.6,23.5846400526234)
--(axis cs:4.4,23.5846400526234)
--(axis cs:4.4,76.1686824702224)
--(axis cs:3.6,76.1686824702224)
--(axis cs:3.6,23.5846400526234)
--cycle;
\addplot [darkslategray63]
table {%
4 23.5846400526234
4 22.7757108791812
};
\addplot [darkslategray63]
table {%
4 76.1686824702224
4 94.6346389584416
};
\addplot [darkslategray63]
table {%
3.8 22.7757108791812
4.2 22.7757108791812
};
\addplot [darkslategray63]
table {%
3.8 94.6346389584416
4.2 94.6346389584416
};
\path [draw=darkslategray63, fill=magma3]
(axis cs:4.6,23.5924370349033)
--(axis cs:5.4,23.5924370349033)
--(axis cs:5.4,63.4135843700026)
--(axis cs:4.6,63.4135843700026)
--(axis cs:4.6,23.5924370349033)
--cycle;
\addplot [darkslategray63]
table {%
5 23.5924370349033
5 22.7300039450824
};
\addplot [darkslategray63]
table {%
5 63.4135843700026
5 76.942881597
};
\addplot [darkslategray63]
table {%
4.8 22.7300039450824
5.2 22.7300039450824
};
\addplot [darkslategray63]
table {%
4.8 76.942881597
5.2 76.942881597
};
\path [draw=darkslategray63, fill=magma4]
(axis cs:5.6,24.5029006068142)
--(axis cs:6.4,24.5029006068142)
--(axis cs:6.4,65.9693176759181)
--(axis cs:5.6,65.9693176759181)
--(axis cs:5.6,24.5029006068142)
--cycle;
\addplot [darkslategray63]
table {%
6 24.5029006068142
6 23.2721194665
};
\addplot [darkslategray63]
table {%
6 65.9693176759181
6 75.0125242275
};
\addplot [darkslategray63]
table {%
5.8 23.2721194665
6.2 23.2721194665
};
\addplot [darkslategray63]
table {%
5.8 75.0125242275
6.2 75.0125242275
};
\addplot [darkslategray63]
table {%
-0.4 48.8577049454672
0.4 48.8577049454672
};
\addplot [darkslategray63]
table {%
0.6 41.9688326081943
1.4 41.9688326081943
};
\addplot [darkslategray63]
table {%
1.6 40.2250078993633
2.4 40.2250078993633
};
\addplot [darkslategray63]
table {%
2.6 43.1863153580994
3.4 43.1863153580994
};
\addplot [darkslategray63]
table {%
3.6 40.3183530786003
4.4 40.3183530786003
};
\addplot [darkslategray63]
table {%
4.6 37.0399971295469
5.4 37.0399971295469
};
\addplot [darkslategray63]
table {%
5.6 38.6309585713213
6.4 38.6309585713213
};

\end{groupplot}

\end{tikzpicture}